\documentclass[twocolumn]{aastex631}
\usepackage{longtable}
\usepackage{graphicx}
\usepackage{tabularx,booktabs}
\usepackage{appendix}
\usepackage{xcolor}
\usepackage{soul,color}
\newcommand{\hi}{H~{\sc i}}
\newcommand{\hii}{H~{\sc ii}}
\newcommand{\ha}{H\,$\alpha$}
\newcommand{\hb}{H\,$\beta$}
\newcommand{\cii}{C~{\sc ii}}
\newcommand{\ciii}{C~{\sc iii}}
\newcommand{\ciia}{C~{\sc ii}$^{*}$}
\newcommand{\nI}{N~{\sc i}}
\newcommand{\nii}{[N~{\sc ii}]}
\newcommand{\oi}{O~{\sc i}}
\newcommand{\oii}{[O~{\sc ii}]}
\newcommand{\oiii}{[O~{\sc iii}]}
\newcommand{\siII}{Si~{\sc ii}}
\newcommand{\siIIa}{Si~{\sc ii}$^{*}$}
\newcommand{\siIII}{Si~{\sc iii}}
\newcommand{\siIV}{Si~{\sc iv}}
\newcommand{\pii}{P~{\sc ii}}
\newcommand{\sii}{S~{\sc ii}}
\newcommand{\feii}{Fe~{\sc ii}}
\newcommand{\feiii}{Fe~{\sc iii}}
\newcommand{\niII}{Ni~{\sc ii}}

\newcommand{\foiii}{[O~{\sc iii}]}

\newcommand{\foi}{[O~{\sc i}]}
\newcommand{\foii}{[O~{\sc ii}]}
\newcommand{\fsii}{[S~{\sc ii}]}
\newcommand{\fsiii}{[S~{\sc iii}]}

\newcommand{\fciv}{C~{\sc iv}}

\newcommand{\fnii}{[N~{\sc ii}]}
\newcommand{\ffeii}{[Fe~{\sc ii}]}
\newcommand{\ffeiii}{[Fe~{\sc iii}]}
\newcommand{\ffeiv}{[Fe~{\sc iv}]}
\newcommand{\ffev}{[Fe~{\sc v}]}
\newcommand{\eld}{$N_{\rm e}$}

\newcommand{\elt}{$T_{\rm e}$}

\shorttitle{Multi-phase gas abundance distribution analysis of NGC~5253}
\shortauthors{Abril-Melgarejo et al.}
\graphicspath{{./}{figures/}}

\begin{document}

\title{Mapping Multi-Phase Metals in Star-forming Galaxies: a spatially resolved UV + Optical Study of NGC~5253}

\author[0000-0002-2764-6069]{Valentina Abril-Melgarejo}
\affiliation{Space Telescope Science Institute 3700 San Martin Drive Baltimore, MD 21218, USA}

\author[0000-0003-4372-2006]{Bethan L. James}
\affiliation{Space Telescope Science Institute 3700 San Martin Drive Baltimore, MD 21218, USA}
\affiliation{ESA for AURA, Space Telescope Science Institute, 3700 San Martin Drive, Baltimore, MD21218, USA}

\author[0000-0003-4137-882X]{Alessandra Aloisi}
\affiliation{Space Telescope Science Institute 3700 San Martin Drive Baltimore, MD 21218, USA}

\author[0000-0003-2589-762X]{Matilde Mingozzi}
\affiliation{Space Telescope Science Institute 3700 San Martin Drive Baltimore, MD 21218, USA}
\affiliation{ESA for AURA, Space Telescope Science Institute, 3700 San Martin Drive, Baltimore, MD21218, USA}

\author[0000-0002-7716-6223]{Vianney Lebouteiller}
\affiliation{Université Paris-Saclay, Université Paris Cité, CEA, CNRS, AIM, 91191, Gif-sur-Yvette, France}

\author[0000-0003-4857-8699]{Svea Hernandez}
\affiliation{ESA for AURA, Space Telescope Science Institute, 3700 San Martin Drive, Baltimore, MD21218, USA}

\author[0000-0002-5320-2568]{Nimisha Kumari}
\affiliation{ESA for AURA, Space Telescope Science Institute, 3700 San Martin Drive, Baltimore, MD21218, USA}


\collaboration{7}{(AAS Journals Data Editors)}



\begin{abstract}
We present a pioneering spatially-resolved, multi-phase gas abundance study on the blue compact dwarf galaxy NGC~5253, targeting 10 star-forming (SF) clusters inside six FUV HST/COS pointings with co-spatial optical VLT/MUSE observations throughout the galaxy. The SF regions span a wide range of ages (1--15 Myr) and are distributed at different radii (50 -- 230 pc). We performed robust absorption-line profile fitting on the COS spectra, covering 1065--1430 \AA\ in the FUV, allowing an accurate computation of neutral-gas abundances for 13 different ions sampling 8 elements. These values were then compared with the ionized-gas abundances, measured using the direct method on MUSE integrated spectra inside analog COS apertures. 
Our multi-phase, spatially resolved comparisons find abundances which are lower in the neutral gas than the ionized gas by 0.22 dex, 0.80 dex and 0.58 dex for log(O/H), log(N/H) and log(N/O), respectively. We modeled the chemical abundance distributions and evaluated correlations as a function of radius and age. It was found that while N, O and N/O abundances decrease as a function of age in the ionized gas, they increase with age in the neutral gas. No strong correlations for N, O or N/O were observed as a function of radius. The N/O and N/H offsets between the phases were found to decrease with age, providing evidence that chemical enrichment happens differentially, first in the ionized-gas phase around 2--5 Myrs (due to N-rich Wolf-Rayet stars) and then mixing out into the cold neutral gas on longer timescales of 10--15 Myr.

\end{abstract}





\section{Introduction} \label{sec:intro}

Metals are fundamental components of galaxy evolution. They are generated by stars forming and evolving within galaxies and, due to gas flows (supernovae explosions, stellar and galactic winds) are redistributed in and around galaxies resulting in a chemical enrichment at various time and spatial scales. Metals regulate the cooling of gas and the transport of momentum, enhancing/quenching star formation \citep[e.g.,][ and references therein]{Maiolino+2019}. Despite this essential role, we have a poor understanding of how metals are distributed among different gas phases through a galaxy. As a galaxy evolves, hot massive stars first ionize the gas around them, then the metals produced in their cores are ejected into the interstellar medium (ISM), and eventually, the enriched gas cools and mingles with the cold neutral gas. Thereby, we need to elucidate how metals mix both between and throughout phases in order to achieve a comprehensive understanding of the complex interplay of metals and their effects on star formation (SF).

In many theoretical models, metals are assumed to be homogeneously distributed in each phase, even though detailed hydrodynamical simulations of star-forming systems and the chemical enrichment of their gas, show significant inhomogeneities (e.g., \citealt{Ritter+2015}, \citealt{Corlies+2018}). Integral-Field Unit (IFU) observations of SFGs have confirmed that the distribution of metals of the hot (T$\sim$ 10$^{5}$ K) gas is not always homogeneous. Such studies reveal chemical inhomogeneities regulated by the interplay of outflows of metal-enriched gas from supernovae (SNe) explosions \citep{James+2013a}, the accretion of metal-poor gas due to interactions/mergers \citep{kumari+2017}, self-enrichment from winds of massive stars (\citealt{James+2009}, \citealt{Kumari+2018}), or starbursts episodes (\citealt{James+2010}, \citealt{Bresolin+2019}). However, these important discoveries remain constrained to the ionized gas phase and hence to the SF regions as opposed to the quiescent neutral gas reservoir.

Local, metal-poor blue dwarf galaxies (BDGs) are ideal laboratories to investigate metal enrichment processes and their distribution since they allow imaging and spectroscopic studies at pc scales. They also possess physical properties reminiscent of high-z systems in terms of stellar masses, clumpy irregular morphology, low metallicities and high SFRs (e.g., I Zw 18, \citealt{Lebouteiller+2013}, DDO 68, \citealt{Sacchi+2016} and the CLASSY survey, \citealt{Berg+2022}), thus the results derived from the study of their properties are useful to understand star-formation in the high-z universe. The bulk of the baryonic mass in BDGs is contained in neutral hydrogen (\hi) regions \citep{Lebouteiller+2013}, so it is expected that most of the metals reside in the neutral gas phase. 

In recent state-of-the-art high-resolution hydrodynamical simulations, the key role of chemical mixing in galaxy evolution scenarios has been probed at galaxy-scale \citep{Emerick+2020}. Such a model was developed by \cite{Emerick+2019} to explore the multi-phase distribution of 15 elements at high spatial resolution (1.8 pc) throughout the BDG Leo P, an isolated extremely metal-poor SFG. This study unveiled significant variations in the amount of global chemical homogeneity between the warm ionized and the cold neutral phases, mainly due to the dependency of ionized gas abundances on the SFR and the rate of chemical enrichment. This opens up a new approach to study chemical mixing processes in local high-z analogs, leading to important findings, like the high dependency between the rate of mixing per element and its nucleosynthetic origin.

Among BDGs, NGC~5253 (3.7 Mpc) is a suitable candidate to study multi-phase gas abundances for several reasons: i) it is a nearby star-forming galaxy (SFG), with observations at high spatial resolution (1$\arcsec$ $\sim$ 15.3 pc, \citealt{Calzetti+2015}), enough to target individual \hii\ regions and star clusters distributed across the galaxy (see Fig. \ref{fig_NGC5253_COS_MUSE}). ii) This galaxy has been observed in radio (VLA \citealt{Kobulnicky+2008}; ATCA, \citealt{Lopez-Sanchez+2012}) and UV \citep[COS,][]{James+2014} bands, which enabled to map the \hi\ distribution from the radio-continuum emission (at kpc scale) and to obtain the \hi\ abundance for two pointings by modeling the Ly$\alpha$ $\lambda$1216 UV absorption profile. iii) Thanks to the considerable amount of ancillary data (imaging and spectroscopic) available at different wavelengths for this galaxy, our knowledge of NGC~5253 is exceptionally detailed. Such data have led to studies finding a large range of stellar (e.g., 1–15 Myr SF clusters, very massive stars, Wolf-Rayet stars) and \hii\ region properties (e.g., localized N-enrichment in the ionized gas in its central clusters, \citealt{Westmoquette+2013}). This BDG has been imaged in more than 13 bands by HST \citep{Calzetti+2015} and mapped by IFS instruments, including VLT-FLAMES \citep{Monreal-Ibero+2012}, GMOS \citep{Westmoquette+2013} and VLT/MUSE. iv) Its properties are reminiscent of a SFG at z=2–3, with a metallicity of $\sim$0.35 Z$_{\odot}$ \citep{Monreal-Ibero+2012} and currently experiencing a starburst episode with a SFR$\sim$0.1 M$_{\odot}$/yr \citep{Calzetti+2015}. Hence, NGC~5253 is an ideal system for performing a comprehensive sampling of chemical mixing scenarios and the relation between the formation and evolution of massive stars and the gas properties. Moreover, NGC~5253 is part of the Centaurus A/M83 galaxy complex \citep{Karachentsev+2007}, in the M83 subgroup that includes the spiral galaxy M83 (NGC~5236) and the dwarf galaxies NGC~5264 and ESO-444-GO84 \citep{Koribalski+2004}, as the nearest neighbors. In this context, environmental implications can also be explored (see Section \ref{subsec:Abu_Dist_neutral_gas}). The main physical properties of NGC~5253 are summarized in Table \ref{Table_gen_NGC5253}.
 
\begin{deluxetable}{lcc}
\label{Table_gen_NGC5253}
\tablecaption{Physical characteristics of NGC~5253}
\tablehead{
\colhead{Property} & \colhead{ NGC~5253} & \colhead{References} } 
\startdata
Morphology                & Im pec \tablenotemark{a}   &  1   \\
z                         & 0.00136 $\pm$ 0.00001      &  2   \\
center position           & 13h 39m 55.9631s           &  3   \\
$\alpha$, $\delta$ (J2000) & -31d 38m 24.388s          &  3   \\
Distance(Mpc)             & 3.709 $\pm$ 0.439          &  4   \\
Inclination (deg)         & 85.4                       &  5   \\
B major, minor axis (')   & 5.13 , 2.34                &  6   \\
M$_{*}\times$10$^{8}$ M$_{\odot}$  & 11.4 $\pm$ 0.5    &  7   \\
M$_{\rm gas}\times$10$^{8}$ M$_{\odot}$  & 1.59        &  7   \\
12+log(O/H)               & 8.28                       &  8   \\
\enddata
\tablerefs{ (1) \citet{deVaucouleur+1991},  (2) \citet{Koribalski+2004}, (3) \citet{Turner+Beck2004}, (4) \citet{Kanbur+2003}, (5) HyperLeda,  (6) \citet{Lauberts+1989}, (7) \citet{Lopez-Sanchez+2012}, (8) \citet{Lopez-Sanchez+2007} } 
\tablenotetext{a}{Irregular/Peculiar, subtype Im: no spiral structure }
\end{deluxetable}

To address fundamental questions about the distribution of metals both between and within gas phases, we have conducted a multi-wavelength spatially resolved analysis on NGC~5253 using HST/COS UV and VLT/MUSE optical spectroscopic observations on six regions containing SF clusters and distributed all along the galaxy disk (see Figure \ref{fig_NGC5253_COS_MUSE}). The multiple UV pointings allow us to derive the neutral gas properties by modeling absorption features in the far UV (e.g., \citealt{James+2014}, \citealt{Hernandez+2020}). For the analysis of the ionized gas, we used optical integral field spectroscopy (IFS) in order to compute chemical abundances from nebular lines, in spatially matched spectra. The targeted SF clusters sample different ages with the aim to accurately explore the impact of stellar feedback, gas mixing timescales \citep[e.g.,][]{Westmoquette+2013} and ejection/enrichment mechanisms.

The structure of the paper is as follows. In Section \ref{sec:clusters_NGC5253}, we summarize the physical properties of the SF clusters targeted in NGC~5253 for this study. The COS-HST and MUSE-VLT observations used in this analysis, are described in Section \ref{sec:observations}. The co-addition procedure applied to COS spectra using different settings, the data analysis performed in order to fit the UV continuum in the spectra, the computation of the total Line Spread Function (LSF) and the rigorous absorption profile fitting are detailed in Section \ref{sec:data_analysis}. Section \ref{sec:results} is focused on the analysis of the distribution of abundances both in the neutral and ionized gas phases. The outputs of the multi-phase analysis and the respective comparison of results are discussed in Section \ref{sec:Discussion}. Concluding remarks are presented in Section \ref{sec:Conclusions}.

\section{Young star clusters in NGC~5253} \label{sec:clusters_NGC5253}

\begin{table*}[htb!]															
\begin{center}																			
\label{Table_Clusters}
\caption{Physical properties of the individual SF clusters in NGC~5253}
\begin{tabular}{cccccccccc}		
\hline\hline																			
ID$_{\rm Cluster}$\tablenotemark{a}	&	COS ID	&	D \tablenotemark{b}	&	Age$\rm_{~SED}$\tablenotemark{a}	&	Av. Age$\rm_{~COS}$	&	M$_{\star}$\tablenotemark{a}	&	Total M$_{\star}$	&	E(B-V)\tablenotemark{a}			&	E(B-V)$\rm_{COS}$			&	E(B-V)$_{\rm H II}$	\\
	&		&	pc	&	Myr	&	Myr	&	10$^4$ M$_{\odot}$	&	10$^4$ M$_{\odot}$	&	(mag)			&	(mag)			&	(mag) \\ \hline 
1	&	OBJ-1234	&	52.91	&	5$^{+1}_{-2}$	&	5.25$^{+2}_{-1}$	&	1.05$^{+0.28}_{-0.22}$	&	4.04$^{+0.68}_{-0.58}$ 	&	0.12	$\pm$	0.03	&	0.15	$\pm$	0.03	&	0.03	$\pm$	0.01	\\
2	&		&		&	5$^{+1}_{-2}$	&		&	0.91$^{-0.31}_{-0.22}$	&		&	0.08	$\pm$	0.03	&				&				\\
3	&		&		&	5$^{+4}_{-2}$	&		&	0.46$^{+0.11}_{-0.10}$	&		&	0.04	$\pm$	0.02	&				&				\\
4	&		&		&	6$^{+1}_{-2}$	&		&	1.62$^{+0.52}_{-0.48}$	&		&	0.32	$\pm$	0.04	&				&				\\
5	&	OBJ-5+11	&	23.11	&	1$^{+1}_{-1}$	&	1$^{+1}_{-1}$	&	7.46$^{+0.20}_{-0.27}$	&	32.96$^{+6.70}_{-4.21}$	&	0.46	$\pm$	0.04	&	0.46	$\pm$	0.05	&	0.27	$\pm$	0.01	\\
11	&		&		&	1$^{+1}_{-1}$	&		&	25.5$^{+6.7}_{-4.2}$	&		&	0.46$^{+0.06}_{-0.02}$			&				&				\\
6	&	OBJ-6	&	118.84	&	10$^{+8}_{-2}$	&	10$^{+8}_{-2}$	&	3.24$^{+1.33}_{-0.94}$	&	3.24$^{+1.33}_{-0.94}$	&	0.12$^{+0.04}_{-0.02}$			&	0.12$^{+0.04}_{-0.02}$			&	0.16	$\pm$	0.01	\\
7	&	OBJ-7	&	224.27	&	10$^{+4}_{-2}$	&	10$^{+4}_{-2}$	&	1.15$^{+0.30}_{-0.56}$	&	1.15$^{+0.30}_{-0.56}$	&	0.05$^{+0.04}_{-0.03}$			&	0.05$^{+0.04}_{-0.03}$			&	0.11	$\pm$	0.01	\\
8	&	OBJ-8	&	179.15	&	15$^{+4}_{-3}$	&	15$^{+4}_{-3}$	&	2.88$^{+1.64}_{-0.84}$	&	2.88$^{+1.64}_{-0.84}$	&	0.16	$\pm$	0.04	&	0.16	$\pm$	0.04	&	0.13	$\pm$	0.01	\\
9	&	OBJ-9+10	&	103.19	&	10$^{+5}_{-2}$	&	9.5$^{+6}_{-2}$	&	5.13$^{+2.12}_{-1.50}$	&	8.76$^{+3.86}_{-2.01}$	&	0.26$^{+0.06}_{-0.04}$			&	0.26$^{+0.05}_{-0.04}$			&	0.14	$\pm$	0.01	\\
10	&		&		&	9$^{+7}_{-2}$	&		&	3.63$^{+3.22}_{-1.34}$	&		&	0.26	$\pm$	0.04	&				&				\\
\hline
\end{tabular}
\end{center}

Summary of physical properties of the individual clusters in NGC~5253 derived by \cite{Calzetti+2015} and average/integrated properties computed for the SF regions within COS apertures. E(B-V)$_{HII}$ corresponds to the color excess determined from the Balmer decrement (F(\ha)/F(\hb)) of the computed optical emission line fluxes (see Section \ref{subsec:em_line_fitting}).\\
\textbf{Notes.}
\tablenotetext{a}{ Cluster ID and SED best-fit parameters (Age, stellar mass and color-excess E(B-V)) from \cite{Calzetti+2015}.}
\tablenotetext{b}{ Distance projected on the sky plane from the galaxy center to each of the COS apertures measured using the archival \textit{HST/}ACS SBC-F125LP UV image of NGC5253.}

\end{table*}																		  
NGC~5253 hosts several young star clusters varying in stellar mass and age, distributed along the stellar disk (see Figure \ref{fig_NGC5253_COS_MUSE}). These are the results of recent starburst episodes that occurred in this dwarf galaxy as inferred by \cite{Calzetti+2015} via spectral energy distribution (SED) fitting of eleven UV-bright star clusters over an ample photometric sample covering from the FUV to the NIR bands (1500 \AA -- 1.9 $\mu$m). The SED modeling performed by \cite{Calzetti+2015} includes the stellar continuum, gas emission, and dust attenuation spectral components, with the aim to robustly obtain best-fit age, stellar mass, and color excess for each cluster (summarized in Table \ref{Table_Clusters}). Since three of our pointings (OBJ-5+11, OBJ-9+10 and OBJ 1234) cover more than one cluster inside the COS apertures (2\farcs{5} in diameter), we computed their average ages, E(B-V) and total stellar mass with their respective averaged uncertainties (reported in Table \ref{Table_Clusters}).

The center of NGC~5253 hosts a giant compact \hii\ region. From radio-continuum emission measurements using 7 mm Very Large Array (VLA) observations, \citealt{Turner+Beck2004} found a compact bright core with dimensions 1.8 pc $\times$ 0.7 pc, a density of 3.5 $\times$ 10$^{4}$ cm$^{-3}$) and an estimated total mass of 4--6$\times$10$^5$ $M_{\odot}$ for the stellar clusters embedded inside the radio nebula. As it is reported in the NASA/IPAC Extragalactic Database (NED) we adopted the coordinates of the super radio nebula from \cite{Turner+Beck2004} as the center position of the galaxy (see Table \ref{Table_gen_NGC5253}). We then measured relative distances (projected onto the sky plane) from the center of the galaxy to the center of each COS aperture, using the archival ACS SBC-F125LP UV-image taken as part of COS observations in program ID 11579 (PI: Aloisi). The measured angular apertures are transformed into distances using the conversion factor 1$\arcsec$ $\sim$ 15.3 pc, as was established by \cite{Calzetti+2015}. Distances to the galactic center for each COS pointing are presented in Table \ref{Table_Clusters}. All observed clusters are very young ($<$ 15 -- 19 Myr) with ages correlating with the distance from the center, being the clusters in the central region the youngest and the farthest the oldest \citep{Calzetti+2015}, suggesting the starburst episodes occur outside-in. This could be related to the supply of fresh gas from the neutral gas structure detected in radio maps of NGC~5253 (ATCA \hi\ bands at high and low resolution), whose maximum \hi\ flux is located between OBJ-6 and OBJ-7, which is slightly shifted to the SE from the optical center \citep{Lopez-Sanchez+2012}. Molecular gas (CO) observations from ALMA were performed along the disk of NGC 5253 by \cite{Miura+2018}, who mapped and characterized 118 molecular clouds. They detected three molecular clouds with IDs 5, 8 and 28 that align with OBJ-1234, OBJ-5+11 and OBJ-9+10, respectfully, which are the youngest SF regions in our sample (see Figure 11 in \citealt{Miura+2018}). 

Clusters 5 and 11 are the youngest ($\sim$ 1 Myr) and the most massive of the sample, with stellar masses of 7.5$\times$10$^4$ $M_{\odot}$ and 25.5$\times$10$^4$ $M_{\odot}$, respectively \citep{Calzetti+2015}. These two clusters are located near the center of the galaxy (at a distance of $\sim$20 pc) and are embedded in a thermal radio emission structure (\citealt{Calzetti+2015}). This radio nebula coincides with the dust lane absorption which strongly affects cluster 11, considerably dimming its light. The other nine clusters are distributed in the starburst UV-bright region with low dust attenuation (A$_{v} \lesssim $ 1 mag), a mass range of 0.5 -- 5$\times$10$^4$ $M_{\odot}$ \cite{Calzetti+2015}. Cluster 3 is the least massive (0.46$\times$10$^4$ $M_{\odot}$), with SED best-fit models in agreement with a structure containing few ($\sim$10) O-type stars. From SED fitting best-fit parameters in \cite{Calzetti+2015}, it is estimated that the supply of ionizing radiation from the system 5+11 is about 50$\%$ of the radio nebula and accounts for $\sim$2/3 of the mechanical energy in this region. While the other 9 clusters supply only 2$\%$ to the total energy of the galaxy, the system 5+11 provides 1/3 of the total ionizing flux. All individual stellar masses, ages and reddening values from best-fit SED modeling are taken from \cite{Calzetti+2015}. From these values, we computed the total stellar mass and average ages and reddening inside each COS aperture (see Table \ref{Table_Clusters}). In Table \ref{Table_Clusters} we also present the color excess values in the \hii\ regions (E(B-V)$\rm _{HII}$) derived from the \ha/\hb\ Balmer decrement measured from the MUSE optical spectra (see description in Section \ref{subsec:em_line_fitting}). This allows us to compare the extinction from the stellar populations and from the ionized gas around the SF regions. In this respect, the single targeted SF regions OBJ-6 and OBJ-7 show the expected higher color excess values for the  \hii\ region component than the stellar component, and very similar values between \hii\ and SED measurements for OBJ-8. However, in contrast to what has been observed in other starburst systems, including NGC~5253 by \cite{Calzetti+1997}, we recover higher E(B-V) values for the stellar component than the \hii\ region component for OBJ-5+11, OBJ-1234 and OBJ 9+10, where we have multiple clusters inside the COS apertures. The discrepancy could be explained by the difficulty of computing the correct weighted average E(B-V) from the integrated optical spectra of multiple sources. As such, the lack of spatial resolution prevents us from correctly contrasting the extinction parameters of the ionized ISM and the stellar component for these specific pointings.

Regarding the detection of massive stars in the context of intense SF in NGC~5253, several studies have reported the presence of Wolf-Rayet (WR) stars near the center of the galaxy (\citealt{Walsh+1987, Walsh+1989}, \citealt{Kobulnicky+1997}, \citealt{Schaerer+1997}; \citealt{Lopez-Sanchez+2007}; \citealt{Lopez-Sanchez+2010A,Lopez-Sanchez+2010B}, \citealt{Monreal-Ibero+2010}, \citealt{Westmoquette+2013}), which are causing an important N local enhancement observed in the ionized gas phase with GMOS-IFU data \citep{Westmoquette+2013}. The location of WR signatures coincide with the position of clusters 1+2+3+4, whose ages of $\sim$5 Myr are in agreement with the expected evolution times for WR episodes. A population of young ($\sim$1 Myr) very massive stars (VMS) with masses $>$100 M$_{\odot}$ was confirmed in the massive compact clusters 5 and 11 (sizes $<$0.6 pc), inside the super radio nebula in the nucleus of the galaxy; they were characterized through a multi-wavelength surface brightness modeling combined with high accuracy astrometry performed by \cite{Smith+2020}.


\section{Observations}\label{sec:observations}

\subsection{COS UV spectroscopy of the neutral gas}

\begin{figure*}[htb!]
\centering
\includegraphics[width=\textwidth]{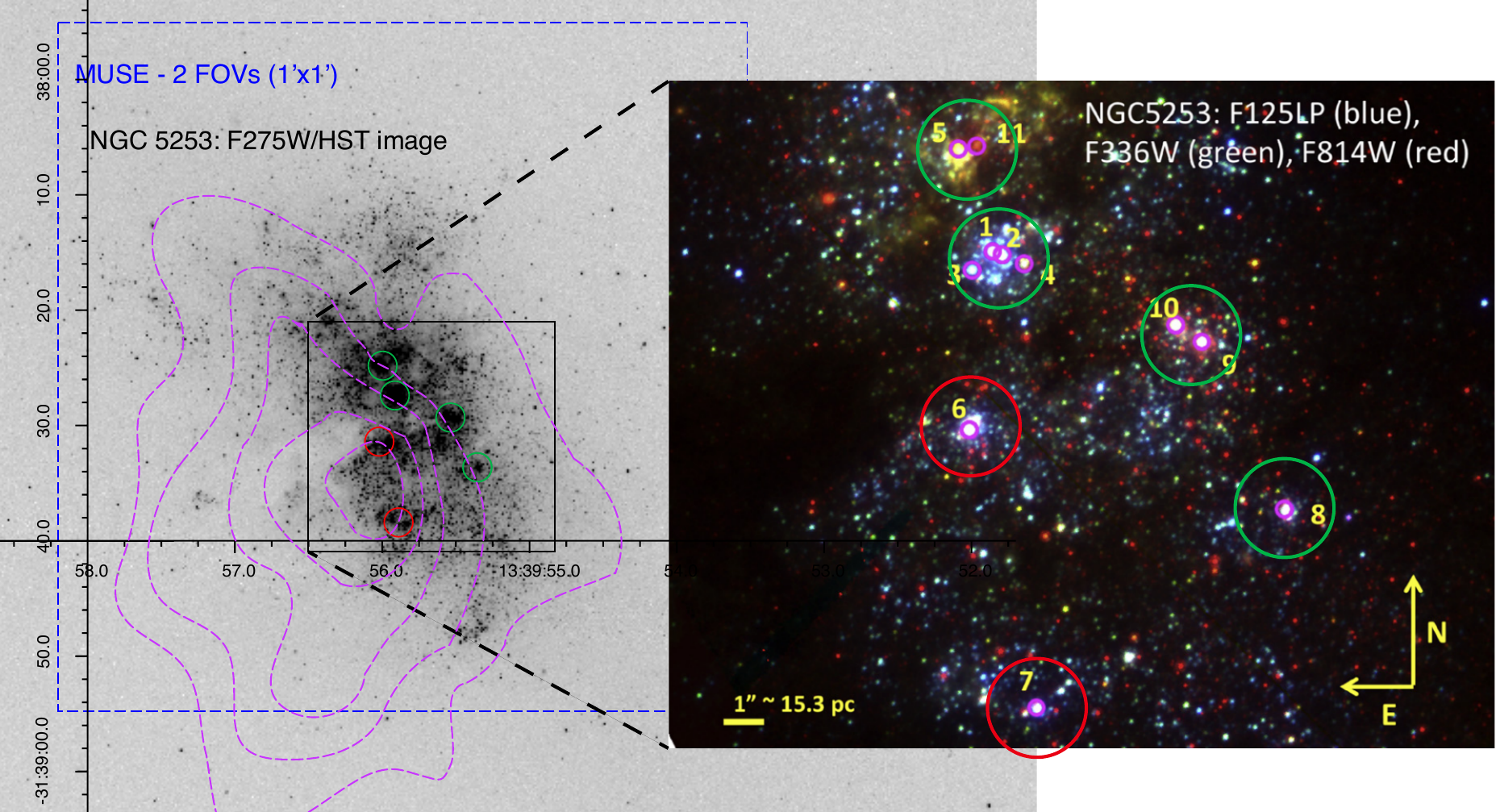}
\figcaption{Spatially matched UV+optical spectroscopic mapping of clusters throughout NGC~5253: HST/WFC3 F275W image with VLT/MUSE optical IFU FoV overlaid in blue. The countours correspond to the high resolution \hi\ flux map from the study performed on ATCA observations by \cite{Lopez-Sanchez+2012} with levels 0.032 (3 $\sigma$), 0.12, 0.22, 0.32 and 0.42 Jy beam$^{-1}$ km s$^{-1}$. The Inset panel shows color composite from HST imaging by \cite{Calzetti+2015}, where each cluster is identified. Red and green circles correspond to COS apertures from archival data (PIDs:11579 and 15193) and new observations (PID: 16240), respectively. \label{fig_NGC5253_COS_MUSE}} 
\end{figure*} 

HST/COS observations were performed by using six pointings towards bright UV young stellar clusters throughout NGC~5253. A total of 31 HST orbits were used for this purpose as part of three different programs with PIDs 11579 and 15193 (PI: Aloisi) containing two pointings, and 16240 (PI: James) containing four additional pointings. A distribution map of all COS pointings on NGC~5253 is presented in Figure \ref{fig_NGC5253_COS_MUSE}. Data were obtained using the G130M medium resolution grating with two different settings centered on $\sim$1222 \AA\ and $\sim$1291 \AA, with a total spectral coverage of $\sim$1066 \AA\ -- 1430 \AA. Observations in program 11579 were acquired using Lifetime Position 1 (LP1), while programs 15193 and 16240 used Lifetime Position 4 (LP4) with average spectral resolutions of R$_{\rm LP1\sim}$[16000 -- 21000] (COS Instrument Handbook\footnote{\url{https://hst-docs.stsci.edu/cosihb}}, \citealt{Osterman+2011}) and R$_{\rm LP4}\sim$ 15000 (COS Instrument Handbook, \citealt{Fox+2018}), respectively.

\begin{table*}[htb!]															
\begin{center}
\caption{Summary of HST/COS Observations on NGC~5253}\label{Table_Obs}
\begin{tabular}																			
{lccccccccc}																			
\hline\hline																			
Target ID 	  & 	R.A.          	& 	Decl.        	& 	PID   	& 	Obs. Date                       	& 	Exp. Time            	& 	Setting     	& 	FWHM$_{\rm Im}$    	& 	S/N &  S/N$\rm_{coadd}$  	\\
          	  & 	(J2000)       	& 	(J2000)      	& 	             	& 	                                	& 	s                	& 	G130M	& 	($\arcsec$)        	& 	    &	\\
\hline																	
OBJ-6     	  & 	13 39 56.02 	& 	-31 38 31.30 	& 	11579	& 	2010-07-03	& 	3038.08	& 	1291-LP1	& 	0.55	& 	11.78	& 	17.88	\\
	  & 		 & 		&	15193	&	2019-05-03	&	1655.97	&	1222-LP4	&	0.70	&	8.43	&		\\
OBJ-7       	& 	13 39 55.89	& 	-31 38 38.34 	& 	11579	& 	2010-07-03	& 	6295.52	& 	1291-LP1	& 	0.92	&   	15.90	&   	28.90	  \\
	& 		& 		& 	15193	& 	2019-05-03	& 	3012.74	& 	1222-LP4	& 	2.06	&   	9.27	&   		  \\
OBJ-9+10    	& 	13 39 55.54	& 	-31 38 29.05	& 	16240	& 	2021-05-11	& 	3595.55	& 	1291-LP4	& 	1.01	&   	9.56	&   	17.90	  \\
	& 		& 		& 		& 	2021-05-11	& 	3215.68	& 	1222-LP4	& 		& 	9.28	& 		  \\
OBJ-8       	& 	13 39 55.35 	& 	-31 38 33.40	& 	16240	& 	2021-02-22	& 	4936.74	& 	1291-LP4	& 	0.93	&   	9.65	&   	18.25	  \\
	& 		& 		& 		& 	2021-02-22 	& 	5236.64	& 	1222 -LP4	& 		& 	10.98	& 		  \\
OBJ-5+11    	& 	13 39 55.98 	& 	-31 38 24.59	& 	16240	& 	2021-05-12	& 	7624.54	& 	1291-LP4	& 	1.45	&   	12.90	&   	25.07	  \\
	& 		& 		& 		& 	2021-03-01	& 	4873.31	& 	1222-LP4	& 		& 	10.93	& 		  \\
OBJ-1234 	& 	13 39 55.91 	& 	-31 38 27.06	& 	16240	& 	2021-05-11	& 	2240.38	& 	1291-LP4	& 	0.85	&   	11.76	&   	16.48	  \\
	& 		& 		& 		& 	 2021-05-11	& 	1782.02	& 	1222-LP4	& 		& 	11.15	& 		  \\
\hline																			
\end{tabular}
\end{center}
\textbf{Notes.}
Summary of COS observations with settings using grating G130M and central wavelengths 1291 \AA\ and 1222 \AA. Sky coordinates correspond to the centers of the COS apertures. COS Lifetime Positions (LP) are indicated in the Observation setting column. Target Acquisition setup corresponds to PSA/Mirror A in all cases. The FWHM corresponds to the extent of the observed source inside the COS aperture (2\farcs{5}). The average S/N values for all the individual and co-added spectra are computed in the range 1155 -- 1165 \AA.
\end{table*}							

Table \ref{Table_Obs} summarizes the COS observations analyzed in this work. The COS spectroscopic exposures in each pointing were executed with 4 FP-POS to mitigate flat-fielding uncertainties, remove grid wire residuals, improve the S/N and reduce the effects of gain sag. We used the standard target acquisition (TA) strategy that consists of performing near ultra-violet (NUV) pre-imaging to properly center the UV sources inside the COS aperture. The ACQ/IMAGE TA was conducted with a lower limit of S/N$\sim$20. For pointings with multiple clusters within the aperture (as in the case of OBJ-9+10, OBJ-5+11 and OBJ-1234), we used a fixed offset target acquisition from a nearby FUV-bright source. All observations were retrieved from the Mikulski Archive at the Space Telescope Science Institute (MAST), and were reduced using the standard \textsc{CalCOS} pipeline (v.3.3.11) as described in detail by \cite{James+2022}.

COS is a spectrograph optimized for observations of point sources. However, star clusters may be extended sources and the three pointings presented above cover more than one cluster per aperture with a complex spatial structure within the aperture. A similar issue arises when dealing with individual clusters that have multiple sightlines toward the OB stars. For the targets with multiple resolved clusters in the aperture, an orient was selected such that the clusters are separated in the cross-dispersion direction, therefore minimizing overlapping. Observations of extended sources with COS are degraded in terms of resolution, due to their extent across the dispersion direction which is accounted for the computation of the LSF, as described in Section \ref{subsec:spec_resol}. We know from a previous study \citep{James+2014} that the two targets already covered by PIDs 11579 and 15193 (OBJ-6 and OBJ-7) were slightly extended ($\sim$0\farcs{5}--0\farcs{7}). The spatial extents measured from the TA images for all the pointings range from 0\farcs{5} to 1\farcs{45} and are reported in Table \ref{Table_Obs}.  


\subsection{MUSE optical spectroscopy of the ionized gas}\label{subsec:MUSE_obs}

Optical integral field spectroscopy (IFS) archival observations of NGC~5253 from VLT/MUSE \footnote{This work makes use of MUSE observations as part of the ESO program ID 095.B-0321 (PI. Vanzi, L.).} are also included in this analysis, covering a field of view (FoV) of 1'$\times$1' with a spectral coverage in the optical and near-infrared ranging from 4750 \AA\ to 9352 \AA\, a spectral sampling of 1.25 \AA\ and a spectral resolution of R$\sim$3027. The processed datacube was downloaded from the ESO Science Portal. These observations correspond to a single exposure performed in February 2015 for a total exposure time of 1620\,s. The data reduction and calibration of the MUSE archival observations were performed using the MUSE standard reduction pipeline \citep[version v1.6]{Weilbacher+2020} which computes the sky subtraction of individual observing blocks (OB), and performs individual pixel alignment, wavelength and flux calibrations on a combined datacube containing all the OBs, resulting in the final output datacube. The MUSE FoV covers the central part of the stellar disk of NGC~5253 (see B diameters in Table \ref{Table_gen_NGC5253}) with an angular sampling of 0\farcs{2} per spaxel which allows us to perform a spatially resolved analysis. Moreover, for our purposes here, MUSE spectroscopic capabilities enable us to perform emission-line modelling
providing the distribution of the ionized gas. Using circular apertures of 2\farcs{5} (12 MUSE spaxels) in diameter, centered on each of the COS pointings, we extracted the integrated MUSE spectra across the equivalent COS apertures. This allows us to derive the chemical properties of the ionized gas in the optical, for an accurate spatial comparison with the neutral gas features derived from the UV spectroscopic analysis of our COS observations. 

\section{Data analysis} \label{sec:data_analysis}


\subsection{Co-addition of COS datasets}\label{subsec:co-add}

\begin{figure*}[htb!]
\centering
\includegraphics[width=1.02\textwidth]{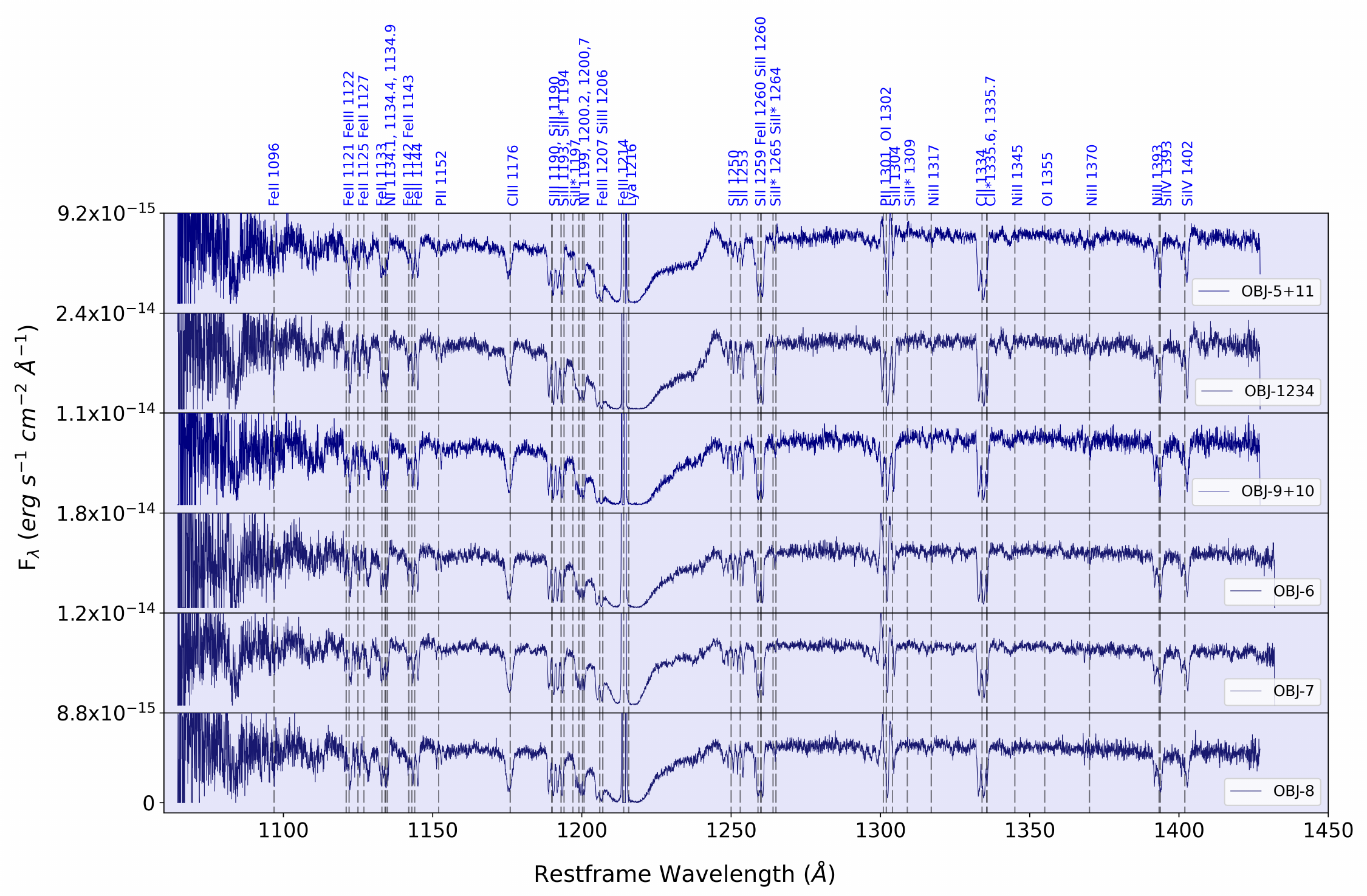}
\figcaption{UV spectral features for each of the COS apertures targeting one or multiple SF clusters in NGC~5253. The observed absorption features show the absorption systems originated in the ISM of both NGC~5253 and the MW, observed along the sight-lines. The co-added spectra (G130M 1222+1291 setting) are organized as a function of age increasing from the top panel which allows to compare different spectral features from the different targets. The emission NV $\lambda$1240 line showing a P-Cygni profile varies as a function of age decreasing its amplitude from the youngest target OBJ-5+11 to the oldest OBJ-8. The most relevant absorption species are indicated for all of the clusters. The shape of the \hi\ absorption profile at 1216 \AA\ is also varying for each target. Absorption features around 1300 \AA\ (\pii\ $\lambda$1301, \oi\ $\lambda$1302 and \siII\ $\lambda$1304) show important differences from target to target, due to geocoronal emission contamination. Stellar absorption feature \ciii\ $\lambda$1176 and stellar winds lines \siIV\ $\lambda\lambda$1393, 1402 are also indicated.\label{fig_All_co-added_spectra} }

\end{figure*} 

Thanks to the unique combination of COS observations in two different settings, G130M/1222 and G130M/1291 with spectral ranges 1065--1365 \AA\ and 1150--1430 \AA\ respectively, the total UV spectral coverage is extended towards a bluer region with respect to the previous study performed by \citet{James+2014} on OBJ-6 and OBJ-7 using the G130M/1291 setting alone, providing access to an additional set of absorption lines. For local galaxies, the COS G130M/1291 configuration covers a wide set of ISM lines (see Figure \ref{fig_All_co-added_spectra}), allowing us to determine the column densities of \cii\ $\lambda$1334, \nI\ $\lambda\lambda\lambda$ 1134.1 1134.4 1134.9, \oi\ $\lambda$1302, \siII\ $\lambda$1304, \pii\ $\lambda$1152, \sii\ $\lambda$1250,$\lambda$1253, \feii\ $\lambda$1142,$\lambda$1143,$\lambda$1144, \niII\ $\lambda$1317,$\lambda$1345,$\lambda$1370; while also covering the Ly$\alpha$ $\lambda$1216 region, essential in computing the neutral hydrogen column density (N(\hi)) and required for the determination of neutral gas abundances. The G130M/1222 setting allows us to cover the \feiii\ $\lambda$1122 line, required to calculate the ionization state of the gas via the \feiii/\feii\ line ratio, an essential component for the ad-hoc photoionization models needed to compute ionization correction factors (ICFs) for each of the pointings, as described in \citet{Hernandez+2020}. Additionally we have detections of \feii\ $\lambda$1096, \feii\ $\lambda$1121, \feii\ $\lambda$1125 and \feii\ $\lambda$1127, which contribute to a robust computation of the \feii\ column density. The UV spectra for all the COS pointings with the main absorption features indicated are shown in Figure \ref{fig_All_co-added_spectra}.

The COS detector is split into two segments (A and B), with a physical gap between them that falls in different regions depending on the grating and central wavelength combination used. For G130M/1222 the gap is located between 1207 and 1223\AA\, while for G130M/1291 the gap covers the range from 1264 to 1291\AA. When using the G130M/1222 spectrum alone, the Ly$\alpha$ is placed in the detector gap, so we can only use part of the wings of the profile in this setting, causing additional uncertainties in N(\hi) measurements. 

\begin{figure*}[htb!]
\centering
\includegraphics[width=\textwidth]{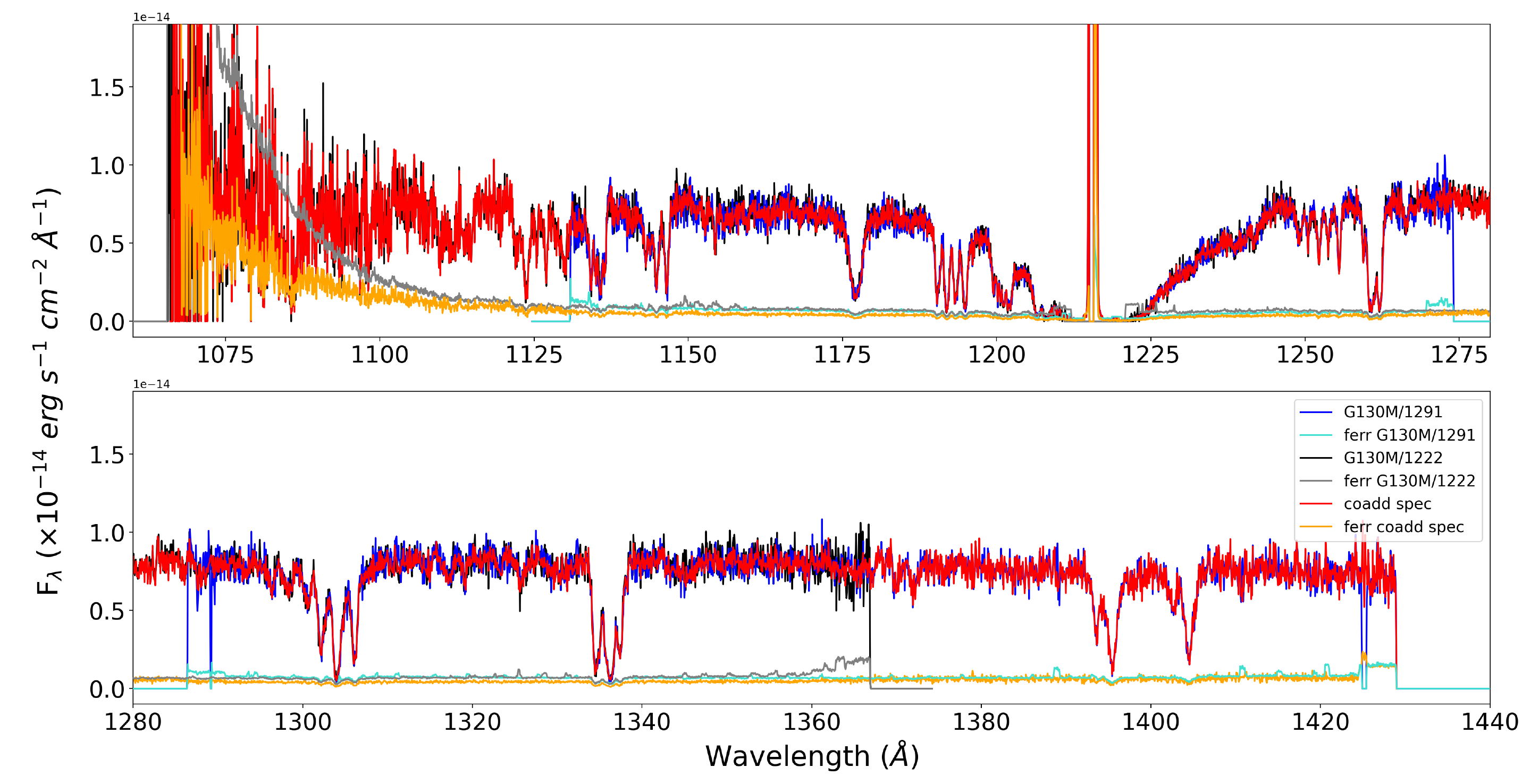}
\figcaption{Comparison of co-added G130M/1291/LP1 and G130M/1222/LP4 spectra and their corresponding spectral errors for OBJ-9+10. All the spectra and errors are overplotted along the total spectral range. The co-added spectrum is shown in red and its corresponding error is indicated in orange. The reduced G130M/1291/LP1 spectrum and its error are shown in blue and cyan, respectively. The G130M/1222/LP4 spectrum is plotted in black with its error in gray. The gaps between the two used settings are visible. The biggest spectral errors correspond to the G130M/1222/LP4, while the resulting error of the co-added spectrum is the lowest in all the spectral range. All spectra are binned in steps of 6 pixels in the dispersion axis, since the COS data are oversampled by that amount. \label{Coadd_spec}}
\end{figure*} 

The spectral gaps for each setting are shifted in a way that configurations with different central wavelengths cover each other's gaps, thus ensuring a continuous spectral coverage (see Figure \ref{fig_All_co-added_spectra}). With the aim to have a continuous spectrum and to increase the average S/N in the overlapping range, we performed a co-addition process for all the pointings. The resulting combined spectra cover the 1065--1430 \AA\ spectral range, the co-added spectrum of OBJ-9+10 is presented as an example in Figure \ref{Coadd_spec}. Co-addition was performed using the IDL routine \textsc{Coadd$\_$X1D}\footnote{\url{https://casa.colorado.edu/~danforth/science/cos/costools.html}} developed by the COS GTO team, which is a generalized co-addition routine with the aim to align and co-add COS FUV observations \citep{Danforth+2010}. We used as input all the \texttt{x1d.fits} files, corresponding to the reduced one-dimensional (1D) spectrum and the error spectrum per exposure. 

The co-addition routine first interpolates the input array from the x1d exposures (flux, flux error, net count rate and gross counts) for individual input \texttt{x1d} exposures along the co-added wavelength array, this to assign a determined $\lambda_{i}$ to the set of spectral features of each original spectra which are not necessarily aligned in wavelength. Since the input spectra are cross-correlated depending on the spectral region around a strong spectral feature in each segment, the software for the $\lambda$ alignment relies on a list of strong ISM features. 

The next step is to determine the co-added flux and its corresponding error. The routine \textsc{Coadd$\_$X1D} has four co-addition methods consisting of a simple mean computation or considering different weighting factors based on data quality flags. The co-addition strategy is selected by giving a numerical value to the \textbf{method} key word as follows:\\
(i) \textbf{method = -1:} Simple mean of flux values per pixel ($ f(i) \neq 0$) of a certain number of exposures $\rm N_{exp}$. The flux error corresponds to the mean of the standard deviation of the flux array ($\sigma(f(i))$/$\rm \sqrt{N_{exp}}$).

To determine the weighted flux and flux error means the code uses different properties of the input exposures as weighting factors, corresponding to the remaining co-addition methods: \\
(ii) \textbf{method = 1:} Modified exposure time weighting ($ w= \rm T_{exp}$), modifies the exposure time of each pixel location to weight their mean contribution to the final flux. \\
(iii) \textbf{method = 2:} Flux error ($f_{err}^{-2}$) weighting ($w= \sqrt{\sum_{i} f(i)^2} /\sum_{i} f_{err}(i)^2 $), allows error tuning but tends to weight toward lower-flux pixels.\\
(iv) \textbf{method = 3:} (S/N$)^{2}$ weighting ($w = \sqrt{\sum_{i} (f(i)/f_{err}(i))^2}$, this method is also adjusted as a function of flux error and tends to weight toward higher-flux pixels.

In the case of pixels for which there is only data from one \texttt{x1d} spectrum, the flux and flux error are adopted from the reduced input spectrum. With the aim of selecting the best method we applied all the methods on targets OBJ-6, OBJ-7 and OBJ-9+10 to test the changes in mean values and in S/N. We noticed that the error-weighted methods (m = 2 and 3) tend to increase the error values of the co-added spectra diminishing the S/N with respect to the \texttt{x1dsum.fits} spectra (containing the sum of all the \texttt{x1d} spectra with the same grating and central $\lambda$). The modified exposure time method (m=1) is effective in increasing the S/N but there were spurious features in the region around Ly$\alpha$ due to a considerable increase of the co-added flux error in this area which is inconvenient for the computation of \hi\ abundances. Finally, the conservative simple mean method (m=-1) gives the best results for our sample, fulfilling the objective of increasing the S/N ($\sim$2 on average) of the co-added spectrum and following the original observed spectral characteristics as it is shown in Figure \ref{Coadd_spec}.


\subsection{Continuum fitting}\label{subsec:continuum}

As described in \cite{James+2014}, using absorption-line measurements requires normalization, so continuum fitting is an essential step to model absorption lines. In order to fit the continuum, an interpolation was applied between flux points located in regions free of strong absorption features (\hi\ damping wings, ISM and stellar absorption lines) and geocoronal emissions. These points act as fixed nodes for the continuum fitting and are selected through a careful visual inspection of the individual spectra. An average flux within an interval of $\sim$ 0.2 \AA\ region either side of each node is calculated, the final interpolation of the continuum is then obtained by applying a spline function among the nodes. The systematic error on the flux estimate is done by computing the variance on the mean flux within a region of 0.4 \AA\ centered on each node. All the co-added spectra were normalized using this method, as well as the individual G130M/1222 and G130M/1291 spectra for OBJ-6 and OBJ-7. The latter was done in order to perform some tests and comparisons in the computation of column densities with the values reported in \cite{James+2014}, using the original \texttt{x1dsum} and the co-added spectra. This comparison is described in more detail in Section \ref{subsec:abs_lines_fit}.

\subsection{Spectral Resolution}\label{subsec:spec_resol}

Here we adopt the single-velocity approximation model to fit the observed profiles. However, it should be noted that the real situation is more complex in that the observed lines arise from a combination of many unresolved velocity components from different absorbing clouds along the many sight lines within the COS aperture, as described in \cite{James+2014}. Line profile fitting requires the estimate of the spectral resolution along the wavelength range. We first take into account the line-spread function (LSF) of the different configurations of the COS instrument. As it is detailed in \citealt[and references therein]{James+2014}, the effective spectral resolution corresponds to the extent of the observed source inside the COS aperture (2\farcs{5}), in the dispersion direction and the observation parameters given by the used setting including grating, central $\lambda$ and LP. The observed spectral resolution is given by two components, the intrinsic resolution ($\rm FWHM_{int}$) or extent of the source as measured in the NUV TA images, and the COS line-spread function (LSF) corresponding to the resolution given by the instrumental response to a point source as a function of wavelength for a certain COS setting ($\rm FWHM_{LSF}$). The extent of the source in the COS dispersion direction is determined from the NUV TA image by subtracting the corresponding PSF as $\rm FWHM_{int}^2= FWHM_{NUVIm}^2 - FWHM_{PSF}^2$, where $\rm FWHM_{PSF} \sim$ 2 pixels \citep{James+2014}. The spatial resolution (FWHM$_{\rm Im}$) for our targets using the different COS settings are listed in Table~\ref{Table_Obs}. The total COS spectral resolution is then computed as follows:
\begin{equation}
   \rm FWHM_{spec}^2 = FWHM_{LSF}^2 + FWHM_{int}^2 .
\end{equation}
This resulted in a broadening of the instrumental LSF to account for the extent of the source inside the COS aperture. 

In the case of the co-added spectra, there are overlapping regions for the used COS configurations in the range [1134--1365 \AA]. Targets OBJ-1234, OBJ-5+11, OBJ-8 and OBJ-9+10 have the same total LSF for the 1222 and 1291 settings, since observations were gathered simultaneously for each target, using the same COS life-time position (LP4). For OBJ-8 and OBJ-5+11 there is more than one TA with very similar LSF values, so we applied a line-fitting test using some absorption features (\hi , \pii , \siII ,  \nI  , \cii  , \ciia , \feii\ and \sii ) and chose those with the best-fit output based on a $\chi^{2}$ minimization applied by \texttt{VoigtFit} (the line fitting code used in this study, see Section \ref{subsec:abs_lines_fit}) and a careful visual inspection of the fits using different sets of input parameters.

For OBJ-6 and OBJ-7 observations in G130M 1291 and 1222 were taken at different lifetime positions (LP1 and LP4, respectively) so the total LSF was measured separately for each setting. We then applied a line fitting test on the co-added spectra (as the one described above), using separately the LSF for each setting in the overlapping regions and chose the LSF configuration providing the best output fits in that wavelength region. We additionally tested line-fitting using the average of the measured LSFs. For OBJ-6 we chose the average of the 1222 and 1291 LSFs while for OBJ-7 we were not able to properly fit the lines in the overlapping area using any of the measured or the average LSF values. For OBJ-7 we then decided not to use the co-added spectrum and instead we analysed separately the  1222 (1040--1133 \AA) and the 1291 (1134-1430 \AA) exposures with their respective LSFs, for the line fitting. With all these considerations, for each of the co-added spectra (except for OBJ-7), we built a composite LSF file containing the 1222 LSF for the blue region [1040--1133 \AA] and the final LSF for the overlapping and red regions. As such, this enabled us to perform the best fit possible to each of the lines and subsequently derive column densities with the highest accuracy available. 

\subsection{Absorption-line profile fitting}\label{subsec:abs_lines_fit}

\begin{figure*}[htb!]
\centering
\includegraphics[width=\textwidth]{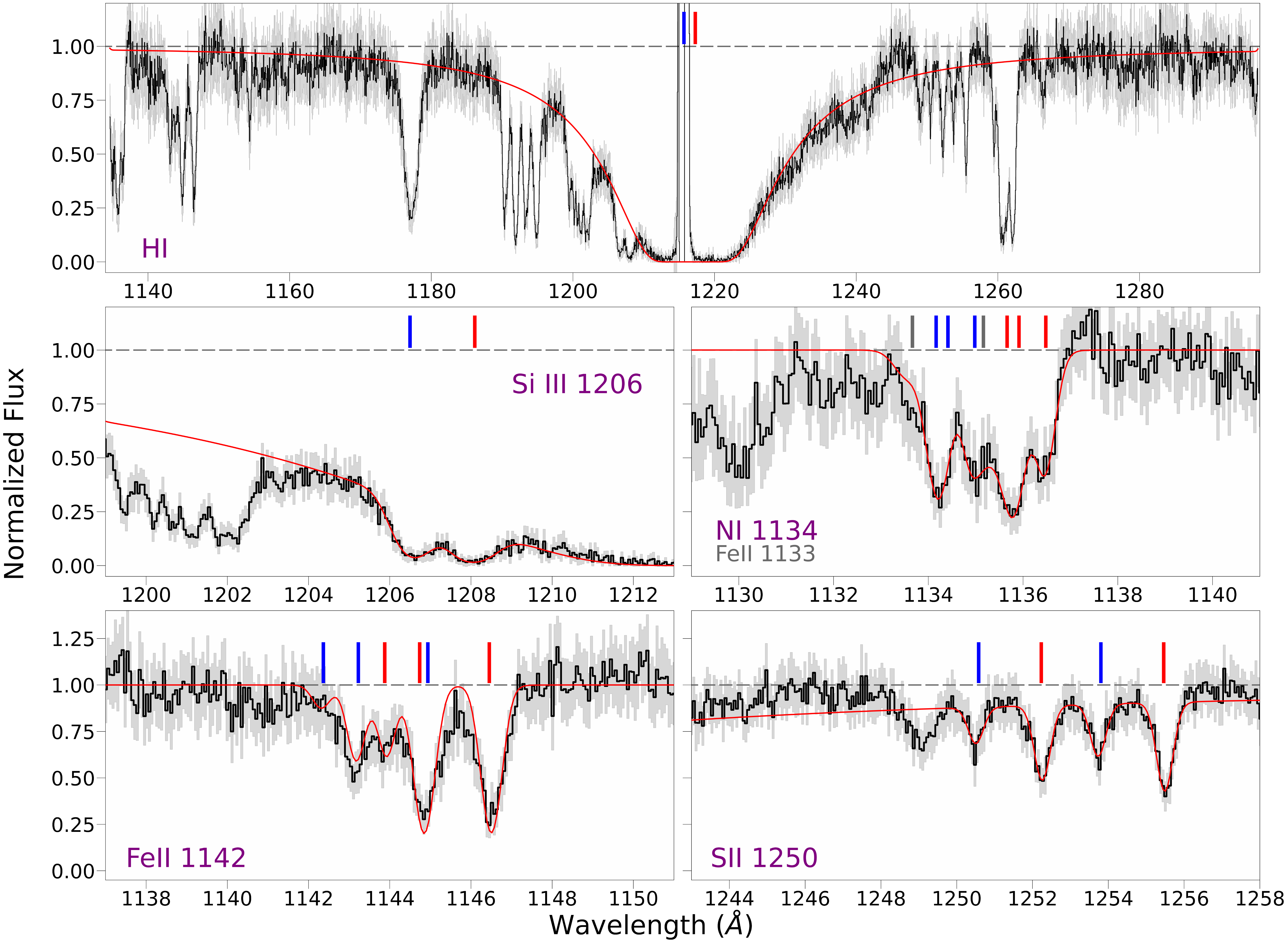}
\figcaption{Absorption line profiles in NGC~5253 for OBJ-9+10. The observed spectra are displayed with black lines and the error spectra correspond to the gray areas. The theoretical line profiles derived using \texttt{VoigtFit} are overplotted in red. The tick marks above the continuum indicate all
the absorption lines considered and their multiple components used for line fitting in each case: with NGC~5253 ISM components in red, ion lines from the MW in blue,  and additional ions in gray. The top panel shows the Ly$\alpha$ profile at z$_{abs}$ = 0.001185 ($v$ = 345.5 km~s$^{-1}$, derived from the best fit HI profile). The other panels display the voigt profiles models of selected metal absorption lines. \label{fig_VoigtFit_ex}} 

\end{figure*} 

We measured column densities of several metals by fitting absorption features produced by different ions present in the cold gas phase of the ISM of our targets following the methodology applied in \cite{James+2014} and \cite{Hernandez+2020}. Theoretical Voigt profiles were used to model the absorption lines using the Python package \texttt{VoigtFit} v.3.13.7 \citep{Krogager2018}\footnote{\url{https://github.com/jkrogager/VoigtFit}}. This code was specifically designed to model absorption profile features taking into account the instrumental line broadening (i.e., LSF) introduced by the spectrograph used to perform the observations. \texttt{VoigtFit} convolves the final COS LSF with the intrinsic source profile, which  is assumed to be single absorption component. This software has a multi-component fitting feature that allows to simultaneously model different components of one or several ions along the line of sight for one or several velocity systems (e.g., NGC~5253, MW or HVCs -high velocity clouds-). The code also has an interactive mode that allows to manually mask out spectral areas and to define velocity components of various elements. The best-fit profile is estimated using a $\chi^2$ minimization approach and an output plot of the profile with residuals allows to visually evaluate the accuracy of the model.

In the analysis presented here, we used the spectrum data file and its corresponding LSF file as main inputs to \texttt{VoigtFit}. In an input script, other parameters can be included, like the object's redshift and the boundaries of the fitting window in velocity or wavelength. Along the line of sight, different components contribute to an observed absorption profile, that includes contamination from the Milky Way (MW) ISM lines which need to be determined in the case of overlapping with the ISM lines of the object. In the case of NGC~5253 at a redshift z$\sim$ 0.00136 (systemic velocity $v_{sys} =$ 417 km s$^{-1}$, \citealt{Koribalski+2004}), MW ISM lines are sometimes overlapped with the object's absorption features, and given the fact that the average LSF of our data is $\sim$ 115 km s$^{-1}$, we are able to de-blend the MW components.  For each target, we did a careful inspection to identify all the MW absorption features that would contaminate our target lines. For each ion we followed a different fitting strategy depending on the number of components (both from NGC~5253 and the MW) and the regions to mask out. We created a python script including all the spectra and LSF input files and a section per ion listing the specific fitting strategy. Once the best-fit model is obtained for a certain fitting window the output contains the column densities (LogN(X)), $b$-parameter (a measurement of the intrinsic line width, or broadening, due to either thermal or turbulent motions of the atoms), redshift and the corresponding errors for each ion and each spectral component. Figure \ref{fig_VoigtFit_ex} shows the followed fitting strategy and the best-fit profiles on top of the normalized observed spectra for OBJ-9+10, inside the corresponding fitting windows for different ions.

We started by testing the software performing line fitting on the original 1291 \texttt{x1dsum} spectra for OBJ-6 and OBJ-7 previously studied in \cite{James+2014} using the IDL software FITLYMAN \citep{Fontana+Ballester1995}. The mean columns densities (Log[N(X)]) obtained from \texttt{VoigtFit} line fitting are in agreement (within 1 -- 2.5$\sigma$) with those reported in \cite{James+2014}, although with larger errors by $\sim$0.04 dex and $\sim$0.03 dex on average for OBJ-6 and OBJ-7, respectively. The $b$-parameter values obtained with \texttt{VoigtFit} are also on average in agreement within 1$\sigma$, and the corresponding errors are larger by $\sim$9 km s$^{-1}$ and $\sim$11 km s$^{-1}$, respectively. 

For co-added spectra we obtained a good agreement (within errors) of our line-fitting outputs (Log[N(X)] and $b$-parameter) and the values obtained by \cite{James+2014} for OBJ-6, while we obtained larger discrepancies for OBJ-7 (2$\sigma$) by 0.64 dex, due to the difference in spectral resolution of the observations for this object (see \ref{subsec:spec_resol}). Based on the outputs of this comparison, we are confident \texttt{VoigtFit} models are robust enough to determine column densities in all the targets.

\subsection{Optical emission-line profile fitting} \label{subsec:em_line_fitting}

For the analysis of the optical MUSE spectra corresponding to each COS pointing described in Section~\ref{subsec:MUSE_obs}, we used a similar fitting method to the one discussed in \cite{mingozzi22}~(Section~3.2), in order to get the emission-line fluxes needed to measure the ionized gas abundances (see Section~\ref{subsec:ionized_abundances}). 
First, we inspected the optical spectra, noticing that in all clusters but OBJ-1234 and OBJ-5+11 (the youngest ones, see Table~\ref{Table_Clusters}), there is a clear \hb\ absorption feature due to the underlying stellar component. Hence, to accurately measure the \hb\ and \ha\ Balmer emission-line fluxes in these cases it was first necessary to fit and subtract the stellar component. 
To do this, we performed the stellar component fit using the Penalized Pixel-Fitting (pPXF; \citealt{cappellari04}) using a linear combination of \citet{vazdekis16} synthetic spectral energy distributions for single stellar population models in the $4800-9300$~\AA\ wavelength range. 
Also, we simultaneously fitted the main emission lines included in the selected wavelength range (i.e., H$\beta$, \oiii~$\lambda\lambda$4959,5007, \nii~$\lambda$5755, \foi~$\lambda\lambda$6300,6364, \fsiii~$\lambda$6312, \nii~$\lambda\lambda$6548,84, H$\alpha$, \fsii~$\lambda\lambda$6717,31, \fsiii~$\lambda$9069) to better constrain the stellar component fit, masking fainter lines and regions affected by sky residuals. 
Finally, we fitted the emission lines of interest in the original spectra of OBJ-1234 and OBJ-5+11 and the stellar subtracted spectra of the other clusters. In particular, apart from the main emission lines already mentioned, we took into account iron (\feiii~$\lambda4883,4986,5270$, \feii~$\lambda$8619) and oxygen lines (\oii~$\lambda\lambda$7320,30).
To obtain the emission-line integrated fluxes, we modeled them with Gaussian profiles, making use of the code MPFIT \citep{markwardt09}, which performs a robust nonlinear least-squares curve fitting. 
As the stronger emission lines (e.g., \oiii, H$\alpha$, \nii, \fsii) showed a narrow profile ($\sigma_{int}\sim40$~km/s) but slight asymmetries, we used the sum of two Gaussian profiles in case the emission line peaks could not be reproduced with a single one. Fluxes were dereddened using the H$\alpha$/H$\beta$ emission line ratios and the \citet{Calzetti+2000} reddening law with a case B recombination ratio. The measured observed and de-reddened fluxes are provided in Table~\ref{tab:emission_fluxes}. 

\section{Results} \label{sec:results}

\subsection{Neutral gas column densities and abundances} \label{subsec:col_dens_abundances}

As explained in \cite{James+2014} and \cite{Hernandez+2020}, for each absorbing ion COS detects a non-linear average combination of unresolved velocity components from different clouds along a sight line towards the UV background sources inside the 2\farcs{5} aperture. \cite{Jenkins1986} showed that complex configurations with multiple unresolved absorbing components can be analyzed jointly by performing a single-velocity profile approximation when the distribution function of line characteristics is not irregular (e.g., bimodal). This approach is valid even when the involved lines have variations in the $b$-parameter and saturation levels (with the exception of very strongly saturated lines). Since the spectral signal within the COS aperture encompasses a combination of different physical unresolved components, an integrated distribution of the kinematic properties in a single component is expected, thus a single-profile approach is valid in the determination of column densities (\citealt{James+2014}, \citealt{Hernandez+2020}). In this scenario, the velocity dispersion parameter ($b$), which is the combination of various line widths and the velocity separations of the different components, does not have an explicit physical meaning while the column density is well-constrained \citep{Jenkins1986}. Due to the limited resolution of the COS observations, we did not introduce additional parameters, such as the number of intervening clouds and multiple velocity components. Accordingly, we adopted a single-velocity approach to model the individual ion absorption features detected in our COS data. 

To compute the neutral hydrogen (\hi) column densities (N(\hi)) we modeled the absorption profiles of the Ly$\alpha$ line at $\lambda$ = 1215.67 \AA\ which is the only \hi\ absorption feature of the Lyman series covered by the COS FUV spectral range. Due to the low redshift of NGC~5253 (at $v_{sys}$ $\sim$ 400 km s$^{-1}$) the MW foreground \hi\ absorption is not de-blended from the intrinsic \hi\ absorption of the target so the blue wing of Ly$\alpha$ is contaminated by that of the Milky Way. However, spectral decomposition of the two Ly$\alpha$ profiles is easily achieved using the method outlined in \cite{James+2014}, allowing for an accurate measurement of N(\hi) towards each cluster.

For the MW foreground absorption we fixed the column density to the value log[N(\hi)$\rm_{MW}$cm$^{-2}$]=20.68, estimated for OBJ-6 and OBJ-7 by \cite{James+2014} using the all-sky Leiden/Dwingeloo survey \citep{Hartmann_Burton1997}. We applied the same value of N(\hi)$\rm_{MW}$ to all targets since the overall \hi\ column density of the MW should be on average the same in the direction of NGC~5253. To model the composite \hi\ foreground+target absorption profile for each target, we then fitted simultaneously Ly$\alpha$ in both the target and the MW by fixing the MW column density as described above, and its velocity using the \feii\ $\lambda$1142 MW line, and by letting the N(\hi) of the object vary freely. In the case of the \hi\ absorption the $b$-parameter is highly degenerated (since this line is in the damped section of the curve of growth –COG– that relates the equivalent width of an absorption line profile to the column density of the absorbing media), so it does not have an important impact on the fitting. We fixed this parameter to $b$ = 60 km s$^{-1}$ for the absorption component intrinsic to each target.

\begin{figure}[htb!]
\centering
\includegraphics[width=1.05\linewidth]{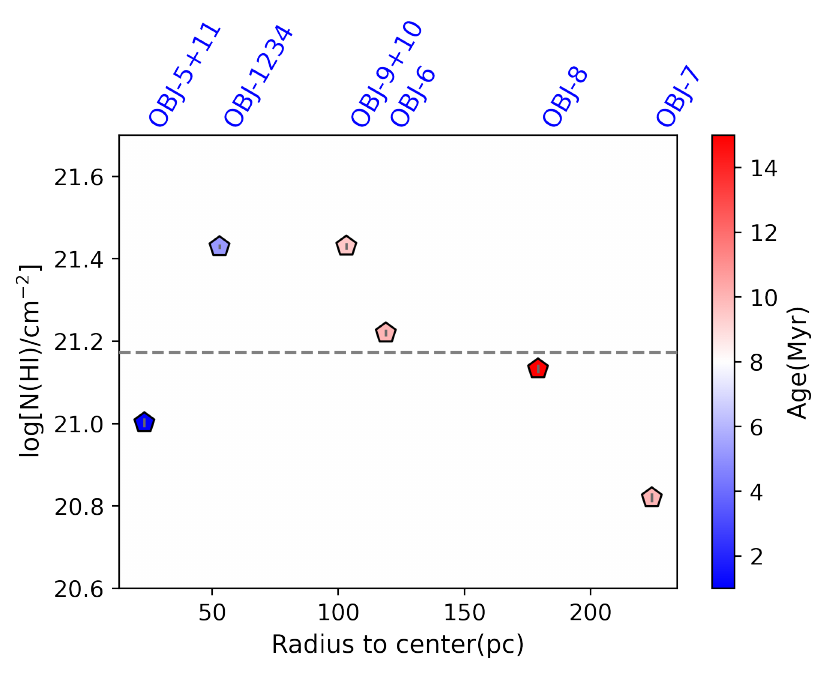}
\figcaption{\hi\ column densities in the neutral gas of the targeted SF clusters in NGC~5253, with an average log[N(\hi)/cm$^{-2}$]$\sim$ 21.17. OBJ-9+10 and OBJ-1234 show the highest column densities (log[N(\hi)/cm$^{-2}$]$ \sim$ 21.43, for both clusters). The error bars for all the targets are inside the pentagon markers and are reported in Table \ref{Table_Column_densities}. The gray dashed line indicates the average \hi\ column density of the sample. \label{fig_H_col_dens}}
\end{figure}

For all our pointings, the blue wing of the Ly$\alpha$ abortion profiles were well fitted using a single-profile and adding the MW component. For OBJ-5+11 and OBJ-1234 we observed the strong stellar wind feature N\textsc{ v} 1240 \AA\, showing a P-Cygni profile with a broad emission on the red wing of the \hi\ profile. This spectral feature arises from strong stellar winds of massive stars in young stellar populations of $<$ 5 Myr \citep{Chisholm+2019}. To check our single-component modeling, we attempted to add a second velocity component to model the intrinsic  Ly$\alpha$ absorption but this did not improve the fit and largely increased the errors. The shape of the total \hi\ profile is well fitted for all the SF regions using Voigt individual profiles for the MW and galaxy components (see upper panel of Figure \ref{fig_VoigtFit_ex}). In Figure \ref{fig_H_col_dens}, the \hi\ column densities are shown as a function of distance from the center and color-coded as a function of the cluster age. The average \hi\ column density for all pointings corresponds to log[N(\hi)/cm$^{-2}$] = 21.17$\pm$0.009, varying between log[N(\hi)/cm$^{-2}$] = 21.43 (for OBJ-1234 and OBJ-9+10) and log[N(\hi)/cm$^{-2}$] = 20.82 (for OBJ-7). \hi\ column density measurements for each of the observed targets are listed in the first row of each target in Table \ref{Table_Column_densities} (see appendix \ref{Appendix_Table_column_densities}). 

\begin{deluxetable}{lcc}[!hb]
\label{Table_E_B_V_Hatom}
\tablecaption{\hi\ column density to color excess ratio for each COS pointing in NGC~5253 and reference values with respect to the MW }
\tablehead{
\colhead{ID} & \colhead{N(HI)/E(B-V)\tablenotemark{a}} & \colhead{$\gamma_{5253}$/$\rm\gamma_{MW}$\tablenotemark{b}} \\
\colhead{} & \colhead{(atoms cm$^{-2}$ mag$^{-1}$)} & \colhead{}}
\startdata
OBJ-1234	&	1.79$\times$10$^{22}$	&	3.00	\\
OBJ-5+11	&	2.18$\times$10$^{21}$	&	0.38	\\
OBJ-6	&	1.38$\times$10$^{22}$	&	2.38	\\
OBJ-7	&	1.32$\times$10$^{22}$	&	2.28	\\
OJB-8	&	8.49$\times$10$^{21}$	&	1.46	\\
OBJ-9+10	&	5.19$\times$10$^{21}$	&	1.79	\\
\enddata
\tablenotetext{a}{N(HI)/E(B-V) for NGC~5253 ($\gamma_{5253}$) in units of atoms cm$^{-2}$ mag$^{-1}$. \hi\ column densities are listed in Table \ref{Table_Column_densities} and E(B-V)$_{COS}$ values in Table \ref{Table_Clusters}. }
\tablenotetext{b}{N(HI)/E(B-V) for the MW corresponds to $\rm\gamma_{MW}$ = 5.8 $\times$10$^{21}$ atoms cm$^{-2}$ mag$^{-1}$ \citep{Bohlin+1978}.}
\end{deluxetable}

The ratio between \hi\ column densities (Table \ref{Table_Column_densities}) and the color excess inside our COS apertures (E(B-V)$_{COS}$ values in Table \ref{Table_Clusters}), allows us to determine the extinction per H atom in a metal-poor system, such as NGC~5253. We provide this column density to extinction ratio ($\gamma_{5253}$) for each of our pointings in Table \ref{Table_E_B_V_Hatom} and a comparison with the same quantity for the MW ($\rm\gamma_{MW}$ = 5.8 $\times$10$^{21}$ atoms cm$^{-2}$ mag$^{-1}$)  measured by \cite{Bohlin+1978} as reference. We see that for OBJ-1234, OBJ-6, OBJ-7, OBJ-9+10, and OBJ-8 the column density to extinction ratio varies between 1.5 to 3 times that of the MW suggesting a lower extinction in these SF regions with respect to the MW. These low extinction values could be related to the low global metallicity of this galaxy. Nevertheless, OBJ-5+11 is exceptionally different showing a stronger extinction per H atom than in the MW; this is in agreement with the high dust extinction values reported in the literature for the center of NGC~5253 and the dust lane observed in the photometry that coincides with the location of clusters 5 and 11 and of the radio nebula in which they are embedded \citep{Calzetti+2015}. These results are evidence of a localized distribution of dust in NGC~5253 in the center of the galaxy where the most recent starburst episode takes place. 

\begin{figure*}[htb!]
\centering
\includegraphics[width=0.9\linewidth]{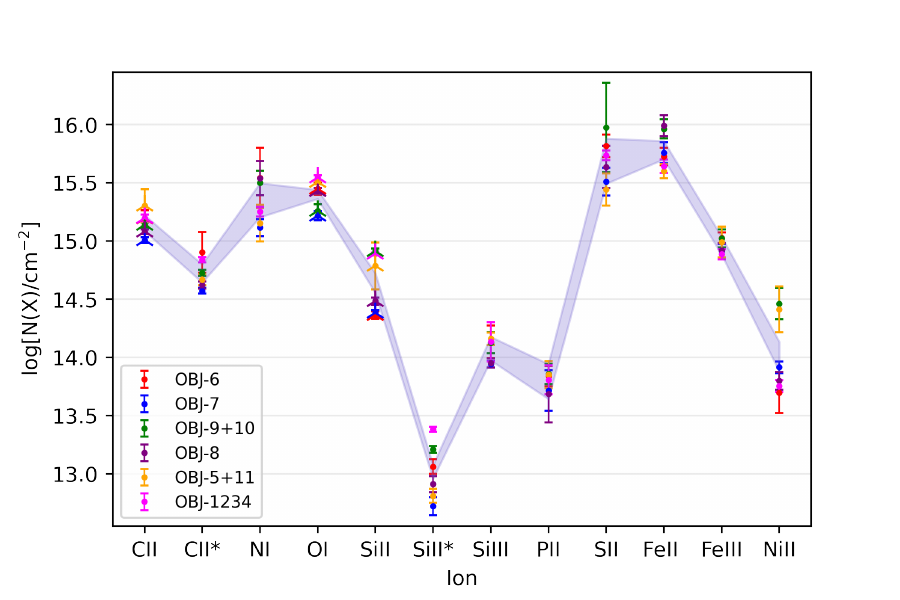}
\figcaption{Distribution of column densities per ion for the neutral gas in different pointings in NGC~5253. The vertical dispersion in the column densities for each ion may be linked to differences in physical properties among the targeted clusters. Each target has an assigned color as indicated in the label. Column densities for \cii\ $\lambda$1334, \siII\ $\lambda$1304 and \oi\ $\lambda$1302 are taken as lower limits (represented by arrows) due to line saturation and geocoronal emission affecting \oi\ and \siII. \label{fig_col_densities}}
\end{figure*}

We modeled the profiles of UV absorption spectral features that arise from atomic transitions of heavy elements in the ISM of NGC~5253 in their neutral, single and double ionized states. All the used neutral and ionized species are listed in Table \ref{Table_Column_densities}. In order to better constrain the column densities, when possible we used multiple lines associated to a certain ion with different values of $f\lambda$, where $f$ is the oscillator strength and $\lambda$ is the rest-frame wavelength of the absorption. The rest-frame wavelengths of the lines used to compute columns densities for each ion and the best-fit outputs from line-profile fitting (column densities, $b$-parameters and systemic velocities) are summarized in Table \ref{Table_Column_densities}. 

Absorption line profiles can be affected by two types of saturation. Classical saturation manifests as an increase of equivalent width (in logarithmic scale) where the line is optically thick, while hidden saturation is observed when multiple components from different absorbers along the same sightline are unresolved and the average column density results in an apparent unsaturated profile. Despite this, we were able to model \cii\ and \ciia\ lines due to the strength of the absorption ($f\sim$1.29$\times$10$^{-1}$), with very prominent profiles. Among the lines affected by classical saturation we can identify the strong lines \cii\ $\lambda$1134.5, \ciia\ $\lambda\lambda$1135.6,1135.7 and \oi $\lambda$1302. Hidden saturation can be identified when fitting lines with different $f\lambda$ of the same ion. An example of hidden saturation is shown in \cite{James+2014} for the triplet \feii\ $\lambda$1142,$\lambda$1143,$\lambda$1144. The strongest absorption \feii\ lines at 1143 and 1144 \AA\ have smaller apparent profiles (apparently unsaturated) than those expected by the best-fit output (saturated profile), obtained by using the column density of the weakest component at 1142 \AA\ least affected by saturation. To tackle hidden saturation in the case of \feii\ we used a new approach performing a simultaneous fitting of seven different lines (\feii\ $\lambda$1096, $\lambda$1121, $\lambda$1125, $\lambda$1127, $\lambda$1142, $\lambda$1143, $\lambda$1144), thanks to the extended $\lambda$ coverage of the co-added spectra in the blue part as can be seen in Figure \ref{fig_VoigtFit_ex}. This allowed us to get better fits for the observed profiles minimizing the hidden saturation effect. A more detailed description of classical and hidden saturation and their impact on UV absorption profiles can be found in \cite{James+2014}. 

Since the broad wings of Ly$\alpha$ in absorption extend along the COS spectra up to $\sim$5000 km s$^{-1}$, some absorption features are observed on top of the \hi\ absorption. In these cases we used the best modeled \hi\ composite profile as a fixed component while modeling the absorptions due to the other ions. This is the case for \siIII$\lambda$1206 and \sii $\lambda\lambda$1250,1253, located in the red and blue wings of the \hi\ profile, respectively. The \siIII $\lambda$1206 profile was modeled fixing the composite \hi\ model (described above) and adding two lines, one corresponding to the absorption of the MW (fixing the velocity of the component) and that of the target with free variable parameters. For \sii$\lambda\lambda$1250,1253 we only used the \hi\ profile when the broadening of the red wing was strongly affecting the \sii\ lines (OBJ-6, OBJ-2 and OBJ-8), otherwise we only used the MW+target components of the \sii\ lines as it is shown in Figure \ref{fig_VoigtFit_ex}. 

Regarding the NI triplet at 1124 \AA\, we added the MW+target components plus the \feii$\lambda$1133 line from the galaxy, resulting in several lines packed in a small wavelength fitting window. For the \feii$\lambda$1133 line we fixed the best-fit parameters for the \feii\ fit obtained by using seven other absorptions (as was described above), and let the fit of the target \nI\ lines vary. This complex profile has sharp overlapping lines, being suitable as visual test for the LSF computation in all the targets. 

Results on column densities (log[N(X)]) per ion (X) and per target, as a product of the line fitting process described above, are presented in Figure \ref{fig_col_densities}, except for \hi\ (which is presented in Figure \ref{fig_H_col_dens}). Different colors represent the different COS pointings and each column corresponds to each of the measured column densities and associated error bars using the components listed in Table \ref{Table_Column_densities} for each ion. In Figure \ref{fig_col_densities}, the vertical dispersion corresponds to physical differences in column densities of the same ion measured for the different targets in our sample. The smallest dispersion in log[N(X)/cm$^{-2}$] $\sim$ 0.14 dex corresponds to \feiii\, while \niII\ shows the largest dispersion of $\sim$ 0.77 dex.

Neutral oxygen abundance is fundamental to assess the distribution of metals in the ISM. Due to the fact that the strongest neutral oxygen absorption \oi\ $\lambda$1302 is saturated and affected by the geocoronal contamination detected by COS around 1304 \AA , we can only use the derived log[N(O)] from line absorption fitting as a lower limit of the intrinsic O column density. Another absorption feature of neutral oxygen in the COS spectral range is \oi\ $\lambda$1355, but this line is extremely faint ($f$=1.248$\times$10$^{-6}$, \citealt{Morton_1991}) and was undetectable for all our targets. An alternative technique to tackle the difficulty in measuring neutral oxygen abundances in the FUV, is the usage of correlations with other elements as proxies to indirectly derive O/H. In particular here we used the relationships presented by \cite{James+2018} between \oi , \sii\ and \pii\ column densities for a sample of nearby and intermediate redshift (z$\sim$0.083--0.321) SFGs spanning a wide range in metallicities (0.03 -- 3.2 Z$_{\odot}$) and \hi\ column densities (18.44 --21.28). The authors found agreement between the column density ratio log(\pii/\sii) in their sample of galaxies with respect to the solar ratio log(\pii/\sii)$_{\odot}$ with a difference of 0.02 dex which is interpreted as a weak depletion of P and S with respect to O in the sample. They found strong correlations in the relations log[N(\pii)] vs. log[N(\oi)] and log[N(\sii)] vs. log[N(\oi)] which are very similar to the solar ratios relations. From this analysis \cite{James+2018} demonstrated that solar ratios log(P/O)$_{\odot}$=--3.28 $\pm$0.06 and log(S/O)$_{\odot}$=-1.57$\pm$0.06 can be used as robust proxies to estimate O/H abundances in the neutral gas of different environments in local and high redshift SFGs. We use S and P abundances to derive a reliable O abundance for our targets, then we used an average of the log[N(O$_{\rm P}$)] and log[N(O$_{\rm S}$)] column densities to derive the final O abundances reported in Table \ref{Table_Abundances_ICF}. All the individual values computed using separately the correlations of O with P and S are presented in Appendix \ref{Appendix_O_PS_abundances}. The comparison between oxygen abundances derived using column densities from direct line fitting of the \oi\ $\lambda$1302 line and from the indirect P/S/O method can be seen in Figure \ref{fig_ICF_final_abundances}. Oxygen abundances from line fitting are underestimated by 0.76 dex on average with respect to the average values computed from the P/S proxies.

\subsection{Ionization corrections in the neutral gas}
\label{subesec:ICF_correction}

\begin{figure*}[htb!]
\includegraphics[width=0.9\linewidth]{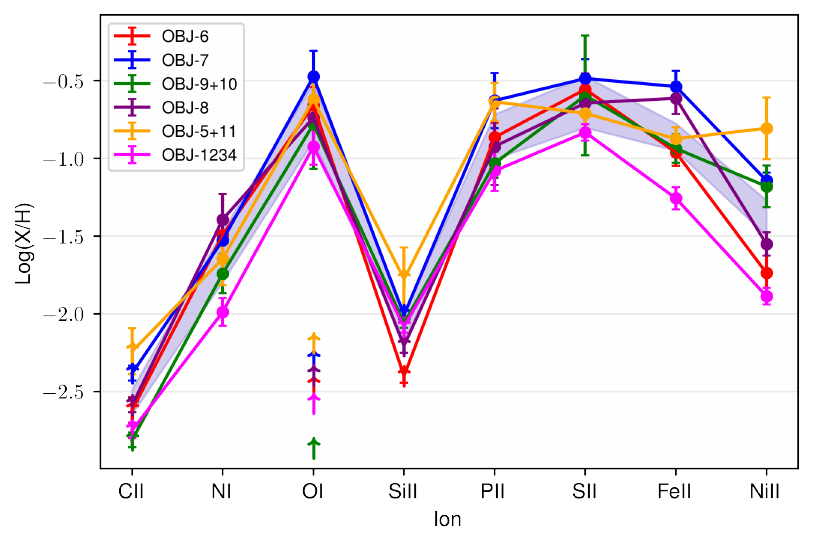}
\figcaption{Neutral gas ionization-corrected abundances per ion for all the SF clusters in NGC~5253. Different colors are assigned to each cluster as indicated in the label. Abundance values are connected by a line for each target with the aim of highlighting trends among the ions from different targets. O abundances are derived from the P/O/S proxy method (\citealt{James+2018}, see Section \ref{subsec:col_dens_abundances}). The abundances of C, O and Si measured from absorption line fitting are displayed as upward arrows, as they are considered lower limits. The shaded area indicates the average abundances within errors for all the targeted SF regions. For all the ions except for \siII\ the values for OBJ-1234 are slightly below the averages.  \label{fig_ICF_final_abundances}}
\end{figure*}

In the determination of chemical abundances of the neutral gas in SFGs from UV absorption-line spectroscopy, it is important to account for the ionization effects from both the neutral and ionized phases of the ISM. As explained in \cite{James+2014}, for N$_{\rm \hi} > $10$^{19}$ cm$^{-2}$ neutral \oi\ and \nI\ as well as elements in the singly ionized stage (\cii, \siII, \pii, \sii, \feii\ and \niII) are the dominant ions in the neutral gas since in this phase the gas is transparent to photons with energies lower than the H ground state potential ($hv<$ 13.6 eV), while their second ionized state requires energies above that threshold. This is the case for all the targets in our sample, so we expect a low impact of ionization corrections in the abundance measurements of our sample. The main supply of ionizing photons comes from UV background sources in SF regions in the observed SF clusters. A fraction of this radiation escapes from the \hii\ regions and is then absorbed by the cold neutral gas, producing the observed absorption UV features and ionizing a part of the gas in the neutral gas clouds. As a consequence, a portion of the ions prevalent in the \hi\ regions are in a higher ionization state and may not be directly detected in the UV regime. The ionization correction factor in the neutral phase (ICF$\rm _{neutral}$) accounts for classical ionization effects in the ISM including the dominant ionization state of a certain element X in the \hi\ regions, and the effects of higher-ionization stages (see \citealt{Hernandez+2020}).

A second effect is related to the contaminating ionized gas from \hii\ regions along the line of sight of each target. This effect is quantified as an additional ionization correction factor (ICF$_{\rm ionized}$). The formation of element absorption lines can be impacted by ionization in both phases, causing an underestimate of the real column densities in the neutral ISM if ICF$\rm _{neutral}$ is not accounted, or an overestimate if ICF$_{\rm ionized}$ is not considered. With the aim of accounting for the ionized gas along the line of sight of COS apertures that contributes to the measured column densities in the cold phase, \citealt{Hernandez+2020} performed an ad-hoc photoionization modeling on targets spanning a wide range of galactic environments. They used the spectral synthesis code \textsc{Cloudy} \citep{Ferland+2017} to simulate the conditions of the observed ISM assuming a simple spherical geometry of the cloud surrounding the sources (assumed to be co-spatial for the sake of the Cloudy models), at a certain effective temperature and UV luminosity. To compute the ionization correction factors (ICF), both ionized and neutral, they used the photoionization simulation results as fixed inputs for different physical characteristics of the neutral gas as metallicity of the source (Z), effective temperature (T$_{\rm eff}$), UV luminosity (LUV), column densities of the \hi\ (log[N(\hi)]) and the iron ionization factor (\feiii/\feii), producing a grid ionization model per parameter and per element. According to results in \citealt{James+2014} and \cite{Hernandez+2020} the strength of the ionization effect is proportional to the number of sources and their geometry inside the detection aperture, as well as to the metallicity of the source. 

The total correction factor is defined as ICF$_{\rm total}$ = ICF$_{\rm ionized}$ -- ICF$_{\rm neutral}$). We directly use the ICF$_{\rm total}$ values obtained by \cite{Hernandez+2020} for OBJ-6 and OBJ-7. In order to compute the ICF factors for OBJs 8, 9+10, 5+11 and 1234, we performed a simple interpolation on the ionization grid models (for both neutral and ionized phases) using the log[N(\hi)] and \feiii/\feii. Then on the interpolated functions using the values obtained from our analysis for all the clusters we computed the ICF factors for each cluster for all the elements. We then proceeded to compute the ICF$_{\rm total}$ values from the results of both interpolations per cluster. Based on a comparison of the ICF$_{\rm total}$ values obtained using the correlations in \cite{Hernandez+2020} and the values computed by \cite{James+2014} for OBJ-6 and OBJ-7, we decided to use the log[N(\hi)] correlation which show the smallest difference. The quantification of the total ionized correction factor (ICF$_{\rm total}$) is presented in Appendix \ref{Appendix_ICF_Computation}, and the ICF$_{\rm total}$ values used for the computation of column densities are listed in Table \ref{Table_ICF_log(N_HI)}.

In order to compute the final abundances we applied the correction to the column density obtained from line-profile fitting as follows,
\begin{equation}
   \rm log[N(X)]_{ICF} = log[N(X)] - ICF_{total},
\end{equation}
where log[N(X)] corresponds to the column density of element X. 

For each cluster observed chemical abundances were determined using the column density measurements for each ion and computing the ratio with respect to the \hi\ column density. We also computed the relative abundances to the solar photospheric abundances adopted from \cite{Asplund+21}. Observed abundances, relative to solar abundances, and ionization corrected abundances are listed in Table \ref{Table_Abundances_ICF}. For the absorption profiles affected by saturation, the derived column densities and abundances are taken as lower limits (marked with $\dag$ symbol in Table \ref{Table_Abundances_ICF} and Table \ref{Table_Column_densities}). The errors for [X/H]$_{\rm ICF}$ correspond to the sum in quadrature of the errors derived from the absorption line fitting, the ICF interpolation and the determination of the solar abundances.

A visualization of the final abundances for each ion in the neutral gas is shown for all the clusters in Figure \ref{fig_ICF_final_abundances}. All the points corresponding to abundances of ions detected in each cluster are connected by lines with the purpose of checking on differences in the abundance trends. We specifically look for crossing lines to check on evolutionary differences among abundances for the different targets due the different ages and location along the galaxy. Overall we see a similar trend for almost all the targets except for OBJ-5+11, specifically for \oi\, \nI\ , \sii\ and \feii\ where the abundances are lower than expected from the other ions where this cluster reports the highest abundances. This is the youngest ($\sim$1 Myr) and most massive target in the center of the galaxy, so these differences could be related to its early evolutionary stage in which the processes that release \nI\ , \sii\ and \feii\ , like WR and SNe events, have not happened yet. The average within errors for all the targeted SF regions is indicated with a shaded area in light blue. Most of the ion abundances are within the average, except for OBJ-1234 for which abundances are all slightly lower than the average.

\startlongtable
\begin{deluxetable*}{lcccccc}

\tablecolumns{7}

 \tablecaption{Abundances of the neutral gas phase in NGC~5253 \label{Table_Abundances_ICF} }
 
\tablehead{
\colhead{Element} & \colhead{Ion} &  \colhead{log(X/H)} & \colhead{log(X/H)$_{\rm ICF}$\tablenotemark{a}}   & \colhead{log(X/H)$_{\odot}$\tablenotemark{b}}  & \colhead{[X/H]\tablenotemark{c}} & \colhead{[X/H]$_{\rm ICF}$\tablenotemark{d}}}
\startdata
\multicolumn{7}{c}{NGC~5253-OBJ-6}\\
\cline{1-7}
C$^{\dag}$	&	\cii	&	-6.03	$\pm$	0.07	&	-6.16	$\pm$	0.07	&	-3.54	$\pm$	0.04	&	-2.49	$\pm$	0.08	&	-2.62	$\pm$	0.08	\\
N	&	\nI	&	-5.68	$\pm$	0.26	&	-5.66	$\pm$	0.26	&	-4.17	$\pm$	0.07	&	-1.51	$\pm$	0.26	&	-1.49	$\pm$	0.27	\\
O$_{\rm P,S}$	&	\oi	&	-3.97	$\pm$	0.12	&	-3.97	$\pm$	0.11	&	-3.31	$\pm$	0.04	&	-0.66	$\pm$	0.12	&	-0.66	$\pm$	0.12	\\
Si$^{\dag}$	&	\siII	&	-6.86	$\pm$	0.03	&	-6.89	$\pm$	0.03	&	-4.49	$\pm$	0.03	&	-2.37	$\pm$	0.04	&	-2.40	$\pm$	0.04	\\
P	&	\pii	&	-7.38	$\pm$	0.09	&	-7.45	$\pm$	0.09	&	-6.59	$\pm$	0.03	&	-0.79	$\pm$	0.09	&	-0.86	$\pm$	0.09	\\
S	&	\sii	&	-5.40	$\pm$	0.10	&	-5.44	$\pm$	0.10	&	-4.88	$\pm$	0.03	&	-0.52	$\pm$	0.10	&	-0.56	$\pm$	0.10	\\
Fe	&	\feii	&	-5.49	$\pm$	0.08	&	-5.50	$\pm$	0.08	&	-4.54	$\pm$	0.04	&	-0.95	$\pm$	0.09	&	-0.96	$\pm$	0.09	\\
Ni	&	\niII	&	-7.53	$\pm$	0.17	&	-7.54	$\pm$	0.17	&	-5.80	$\pm$	0.04	&	-1.73	$\pm$	0.18	&	-1.74	$\pm$	0.17	\\
\cline{1-7}
\multicolumn{7}{c}{NGC~5253-OBJ-7} \\
\cline{1-7}
C$^{\dag}$	&	\cii	&	-5.81	$\pm$	0.03	&	-5.92	$\pm$	0.03	&	-3.54	$\pm$	0.04	&	-2.27	$\pm$	0.05	&	-2.38	$\pm$	0.05	\\
N	&	\nI	&	-5.70	$\pm$	0.07	&	-5.70	$\pm$	0.07	&	-4.17	$\pm$	0.07	&	-1.53	$\pm$	0.10	&	-1.53	$\pm$	0.10	\\
O$_{\rm P,S}$	&	\oi	&	-3.78	$\pm$	0.17	&	-3.78	$\pm$	0.16	&	-3.31	$\pm$	0.04	&	-0.47	$\pm$	0.17	&	-0.47	$\pm$	0.17	\\
Si$^{\dag}$	&	\siII	&	-6.44	$\pm$	0.02	&	-6.50	$\pm$	0.02	&	-4.49	$\pm$	0.03	&	-1.95	$\pm$	0.04	&	-2.01	$\pm$	0.04	\\
P	&	\pii	&	-7.11	$\pm$	0.17	&	-7.22	$\pm$	0.18	&	-6.59	$\pm$	0.03	&	-0.52	$\pm$	0.18	&	-0.63	$\pm$	0.18	\\
S	&	\sii	&	-5.31	$\pm$	0.12	&	-5.37	$\pm$	0.12	&	-4.88	$\pm$	0.03	&	-0.43	$\pm$	0.12	&	-0.49	$\pm$	0.12	\\
Fe	&	\feii	&	-5.06	$\pm$	0.09	&	-5.08	$\pm$	0.09	&	-4.54	$\pm$	0.04	&	-0.52	$\pm$	0.10	&	-0.54	$\pm$	0.10	\\
Ni	&	\niII	&	-6.91	$\pm$	0.05	&	-6.94	$\pm$	0.05	&	-5.80	$\pm$	0.04	&	-1.11	$\pm$	0.07	&	-1.14	$\pm$	0.05	\\
\cline{1-7}
\multicolumn{7}{c}{NGC~5253-OBJ-9+10} \\
\cline{1-7}
C$^{\dag}$	&	\cii	&	-6.29	$\pm$	0.02	&	-6.35	$\pm$	0.02	&	-3.54	$\pm$	0.04	&	-2.75	$\pm$	0.09	&	-2.81	$\pm$	0.05	\\
N	&	\nI	&	-5.93	$\pm$	0.11	&	-5.91	$\pm$	0.11	&	-4.17	$\pm$	0.07	&	-1.76	$\pm$	0.14	&	-1.74	$\pm$	0.13	\\
O$_{\rm P,S}$	&	\oi	&	-4.09	$\pm$	0.29	&	-4.09	$\pm$	0.28	&	-3.31	$\pm$	0.04	&	-0.78	$\pm$	0.29	&	-0.78	$\pm$	0.29	\\
Si$^{\dag}$	&	\siII	&	-6.52	$\pm$	0.02	&	-6.55	$\pm$	0.01	&	-4.49	$\pm$	0.03	&	-2.03	$\pm$	0.05	&	-2.06	$\pm$	0.03	\\
P	&	\pii	&	-7.57	$\pm$	0.09	&	-7.62	$\pm$	0.09	&	-6.59	$\pm$	0.03	&	-0.98	$\pm$	0.09	&	-1.03	$\pm$	0.09	\\
S	&	\sii	&	-5.46	$\pm$	0.38	&	-5.48	$\pm$	0.38	&	-4.88	$\pm$	0.03	&	-0.58	$\pm$	0.41	&	-0.60	$\pm$	0.38	\\
Fe	&	\feii	&	-5.47	$\pm$	0.08	&	-5.48	$\pm$	0.08	&	-4.54	$\pm$	0.04	&	-0.93	$\pm$	0.12	&	-0.94	$\pm$	0.09	\\
Ni	&	\niII	&	-6.97	$\pm$	0.13	&	-6.98	$\pm$	0.13	&	-5.80	$\pm$	0.04	&	-1.17	$\pm$	0.17	&	-1.18	$\pm$	0.13	\\
\cline{1-7}
\multicolumn{7}{c}{NGC~5253-OBJ-8} \\
\cline{1-7}
C$^{\dag}$	&	\cii	&	-6.05	$\pm$	0.02	&	-6.13	$\pm$	0.02	&	-3.54	$\pm$	0.04	&	-2.51	$\pm$	0.06	&	-2.59	$\pm$	0.05	\\
N	&	\nI	&	-5.59	$\pm$	0.15	&	-5.56	$\pm$	0.15	&	-4.17	$\pm$	0.07	&	-1.42	$\pm$	0.22	&	-1.39	$\pm$	0.16	\\
O$_{\rm P,S}$	&	\oi	&	-4.05	$\pm$	0.23	&	-4.05	$\pm$	0.22	&	-3.31	$\pm$	0.04	&	-0.74	$\pm$	0.23	&	-0.74	$\pm$	0.23	\\
Si$^{\dag}$	&	\siII	&	-6.65	$\pm$	0.03	&	-6.70	$\pm$	0.03	&	-4.49	$\pm$	0.03	&	-2.16	$\pm$	0.06	&	-2.21	$\pm$	0.04	\\
P	&	\pii	&	-7.45	$\pm$	0.24	&	-7.51	$\pm$	0.24	&	-6.59	$\pm$	0.03	&	-0.86	$\pm$	0.27	&	-0.92	$\pm$	0.25	\\
S	&	\sii	&	-5.50	$\pm$	0.18	&	-5.52	$\pm$	0.18	&	-4.88	$\pm$	0.03	&	-0.62	$\pm$	0.21	&	-0.64	$\pm$	0.18	\\
Fe	&	\feii	&	-5.14	$\pm$	0.09	&	-5.15	$\pm$	0.09	&	-4.54	$\pm$	0.04	&	-0.60	$\pm$	0.13	&	-0.61	$\pm$	0.10	\\
Ni	&	\niII	&	-7.33	$\pm$	0.08	&	-7.35	$\pm$	0.08	&	-5.80	$\pm$	0.04	&	-1.53	$\pm$	0.12	&	-1.55	$\pm$	0.08	\\
\cline{1-7}
\multicolumn{7}{c}{NGC~5253-OBJ-5+11} \\
\cline{1-7}
C$^{\dag}$	&	\cii	&	-5.70	$\pm$	0.14	&	-5.78	$\pm$	0.14	&	-3.54	$\pm$	0.04	&	-2.16	$\pm$	0.18	&	-2.24	$\pm$	0.15	\\
N	&	\nI	&	-5.85	$\pm$	0.16	&	-5.81	$\pm$	0.16	&	-4.17	$\pm$	0.07	&	-1.68	$\pm$	0.23	&	-1.64	$\pm$	0.17	\\
O$_{\rm P,S}$	&	\oi	&	-3.93	$\pm$	0.15	&	-3.93	$\pm$	0.14	&	-3.31	$\pm$	0.04	&	-0.62	$\pm$	0.15	&	-0.62	$\pm$	0.15	\\
Si$^{\dag}$	&	\siII	&	-6.22	$\pm$	0.20	&	-6.27	$\pm$	0.20	&	-4.49	$\pm$	0.03	&	-1.73	$\pm$	0.23	&	-1.78	$\pm$	0.20	\\
P	&	\pii	&	-7.15	$\pm$	0.12	&	-7.23	$\pm$	0.12	&	-6.59	$\pm$	0.03	&	-0.56	$\pm$	0.15	&	-0.64	$\pm$	0.12	\\
S	&	\sii	&	-5.56	$\pm$	0.14	&	-5.59	$\pm$	0.14	&	-4.88	$\pm$	0.03	&	-0.68	$\pm$	0.17	&	-0.71	$\pm$	0.14	\\
Fe	&	\feii	&	-5.40	$\pm$	0.06	&	-5.41	$\pm$	0.06	&	-4.54	$\pm$	0.04	&	-0.86	$\pm$	0.10	&	-0.87	$\pm$	0.08	\\
Ni	&	\niII	&	-6.59	$\pm$	0.20	&	-6.61	$\pm$	0.20	&	-5.80	$\pm$	0.04	&	-0.79	$\pm$	0.24	&	-0.81	$\pm$	0.20	\\
\cline{1-7}
\multicolumn{7}{c}{ NGC~5253-OBJ-1234 } \\
\cline{1-7}
%
C$^{\dag}$	&	\cii	&	-6.23	$\pm$	0.03	&	-6.29	$\pm$	0.03	&	-3.54	$\pm$	0.04	&	-2.69	$\pm$	0.07	&	-2.75	$\pm$	0.05	\\
N	&	\nI	&	-6.18	$\pm$	0.04	&	-6.16	$\pm$	0.06	&	-4.17	$\pm$	0.07	& -2.01	$\pm$	0.11	& -1.99	$\pm$	0.09	\\
O$_{\rm P,S}$	&	\oi	&	-4.23	$\pm$	0.12	&	-4.23	$\pm$	0.11	&	-3.31	$\pm$	0.04	&	-0.92	$\pm$	0.12	& -0.92	$\pm$	0.12	\\
Si$^{\dag}$	&	\siII	&	-6.54	$\pm$	0.01	&	-6.57	$\pm$	0.03	&	-4.49	$\pm$	0.03	&	-2.05	$\pm$	0.04	&	-2.08	$\pm$	0.04	\\
P	&	\pii	&	-7.62	$\pm$	0.13	&	-7.67	$\pm$	0.13	& -6.59	$\pm$	0.03	&	-1.03	$\pm$	0.16	&	-1.08	$\pm$	0.13	\\
S	&	\sii	&	-5.69	$\pm$	0.04	&	-5.71	$\pm$	0.04	&	-4.88	$\pm$	0.03	&	-0.81	$\pm$	0.07	&	-0.83	$\pm$	0.05	\\
Fe	&	\feii	&	-5.79	$\pm$	0.06	&	-5.80	$\pm$	0.06	&	-4.54	$\pm$	0.04	&	-1.25	$\pm$	0.10	&	-1.26	$\pm$	0.07	\\
Ni	&	\niII	&	-7.68	$\pm$	0.05	&	-7.69	$\pm$	0.05	&	-5.80	$\pm$	0.04	&	-1.88	$\pm$	0.09	&	-1.89	$\pm$	0.05	\\
\enddata
\vspace{0.2cm}
\textbf{Notes.}
\tablenotetext{a}{ Ionization-corrected ratios in logarithmic scale, computed using interpolation on correction factor from \cite{Hernandez+2020}. The ICF values are listed in Table \ref{Table_ICF_log(N_HI)}. Ions affected by saturation are lower limits indicated with a $\dagger$ symbol. Oxygen abundances are computed using the O column densities derived from the correlation with P and S (N[O$_{\rm P,S}$]) described in Section \ref{subsec:col_dens_abundances}. }
\tablenotetext{b}{ Solar photospheric abundances from \cite{Asplund+21}}
\tablenotetext{c}{ [X/H] = log(X/H)--log(X/H)$_{\odot}$}
\tablenotetext{d}{ [X/H]$_{\rm ICF}$ = log(X/H)$_{\rm ICF}$--log(X/H)$_{ \odot}$}
\end{deluxetable*}

\subsection{Ionized gas abundances from optical emission lines} \label{subsec:ionized_abundances}
   
\begin{deluxetable*}{l|cccccc}												
\tablecolumns{7}								
 \tablecaption{Abundances of the ionized gas \label{Table_Abu_ion_gas} }			
\tablehead{							
\colhead{Quantity}	&	\colhead{OBJ-6}			&	\colhead{OBJ-7}			&	\colhead{OBJ-9+10}			&	\colhead{OBJ-8} 			&	\colhead{OBJ-5+11} 			&	\colhead{OBJ-1234} }			
\startdata																									
N$_{\rm e}$	&	73.19	$\pm$	1.38	&	38.78	$\pm$	4.79	&	48.27	$\pm$	1.50	&	45.20	$\pm$	6.57	&	354.28	$\pm$	1.12	&	157.54	$\pm$	1.24	\\
T$_{\rm e}$(\nii)	&	10121	$\pm$	568.13	&	10020	$\pm$	828.55	&	$>$8093		&	12948	$\pm$	469.01	&	10744	$\pm$	23.77	&	10174	$\pm$	65.97	\\
T$_{\rm e}$(\oiii)$^\dagger$	&	10173	$\pm$	568.13	&	10029	$\pm$	828.55	&	$>$ 7276		&	14212	$\pm$	469.01	&	11063	$\pm$	23.77	&	10248	$\pm$	65.97	\\
	&				&				&				&				&				&				\\
O$^{+}$/H$^{+}\times$  10$^{4}$	&	0.79	$\pm$	0.27	&	0.68	$\pm$	0.39	&	$<$ 1.55		&	0.40	$\pm$	0.09	&	0.31	$\pm$	0.012	&	1.23	$\pm$	0.05	\\
O$^{++}$/H$^{+}\times$ x 10$^{4}$	&	1.10	$\pm$	0.23	&	1.23	$\pm$	0.40	&	$<$ 3.97		&	0.36	$\pm$	0.03	&	1.60	$\pm$	0.011	&	1.53	$\pm$	0.03	\\
O/H $\times$ 10$^{4}$	&	1.89	$\pm$	0.76	&	1.91	$\pm$	1.25	&	$<$ 5.52		&	0.76	$\pm$	0.18	&	1.91	$\pm$	0.076	&	2.76	$\pm$	0.12	\\
12+log(O/H)	&	8.28	$\pm$	0.18	&	8.28	$\pm$	0.29	&	$<$ 8.74		&	7.88	$\pm$	0.10	&	8.28	$\pm$	0.017	&	8.44	$\pm$	0.02	\\
	&				&				&				&				&				&				\\
N$^{+}$/H$^{+}\times$  10$^{5}$	&	0.43	$\pm$	0.07	&	0.39	$\pm$	0.09	&	$<$ 0.88		&	0.27	$\pm$	0.02	&	0.42	$\pm$	0.002	&	0.39	$\pm$	0.01	\\
N/H$\times$  10$^{5}$	&	1.03	$\pm$	0.45	&	1.10	$\pm$	0.77	&	$<$ 3.14		&	0.52	$\pm$	0.13	&	2.55	$\pm$	0.102	&	0.87	$\pm$	0.04	\\
12+log(N/H)	&	7.01	$\pm$	0.19	&	7.04	$\pm$	0.30	&	$<$ 7.50		&	6.71	$\pm$	0.01	&	7.41	$\pm$	0.008	&	6.94	$\pm$	0.01	\\
log(N/O)	&	-1.26	$\pm$	0.26	&	-1.24	$\pm$	0.42	&	 -1.25		&	-1.17	$\pm$	0.15	&	-0.87	$\pm$	0.024	&	-1.50	$\pm$	0.03	\\
	&				&				&				&				&				&				\\
S$^{+}$/H$^{+}\times$  10$^{6}$	&	2.58	$\pm$	0.38	&	2.15	$\pm$	0.53	&	$<$ 5.23		&	1.55	$\pm$	0.12	&	0.87	$\pm$	0.005	&	1.77	$\pm$	0.03	\\
S$^{2+}$/H$^{+}\times$  10$^{6}$	&	1.82	$\pm$	0.04	&	1.57	$\pm$	0.04	&	$<$ 1.66		&	1.43	$\pm$	0.08	&	1.87	$\pm$	0.007	&	5.03	$\pm$	0.03	\\
S/H$\times$  10$^{6}$	&	5.46	$\pm$	0.66	&	4.77	$\pm$	0.92	&	$<$ 9.49		&	3.54	$\pm$	0.28	&	4.84	$\pm$	0.018	&	8.34	$\pm$	0.12	\\
12+log(S/H)	&	6.74	$\pm$	0.05	&	6.68	$\pm$	0.08	&	$<$ 6.98		&	6.55	$\pm$	0.03	&	6.69	$\pm$	0.002	&	6.92	$\pm$	0.01	\\
log(S/O)	&	-1.54	$\pm$	0.18	&	-1.60	$\pm$	0.30	&	 -1.77		&	-1.33	$\pm$	0.11	&	-1.60	$\pm$	0.02	&	-1.52	$\pm$	0.02	\\
	&				&				&				&				&				&				\\
Fe$^{+}$/H$^{+}\times$  10$^{6}$	&	2.41	$\pm$	0.33	&	1.93	$\pm$	0.43	&	$<$ 4.82		&	2.01	$\pm$	0.15	&	0.31	$\pm$	0.003	&	0.59	$\pm$	0.01	\\
Fe$^{2+}$/H$^{+}\times$  10$^{6}$	&	0.75	$\pm$	0.15	&	0.52	$\pm$	0.19	&	$<$ 2.86		&	0.35	$\pm$	0.03	&	0.19	$\pm$	0.002	&	0.26	$\pm$	0.01	\\
Fe/H$\times$  10$^{6}$	&	7.53	$\pm$	3.20	&	7.12	$\pm$	4.93	&	$<$ 22.64		&	5.03	$\pm$	1.23	&	2.48	$\pm$	0.10	&	1.74	$\pm$	0.08	\\
12+log(Fe/H)	&	6.88	$\pm$	0.18	&	6.85	$\pm$	0.30	&	$<$ 7.35		&	6.70	$\pm$	0.11	&	6.40	$\pm$	0.02	&	6.24	$\pm$	0.02	\\
log(Fe/O)	&	-1.40	$\pm$	0.25	&	-1.43	$\pm$	0.41	&	 -1.39		&	-1.18	$\pm$	0.15	&	-1.89	$\pm$	0.02	&	-2.20	$\pm$	0.03	\\
\enddata							
\vspace{0.2cm}								
\textbf{Notes. }
Direct method measured abundances in the ionized gas phase. Integrated spectra were extracted from 6 COS equivalent circular apertures (2\farcs{5} diameter) on MUSE observations centered on each of the targeted stellar clusters in NGC~5253. $^\dagger$Derived from T$_{\rm e}$(\nii) using the relationship of \citet{Garnett+1992}, as described in Section~\ref{subsec:ionized_abundances}.
\end{deluxetable*}	
In order to make an accurate comparison between the neutral and ionized gas throughout NGC~5253, it was important to extract the optical spectra using the exact pointing and aperture size used to collect the UV spectra. To do this, we utilized the VLT/MUSE observations discussed in Section~\ref{sec:observations}, integrating the spectral cube across a circular 2\farcs{5} aperture centered on each of the clusters. Using these spectra, we calculated chemical abundances following the methodology outlined in \citet{James+2017} and \citet{James+2020}. To summarize here, chemical abundances are calculated using the `direct method', where abundance measurements are based on the physical conditions of the gas (i.e., utilizing electron temperature, \elt, and electron density, \eld) and attenuation corrected line fluxes. Each \hii\ region is modelled by three separate ionization zones (low, medium, and high), and the abundance calculations for ions within each zone are made using the temperature within the respective zone. We refer the reader to \citet{James+2017} for further details and \citet{Berg+2021} for a clear demonstration of the zones, the ions that reside within them, and the temperature and density-sensitive lines utilized for each zone. 

\begin{figure*}[htb!]
\includegraphics[width=0.9\linewidth]{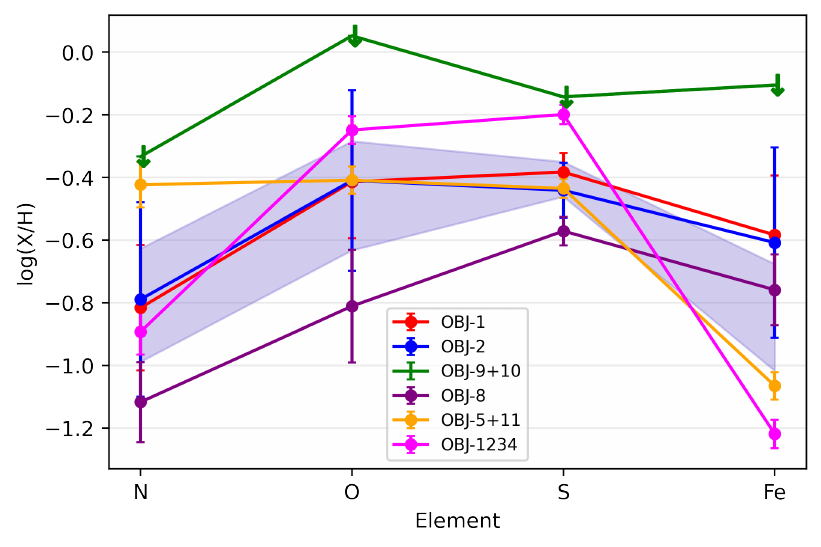}
\figcaption{Ionized gas ionization-corrected abundances per ion for all the SF clusters in NGC~5253, derived using the direct method. Different colors are assigned to each cluster as indicated in the label, coinciding with the colors used in Figure \ref{fig_ICF_final_abundances}. The shaded area indicates the average abundances within errors for all the targeted SF regions. Abundance values for OBJ-9+10 are represented by downward arrows since they correspond to upper limits, due to low S/N limitations in the $T_{e}$ estimates.  \label{fig_ion_abundances}}

\end{figure*}

The only difference between the methods used here and that of \citet{James+2017} is that due to MUSE's restricted wavelength coverage in the blue, here we do not have access to the \foiii$\lambda$4363 auroral line to derive the electron temperature (\elt) of the high ionization zone. Instead, we have the \fsiii~$\lambda$6312 auroral line to derive \elt(\fsiii), which pertains to the intermediate ionization zone and the \fnii~$\lambda$5755 auroral line to derive \elt(\fnii), which pertains to the low ionization zone. In order to derive the temperature of the high ionization zone (where O$^{2+}$ resides) we use the relationship between \elt(\fnii) and \elt(\foiii) from \citet{Garnett+1992}. We chose not to use \elt(\fsiii) to derive \elt(\foiii) because the \elt(\fsiii) values are found to be rather high (compared to values in the literature for this galaxy), which may be a result of the \fsiii~$\lambda$9069 line being contaminated by poor sky subtraction \citep[as seen in][]{James+2020,Arellano-Cordova+2022}. In all cases, the \fnii~$\lambda$5755 line was detected witn S/N$>3$, thus allowing for reliable \elt(\fnii) and \elt(\foiii) estimates, with the exception of OBJ-9+10. For this pointing, we use an upper limit on the \fnii~$\lambda$5755 line flux to derive a lower limit on the \elt(\fnii) and \elt(\foiii) values, and subsequent upper limits on all ionic and elemental abundances. As a result, we are unable to provide limits or errors on the the elemental abundance ratios for OBJ-9+10 (N/O, S/O, Fe/O) and should thus be treated with caution.

Overall, we follow the ICF recommendations of \citet{Berg+2021}, who cover ICF(S$^+$+S$^{2+}$) from \citet{Thuan+1995} and ICF(N$^+$) of \citet{Peimbert+1969} in their analysis. Regarding iron in particular, in this work we derive Fe$^{2+}$/H$^+$ and Fe$^{+}$/H$^+$ using \ffeiii~$\lambda$4986 and \ffeii~$\lambda$8619, respectively. As shown in \citet{Rodriguez+2003}, \ffeii~$\lambda$8619 is known to be almost completely insensitive to the effects of UV pumping (unlike most other \ffeii\ emission lines), and can therefore lead to a reliable measurement of Fe$^{+}$/H$^+$. For Fe$^{2+}$/H$^+$ abundances, we do not have wavelength coverage of the typically used \ffeiii\ lines, with \ffeiii~$\lambda$4986 being the strongest available. Since we do not detect any \ffeiv\ or \ffev\ lines in our spectra, it was necessary to adopt an ICF for iron to account for unseen ionic abundances. Here we adopt the ICF(Fe$^{2+}$) from \citet{Izotov+2006}. It should be noted that several works have been dedicated to exploring the most accurate ICF for iron, which can lead to large amounts of uncertainty in Fe/H as shown by \cite{Rodriguez+2005} and \cite{Delgado-Inglada+2019}. The \eld, \elt, ionic and elemental oxygen abundances, nitrogen abundances, iron abundances and nitrogen-to-oxygen abundance ratios for each \hii\ region are shown in Table \ref{Table_Abu_ion_gas}. Figure \ref{fig_ion_abundances} shows the distribution of ionized gas abundances per element (N, O, S and Fe), per SF region with respect to the average plus average errors of the sample (excluding OBJ-9+10), represented with a shaded region. The ionized phase abundance values for OBJ-9+10 are represented as downward arrows indicating upper limits. OBJ-1234 has the highest O and S ionized gas abundances while OBJ-8 shows the lowest N, o and S abundances, all lower than the average of the sample. OBJ-5+11 in the center of NGC~5253 shows the highest N abundance.
	

\section{Discussion} \label{sec:Discussion}

As an attempt to determine the bulk of metals between gas phases and the mixing timescales, in recent theoretical studies using simulations, the abundance features of present-day stellar populations are contrasted with different models of metal mixing/diffusion in the ISM \citep[e.g.,][]{Escala+2018}. Current simulations have shown that the majority of metals may reside in the cold gas, although it has lower metallicities than the hot ionized gas \citep[e.g.,][]{Emerick+2019, Arabsalmani+2023}. By quantifying abundance offsets among different phases, we can ascertain the primary reservoir containing the highest concentration of metals and the total metal content. Observations have also been performed in order to explore offsets between gas phases, as in the study performed by \cite{Lebouteiller+2009} to determine metal enrichment in the galaxy Pox 36 using FUSE UV spectroscopic data. They found the neutral gas phase of Pox 36 is 7 times metal-deficient compared to the ionized gas in \hii\ regions. Another example of such a study was performed by \cite{Lebouteiller+2013} on the extremely metal-poor local galaxy I Zw 18, with the aim to investigate the relation between star-formation history (SFH) and metallicity evolution and to constrain the spatial and temporal scales at which \hi\ and \hii\ regions are enriched. Metals are released mainly by massive stars during starburst episodes, enriching first the hot phase (10$^{6}$ K) constrained to small spatial scales and mixing in short timescales of the order of few Myrs \cite{Kunth+Sargent_1986}, while mixing timescales in the neutral phase could be as long as 10$^{9}$ yr \cite{Lebouteiller+2013}. The authors found lower abundances (by a factor of $\sim$2) and metallicities (1/46 Z$_{\odot}$ vs. 1/31 Z$_{\odot}$) in the \hi\ regions compared to the ionized gas regions. Another multi-phase co-spatial abundance study on the metal-rich spiral galaxy M83 was performed by \cite{Hernandez+2021}, showing comparable abundances of the multi-phase gas and stellar components in the galaxy disk, implying that the mixing of newly synthesized elements from massive stars takes longer than their average lifetimes $\sim$10 Myr at spatial scales of $\sim$100 pc. The accuracy of the total metal mass estimation of a galaxy from single-phase metallicity measurements – as is commonly done in galaxy evolution studies – may strongly rely on the magnitude of this offset. Regarding the distribution of metals, it is important to investigate the presence of localized variations between the two phases and what processes regulate/produce those variations. 

The multi-region detailed abundance analysis performed on NGC~5253 using UV and optical data-sets, allows us to simultaneously explore the distribution of different elements in the neutral and ionized gas phases as a function of the age of the targeted clusters and their distance from the center of the galaxy. This in order to search for abundance in-homogeneities and differences related to gas phases, spatial distribution, ages of the stellar populations and evolution timescales of the processes responsible for chemical enrichment.

\subsection{Abundance distribution in the neutral gas} \label{subsec:Abu_Dist_neutral_gas}

\begin{figure*}[htb!]
\centering
 \gridline{\fig{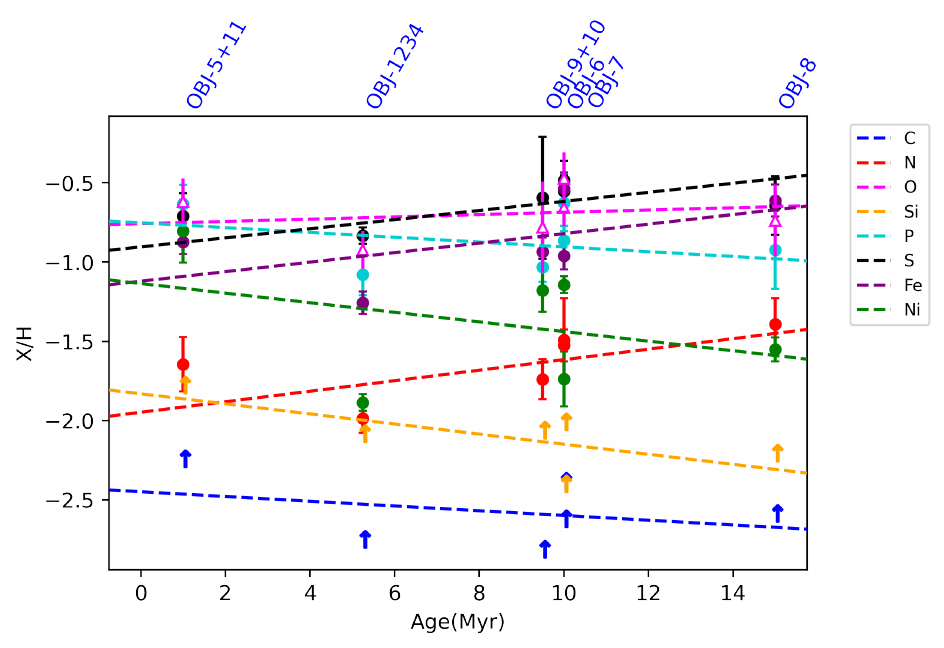}{0.7\textwidth}{}}
 \vspace{-1cm}
 \gridline{\fig{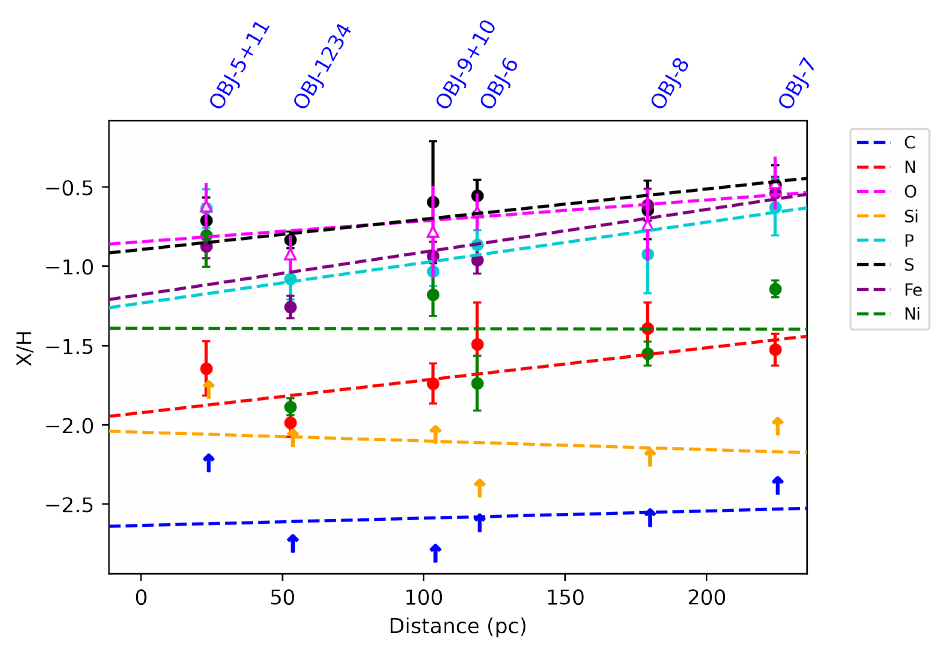}{0.7\textwidth}{}}
 \vspace{-1cm}
\figcaption{Ionization-corrected neutral gas abundances per ion as a function of age (upper panel) and distance from center (lower panel) of NGC~5253. Abundances are represented by dots and triangles with a color per ion. The triangles correspond to the O abundances computed using the P/S/O proxy described in Section \ref{subsec:col_dens_abundances}. The best LTS linear regression fits are represented by the dashed lines with an assigned color to each ion. The regression numerical parameters linked to these dashed lines are presented in Table \ref{Table_lin_reg_param_X_H_neu}. Abundance distributions for \cii\ and \siII\ are represented by arrows indicating lower limits. \label{fig_Average_Abundances_Age_radius}}

\end{figure*}

In the case of the neutral gas, elemental abundances are shown as a function of age and distance from center in the upper and lower panels of Figure \ref{fig_Average_Abundances_Age_radius}, respectively. In order to quantify the distribution of the different abundances as a function of age and radius we applied a robust least trimmed squares (LTS) linear regression method using the \texttt{ltsfit} code \citep{Cappellari+2013a}. This linear fitting method takes into account the error of the variables involved and offers the possibility to clip outlier points. In the case of our somewhat limited sample size, we decided to conserve all the points for the line regressions. As outputs of linear regression fitting we recover the slope ($m$), intercept ($b_{0}$) and the root mean square (RMS) error. We additionally performed a Pearson correlation test to asses the level of correlation between the abundances and the ages of the SF regions or radius. Pearson parameters $r$ and $p$ denote the probability of correlation and the probability of a random distribution, respectively. We consider there is a correlation when the combination of Pearson coefficients fulfill the condition: $r_{\rm P} >$ 60 and $p_{\rm P} <$ 25. All the numerical values derived from the linear regression fits and the Pearson correlation test for X/H are presented in Table \ref{Table_lin_reg_param_X_H_neu}. 
 
Figure \ref{fig_Average_Abundances_Age_radius} shows the best linear models for the abundance distributions as dashed lines. Strong correlations as a function of radius are observed for N/H, S/H and Fe/H ranging with an r Pearson value ($r_{\rm P}$) between 0.65 and 0.79. However, the slopes themselves of X/H as a function of radius are negligible in size ($<$0.003) evidencing a low variation of abundances along the radius of the galaxy. In contrast, slopes of X/H as a function of age are one order magnitude larger around 0.03. A correlation for N/H ($r_{\rm P}$ = 0.60) and anti-correlation for Si/H ($r_{\rm P}$ = -0.69) are observed as a function of age. As was discussed above, abundances obtained for \cii\ and \siII\ are lower limits since these lines are impacted by saturation and geocoronal emission (for \siII $\lambda$1304).

Average abundances for C, N, O, Si, P, S, Fe and Ni correspond to -2.56 dex, -1.63 dex, -0.70 dex, -2.09 dex, -0.86 dex, 0.64 dex, -0.86 dex and -1.38 dex, respectively. For OBJ-1234 most abundance values are sightly lower than the sample average; this could be due to high \hi\ column densities, as shown in Figure \ref{fig_H_col_dens}. Nevertheless, OBJ-9+10 with the same log(N(\hi)) shows column densities inside the average of the sample, so OBJ-1234 could be a region of intrinsic lower metallicity with respect to the average of the galaxy. \cite{Lopez-Sanchez+2007} detected and characterized an infalling diffuse, low metallicity \hi\ cloud from radio emission observations of NGC~5253. According to \cite{van_den_Bergh_1980}, this neutral gas structure might have been caused by a tidal interaction between M83 and NGC~5253 (with a sky projected separation of $\sim$100 Kpc) that probably started around 100 Myr ago. The kinematics of the \hi\ in-falling cloud was studied in \cite{Lopez-Sanchez+2012}, who found a very turbulent gas with complex kinematics not following a rotating disk. This could be due to the combination of intense SF activity in the center of the galaxy, the release of strong stellar winds and gas losses due to ram-pressure stripping from the interaction with the dense IGM of the galaxy group where NGC~5253 resides \citep{Lopez-Sanchez+2012}. From the \hi\ contours (ATCA data) shown in Figure \ref{fig_NGC5253_COS_MUSE}, we see that the maximum density of the cloud coincides with the location of OBJ-6 and OBJ-7. However, the radio data has a much lower resolution (beam size 13\farcs{6} $\times$ 7\farcs{5}) so it is not directly comparable to the pc scales at which we are analysing UV and optical data. As outlined by \cite{Lopez-Sanchez+2012}, this low metallicity cloud is the likely source of the starburst episodes in NGC~5253 and might be altering the ISM abundance distribution and the kinematics of the gas, but the radio data itself is not enough to evaluate localized changes, so we rely on our results from the UV Ly$\alpha$ absorption profile column densities to map the distribution of the H\,{\rm I} in our sample. In this respect, we do not see a relation between the neutral gas cloud in radio and the lower abundances found for OBJ-1234.

\begin{deluxetable}{cccccc}[htb!]
\label{Table_lin_reg_param_X_H_neu}
\tablecaption{Linear regression parameters and correlation coefficients of neutral gas abundance X/H distributions as a function of Age and Radius.}
\tablehead{
 \colhead{X/H$_{neu}$} & \colhead{$m$\tablenotemark{a}} &\colhead{$b_{0}$\tablenotemark{a}} &  \colhead{RMS\tablenotemark{a}} & \colhead{$r_{\rm P}$\tablenotemark{b}} & \colhead{$p_{\rm P}$\tablenotemark{b}}  }
\startdata
\multicolumn{6}{c}{ Age } \\
\hline 
C/H	&	-0.015	$\pm$	0.022	&	-2.45	$\pm$	0.22	&	0.22	&	-0.39	&	0.44	\\
N/H	&	0.033	$\pm$	0.020	&	-1.95	$\pm$	0.18	&	0.19	&	0.60	&	0.21	\\
O/H	&	0.007	$\pm$	0.018	&	-0.76	$\pm$	0.15	&	0.17	&	0.06	&	0.91	\\
Si/H	&	-0.032	$\pm$	0.020	&	-1.83	$\pm$	0.19	&	0.17	&	-0.69	&	0.13	\\
P/H	&	-0.015	$\pm$	0.021	&	-0.75	$\pm$	0.18	&	0.21	&	-0.26	&	0.61	\\
S/H	&	0.029	$\pm$	0.014	&	-0.90	$\pm$	0.11	&	0.14	&	0.52	&	0.29	\\
Fe/H	&	0.030	$\pm$	0.023	&	-1.12	$\pm$	0.22	&	0.25	&	0.52	&	0.29	\\
Ni/H	&	-0.030	$\pm$	0.041	&	-1.14	$\pm$	0.39	&	0.42	&	-0.38	&	0.46	\\
\hline															
\multicolumn{6}{c}{ Radius } \\															
\hline 															
C/H	&	4.58$\times$10$^{-4}$	$\pm$	0.001	&	-2.64	$\pm$	0.19	&	0.24	&	0.03	&	0.95	\\
N/H	&	0.002	$\pm$	0.001	&	-1.92	$\pm$	0.14	&	0.18	&	0.65	&	0.16	\\
O/H	&	0.001	$\pm$	0.001	&	-0.85	$\pm$	0.11	&	0.15	&	0.52	&	0.29	\\
Si/H	&	-0.001	$\pm$	0.001	&	-2.05	$\pm$	0.18	&	0.22	&	-0.37	&	0.46	\\
P/H	&	0.003	$\pm$	0.001	&	-1.23	$\pm$	0.15	&	0.10	&	0.22	&	0.67	\\
S/H	&	0.002	$\pm$	0.001	&	-0.89	$\pm$	0.07	&	0.11	&	0.75	&	0.08	\\
Fe/H	&	0.003	$\pm$	0.001	&	-1.18	$\pm$	0.14	&	0.18	&	0.79	&	0.06	\\
Ni/H	&	-2.51$\times$10$^{-4}$	$\pm$	0.003	&	-1.39	$\pm$	0.37	&	0.46	&	-0.06	&	0.90	\\
\enddata
\tablenotetext{a}{Linear regression parameters: slope ($m$), intercept ($b_{0}$) and RMS derived with \texttt{ltsfit}. Figure \ref{fig_Average_Abundances_Age_radius} shows the X/H distributions for the neutral gas and their respective linear best fit models as a function of age and radius. }
\tablenotetext{b}{Pearson correlation $r_{\rm P}$ and $p_{\rm P}$ coefficients.}
\end{deluxetable}


Since NGC~5253 is known for harboring areas of intriguing N enrichment in its ionized gas, here we also explore the relative abundance ratio N/O of the neutral phase. In order to quantify the variation in the distribution of the N/O relative abundance we applied a LTS linear regression as a function of age and radius complemented with a correlation analysis. Linear fit parameters and correlation coefficients are shown in Table \ref{Table_lin_reg_param_X_Y_ion_neu}, and further discussed in Section \ref{sec:comparison_neu_ion}. 

From the individual O and N element abundance distributions in Table \ref{Table_lin_reg_param_X_H_neu}, O/H shows low correlation coefficients ($r_{\rm P}$= 0.06 and 0.52 for age and radius, respectively) and a uniform trend as a function of age and radius. On the other hand, N/H shows a strong correlation as a function of age ($r_{\rm P}$= 0.75), with a subtle increase ($m$ = 0.033), while radially N/H shows a strong correlation ($r_{\rm P}$= 0.65) although with a flat gradient ($m$ = 0.002). This suggests slightly lower neutral N abundances in the center of the galaxy where the youngest clusters are located. 

Through the analysis of the N/O relative abundances we can trace the variation of the N content contrasted with a relatively constant oxygen distribution. While no correlation was found for the radial distribution of N/O ($r_{\rm P}$= 0.36), a strong correlation was obtained for N/O with age ($r_{\rm P}$= 0.75). From the linear best fits for the neutral N/O values we observe an increase as a function of age with a slope of $m$ = 0.023. The observed increment of the N/O abundance as a function of age ($\sim$0.41 dex) might be related to the presence of Nitrogen rich WR stars (WN type) around 2--5\,Myr which enrich the ISM with mainly N (\citealt{Schaerer+1997}, \citealt{Brinchmann+2008}, \citealt{James+2013b}, \citealt{Westmoquette+2013}). However, this has only ever been recorded in the ionized gas phase. For older clusters the time elapsed from the WR episodes may have been enough to detect the enrichment in the cold phase, i.e., of the order of the mixing timescale for the neutral phase. Since the youngest clusters are located in the center of the galaxy and the oldest in the outskirts we can evidence the increase of N abundance in both relations. The oldest cluster OBJ-8 (15 Myr) has the highest N/O relative abundance in the neutral phase while the youngest OBJ-5+11 and OBJ-1234 (1 -- 5 Myr) show the lowest. This is particularly interesting because from the study performed by \cite{Westmoquette+2013}, the WR spectral signatures of both types WC and WN stars were detected in a region coinciding with OBJ-1234 with an average age of 5\,Myr \citep{Calzetti+2015}. They attribute these stars as the cause of the high N/O measurement in the ionized gas within the same region (Section~\ref{sec:ionized_gas}). However, in the neutral phase, we do not see yet evidence of N enrichment in OBJ-5+11 (or OBJ-1234).  Interestingly, this result is in contrast with previous optical and NIR studies on the warm ionized phase confirming a N/O enhancement in the central region of NGC~5253 ($\sim$50 pc, \citealt{Monreal-Ibero+2012}, \citealt{Westmoquette+2013}), which we further describe below. This may be due to longer timescales for the chemical mixing transition from the ionized warm phase to the cold neutral phase to be detectable.

\subsection{Abundance distribution in the ionized gas}\label{sec:ionized_gas}

The metal content of the ionized gas throughout NGC~5253 has been studied extensively in the literature \citep[see ][for a complete review]{Westmoquette+2013}, mostly owing to its enhanced nitrogen abundance at the center of the galaxy. As such, here we provide a brief description of the ionized gas abundance distributions throughout NGC~5253 (numerical values presented in Table \ref{Table_Abu_ion_gas}), along with a quantitative analysis in regards to their relation with the age and radial distributions of the studied SF regions. This with the aim of providing context and background for the subsequent comparative analysis of the neutral and ionized gas abundances.  

A degree of inhomogeneity is seen in the ionized gas abundances of all the elements assessed here - Fe, S, O, and N. In Figure~\ref{fig_metal_N_H_N_O_ion_neu_radius_age_cm} of Section \ref{sec:comparison_neu_ion}, we show the radial distribution of O/H, N/H and N/O of the ionized gas for each of the COS pointings (star symbols). LTS linear regressions were applied to characterize these distributions as a function of age and radius. The distribution of the ionized phase O/H shows an anti-correlation with age and radius, characterized by the equations O/H$_{ion}$ = $-$0.028$\times$age(Myr) + 8.44 and O/H$_{ion}$ = $-$0.002$\times$R(pc) + 8.42, respectively, with OBJ-1234 and OBJ-8 showing the highest and lowest ionized gas O/H abundances (difference of 0.56 dex), respectively. The negative correlation with age is strong ($r_{\rm P}$= $-$0.72), inferring that at higher ages O/H$_{ion}$ abundances diminish in the ionized phase. However the correlation of ionized O/H with radius is far weaker ($r_{\rm P}$= $-$0.49 with $p_{\rm P}$ = 0.40). This inhomogeneity is echoed in the N/H ionized gas abundances, albeit with a slightly lower level of dispersion, with OBJ-5+11 and OBJ-8 showing the highest and lowest values (difference of 0.69 dex), respectively. The N/H$_{ion}$ distribution shows a strong negative correlation with age ($r_{\rm P}$= $-$0.86), while an ambiguous correlation for the radial O/H distribution ($r_{\rm P}$= $-$0.57 with $p_{\rm P}$ = 0.32). The linear models for the N/H distribution with age can be represented by N/H$_{ion}$ = $-$0.048$\times$age(Myr) + 7.37. 

Regarding the ionized gas N/O distribution, while overall negative slopes can be fit as a function of age and radius, there is no significant correlation ($r_{\rm P}$=-0.17 and $r_{\rm P}$=-0.29, respectively). Considering that the O abundance of the ionized gas for OBJ-5+11 is close to the average O/H of the sample, the detected high N/O relative abundance is likely due to an ongoing N-enhancement process in this region. According to these results we detect a N-enrichment for the ionized gas N/H (higher slopes, larger variation gradients and strong correlation coefficients with respect to a flat O/H distribution), that shows a peak at the center of the galaxy (around OBJ-5+11 and OBJ-1234) and decreases progressively with age and radius.

To further validate our ionized gas calculations for N/O, in Figure \ref{fig_N_O_abu_ion_neu_radius_center_age_lit_cm_A} we also compare our findings with those from multiple abundance studies of the ionized phase of NGC~5253 in the literature (\citealt{Walsh+1987, Walsh+1989}, \citealt{Kobulnicky+1997}, \citealt{Lopez-Sanchez+2007}, \citealt{Monreal-Ibero+2012}, \citealt{Westmoquette+2013}). Our ionized phase N/O values agree within errors at all locations with those values in the literature. Adding the literature values, the decreasing behavior of the N/O abundance in the ionized gas as a function of radius is more evident.

N and O are the most relevant elements of this study. They arise from multiple sources such as winds from massive stars, or the instabilities of evolved medium mass stars, which happen at different timescales. N is produced mainly by two types of stellar phases via the CNO process: i) stellar winds from massive ($>$ 50 M$_{\odot}$) WR type WN stars releasing N around 2 -- 5 Myr; ii) evolved stars of intermediate mass (3  -- 8 M$_{\odot}$) in the Asymptotic Giant branch (AGB) stage (with main sequence timescales of 10$^{7}$ -- 10$^{9}$ yrs), where N is synthesised from carbon in the helium burning shell of these stars \citep[and references therein]{Walsh+1987}. A photometric SED fitting study performed by \cite{Grijs+2013} in several SF regions in NGC~5253 shows that the contribution of the AGB component to the SED is important in stellar populations around 10$^{8}$ to 10$^{9}$ yrs when AGB stars account for a bolometric luminosity fraction of 25 - 40 $\%$ (\citealt{Charlot1996}; \citealt{Schulz+2002}). They found stellar populations with metallicities Z $<$ 0.5 Z$_{\odot}$ are less affected by the AGB component. The presence of WR of types WC and WN in NGC~5253 has been reported several times in the literature (see, \citealt{Walsh+1987, Schaerer+1997, Westmoquette+2013}) due to the detection of broad emission of N~{\sc iii} $\lambda$4650 \AA\ and \fciv\ $\lambda$5808 \AA (the blue and red `WR bumps', respectively). The very young ages of the SF regions targeted in this study (1 -- 15 Myr) and the lack of evidence for an O enhancement (mainly produced by supernovae Type II; see next paragraph) suggest that AGB episodes are not happening yet or are not yet detectable due to the low fraction of the luminosity function they contribute at these early stages, as was discussed above. Therefore, the N enrichment observed in our results and reported in the literature for the ionized phase is mainly produced by WN stars in the youngest clusters (OBJ-5+11, OBJ-1234 and OBJ-9+10).

Another source of chemical enhancement are supernovae Type II (SNe II)  explosions of massive stars (8 -- 40 M$_{\odot}$); producing O, Ne, Mg, Al, and other nucleosynthesis products, with reference timescales of 10$^{7}$ years for a 15 M$_{\odot}$ star \citep{Woosley_Janka2005}. SNe II explosions are highly energetic episodes that release metals in the hot diffuse phase (T $\sim$ 10$^{6 - 7}$K), which are detectable in radio and X-ray observations. \cite{Beck+1996} studied the centimeter-wavelength radio emission of NGC~5253 that shows a flat continuum of thermal free-free emission originating from the \hii\ regions, contrasted with a low fraction of synchrotron emission from SNe remnants at the center of the galaxy \citep{Turner+1998}. This suggests evidence of only a few SNe episodes in these young SF regions, thus their contribution to metal enrichment is expected to be very low. \cite{Summers+2004} performed a study of the X-ray emission in NGC~5253 from Chandra and XMM Newton observations, identifying both discrete and diffuse emission sources located in a region of 1 kpc in diameter from the center with a spatial distribution coinciding with the \ha\ emission map. These X-ray sources are attributed to OB associations or individual star clusters producing bubbles and superbubbles of ionized gas.

With regards to Fe/H, perhaps unsurprisingly, the youngest clusters (OBJ-5+11 and OBJ-1234) show the lowest Fe/H abundances in the ionized gas, lower by almost 0.6 dex compared to the cluster with the highest Fe/H (OBJ-6). Within the uncertainties, OBJ-6, 7 and 8 have similar Fe/H values in the ionized gas. 

As another alpha element, the picture for S/H is similar to O/H, with a slightly lower gradient of $\sim$0.4~dex among the clusters (compared to 0.56 dex for O/H). OBJ-1234 shows the highest S/H ionized gas abundance, whereas the oldest cluster (OBJ-8) has the lowest. 


\subsection{Comparison of neutral and ionized gas abundances}\label{sec:comparison_neu_ion}

A direct comparison between ionized and neutral gas allows to confirm or discard the presence of abundance differences among phases with the purpose of determining the complete context of chemical evolution and the physical processes involved in global and localized enrichment.

\begin{figure}[htb!]
\centering
 \gridline{\fig{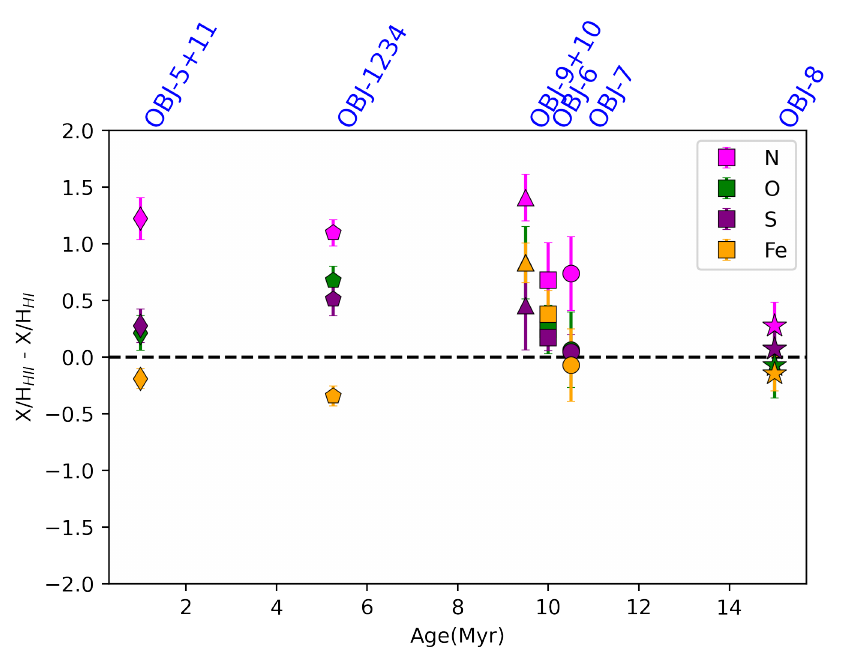}{0.49\textwidth}{}}
 \vspace{-1cm}
 \gridline{\fig{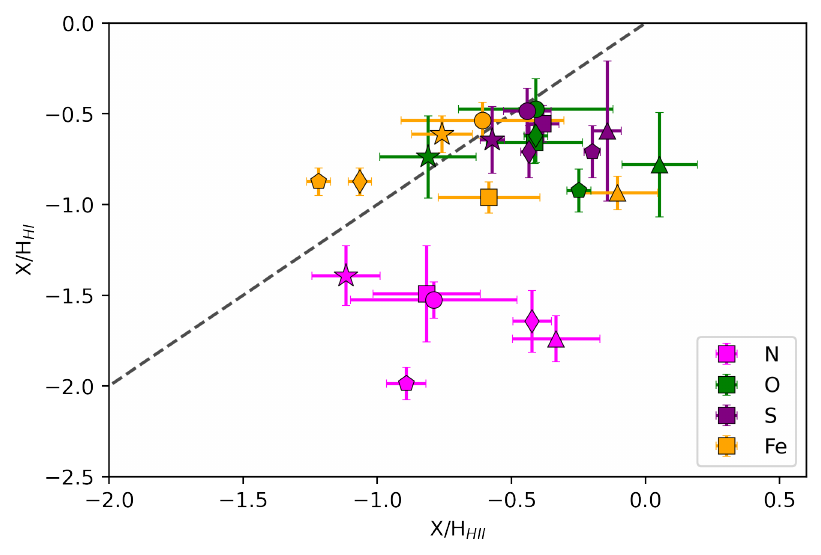}{0.49\textwidth}{}}
 \vspace{-1cm}

\figcaption{Comparison of the ionized and neutral gas abundances. Different symbols correspond to different targets: squares (OBJ-6), circles (OBJ-7), triangles (OBJ-9+10), stars (OBJ-8), diamonds (OBJ-5+11) and pentagons (OBJ-1234). Different colors correspond to different elements. Upper panel: Phase abundance difference X/H$\rm _{HII} -$ X/H$\rm _{HI}$ as a function of age. The dashed line indicates zero difference. Bottom panel: Ionized gas abundances vs. neutral gas abundances. The dashed line indicates the 1:1 line relation. These diagrams show higher abundances for the ionized phase, except for Fe which shows similar abundance values in both gas phases. \label{Diff_ionized_neutral_abundances}} 

\end{figure}

In order to quantify the variation in abundances between gas phases we apply a simultaneous co-spatial analysis obtaining abundances for a set of elements (N, O, S and Fe) measured in both the ionized and neutral gas phases. In Figure \ref{Diff_ionized_neutral_abundances} we present the abundance offset X/H$_{HII}$--X/H$_{HI}$ as a function of cluster age (upper panel) and the comparison between abundances in both gas phases X/H$_{HI}$ vs. X/H$_{HII}$ (bottom panel). Overall, ionized gas abundances are larger than neutral gas abundances. The largest difference of 0.71 dex on average corresponds to N, as well as the largest dispersion in the diagram with respect to the 1-1 line (see the lower panel of Figure \ref{Diff_ionized_neutral_abundances}). The youngest clusters, OBJ-5+11 and OBJ-1234, as well as OBJ-9+10 show the largest offsets for N, O and S. OBJ-9+10 has the largest variation for Fe. On average Fe offsets are the smallest having similar abundances in both phases, this is related to the fact that Fe is only produced in minimal amount by SNe II events and the bulk of Fe comes from supernovae Type Ia (SNe Ia) episodes (generating Fe-peak elements) which happen on timescales of the order of a few Gyr, much longer than the age of the sampled clusters. N, instead, is already produced by WR stars ($\sim$2 -- 5 Myr). The distributions of S and O are similar in the X/H$_{HI}$ vs. X/H$_{HII}$ diagram, this is expected due to the common origin of these alpha elements. Few targets have similar abundance values in both gas phases for Fe, S and O. For OBJ-5+11, OBJ-1234 and OBJ-8, Fe has a slightly higher abundance in the neutral phase, this could be due to the high uncertainty in determining the ICF factor for Fe in \hii\ regions. For Fe, O and S there is a larger dispersion in the ionized gas abundances compared to those in the neutral gas (see the bottom panel of Figure \ref{Diff_ionized_neutral_abundances}). 

Larger abundances in the ionized gas might be a consequence of different mixing/cooling timescales of metals taking into account the layered structure of the ionized regions in terms of temperature, densities and ionization level \citep{Berg+2021}. Ionized gas directly surrounds massive stars which are responsible for the production of metals throughout their evolution, so we would expect to see increased levels of metals in that phase first relative to the neutral phase, where the injection of metals depends on the conditions of the gas transition and the time needed for cooling, and eventual mixing. 

\begin{deluxetable}{cccc}[htb!]					
\label{Table_lin_reg_param_X_Y_ion_neu}									
\tablecaption{Linear regression parameters of multi-phase 12+log(O/H), 12+log(N/H) and N/O distributions as a function of Age and Radius.}															
\tablehead{															
 \colhead{X/Y} & \colhead{$m$\tablenotemark{a}} &\colhead{$b_{0}$\tablenotemark{a}}& \colhead{RMS\tablenotemark{a}} }
\startdata															
\multicolumn{4}{c}{ Age } \\			
\hline 															
12+log(O/H)$_{\rm ion}$	&	-0.028	$\pm$	0.015	&	8.44	$\pm$	0.13	&	0.17	\\
12+log(O/H)$_{\rm neu}$	&	0.007	$\pm$	0.018	&	7.93	$\pm$	0.15	&	0.17	\\
12+log(N/H)$_{\rm ion}$	&	-0.048	$\pm$	0.017	&	7.37	$\pm$	0.12	&	0.15	\\
12+log(N/H)$_{\rm neu}$	&	0.033	$\pm$	0.020	&	5.88	$\pm$	0.18	&	0.19	\\
N/O$_{\rm ion}$	&	-0.013	$\pm$	0.024	&	-1.10	$\pm$	0.22	&	0.25	\\
N/O$_{\rm neu}$	&	0.023	$\pm$	0.021	&	-2.02	$\pm$	0.17	&	0.12	\\
\hline											
\multicolumn{4}{c}{ Radius } \\											
\hline 											
12+log(O/H)$_{\rm ion}$	&	-0.002	$\pm$	0.001	&	8.42	$\pm$	0.14	&	0.22	\\
12+log(O/H)$_{\rm neu}$	&	0.001	$\pm$	0.001	&	7.84	$\pm$	0.11	&	0.15	\\
12+log(N/H)$_{\rm ion}$	&	-0.002	$\pm$	0.002	&	7.27	$\pm$	0.18	&	0.25	\\
12+log(N/H)$_{\rm neu}$	&	0.002	$\pm$	0.001	&	5.91	$\pm$	0.14	&	0.18	\\
N/O$_{\rm ion}$	&	-5$\times$10$^{-4}$	$\pm$	0.002	&	-1.15	$\pm$	0.22	&	0.20	\\
N/O$_{\rm neu}$	&	5$\times$10$^{-4}$	$\pm$	0.001	&	-1.91	$\pm$	0.13	&	0.17	\\
\enddata
\tablenotetext{a}{Linear regression parameters: slope ($m$), intercept ($b_{0}$) and RMS derived with \texttt{ltsfit} for 12+log(O/H) and 12+log(N/H) and N/O abundances. Figures \ref{fig_metal_N_H_N_O_ion_neu_age_radious_cm} and \ref{fig_metal_N_H_N_O_ion_neu_radius_age_cm} show the multi-phase X/H distributions for 12+log(O/H), 12+log(N/H) and N/O distributions and their respective linear best fit models as a function of age and radius. }
\end{deluxetable}

\begin{figure*}[htb!]

\centering
 \gridline{\fig{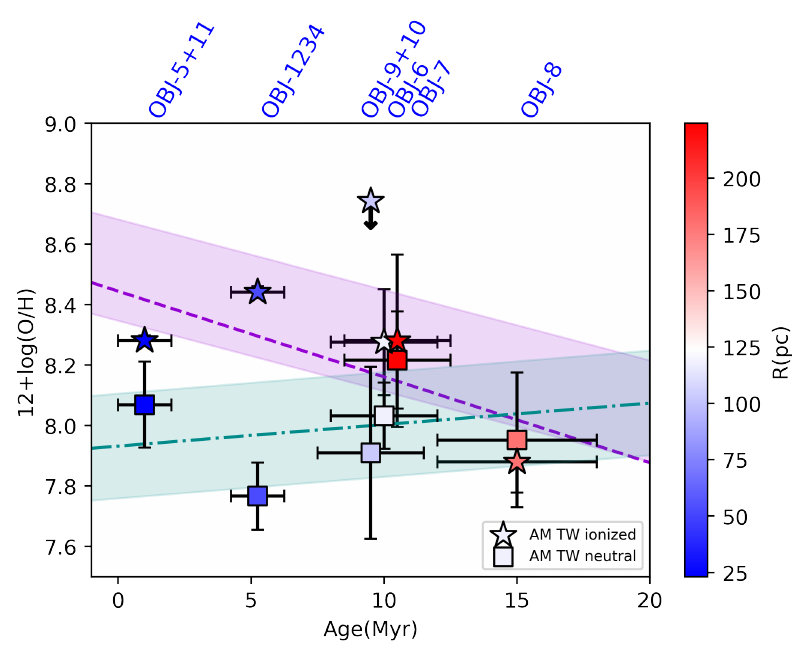}{0.55\textwidth}{}}
 \vspace{-1cm}
 \gridline{\fig{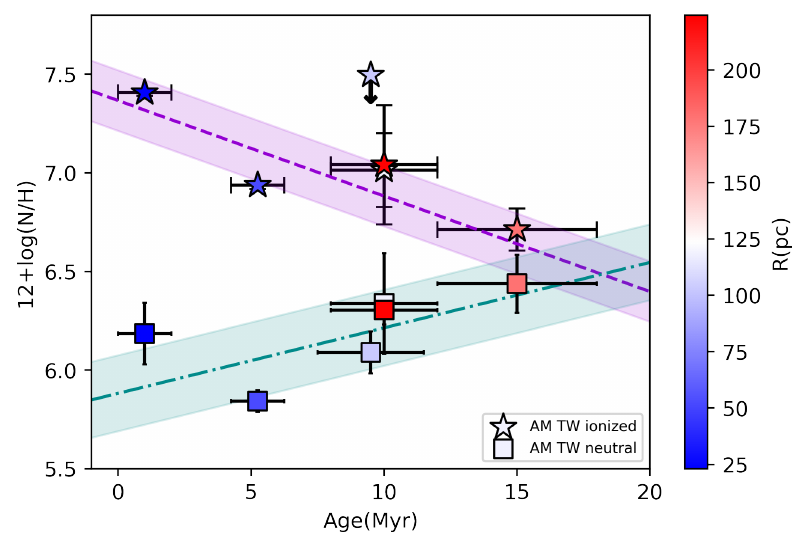}{0.55\textwidth}{}} 
 \vspace{-1cm}
 \gridline{\fig{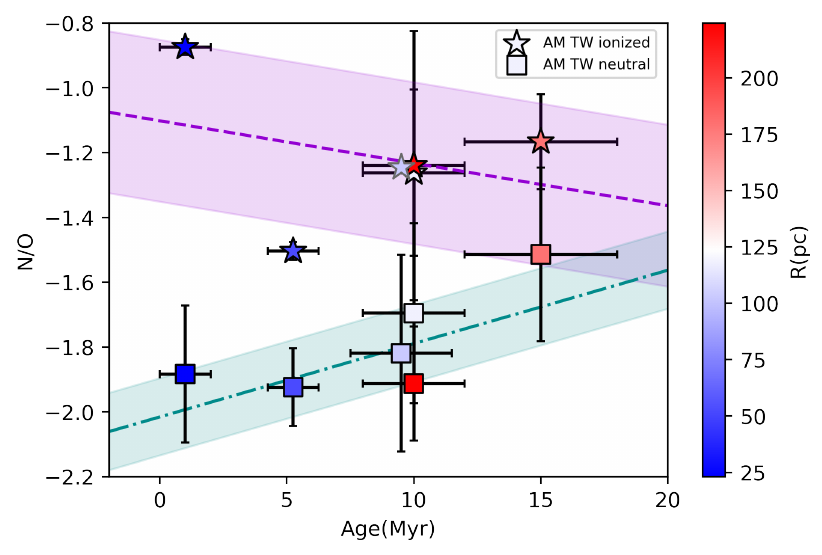}{0.55\textwidth}{}}  
 \vspace{-1cm}

\figcaption{Co-spatial 12+log(O/H) (upper panel), 12+log(N/H) (middle panel) and N/O (lower panel) abundances in the ionized and neutral gas phases, as a function of age, color-coded as a function of the distance to the center of the targeted clusters. Stars indicate ionized gas abundances derived from optical emission lines while squares correspond to neutral gas abundances from UV absorption lines. We performed least trimmed squares linear fit regressions using \texttt{ltsfit}. The best fit line models are shown as violet dashed and dot-dashed turquoise lines for the ionized and neutral distributions respectively. Shaded areas indicate the errors of each line regression. Linear regression best fit model parameters are reported in Table \ref{Table_lin_reg_param_X_Y_ion_neu}. Ionized phase abundance values for OBJ-9+10 are upper limits (gray contour stars) and were excluded for the linear regression computations.\label{fig_metal_N_H_N_O_ion_neu_age_radious_cm}} 

\end{figure*}

\begin{figure*}[ht!]

\centering
 \gridline{\fig{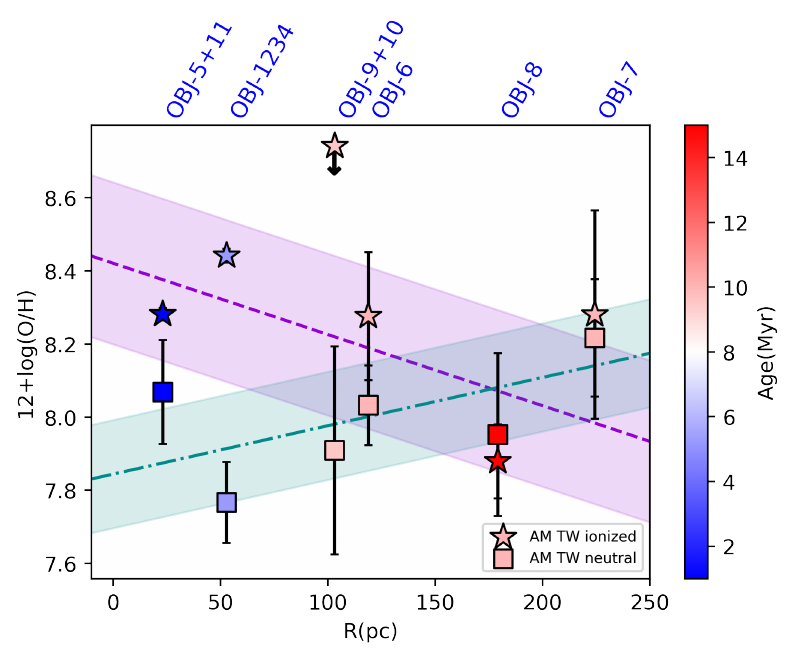}{0.55\textwidth}{}}
 \vspace{-1cm}
 \gridline{\fig{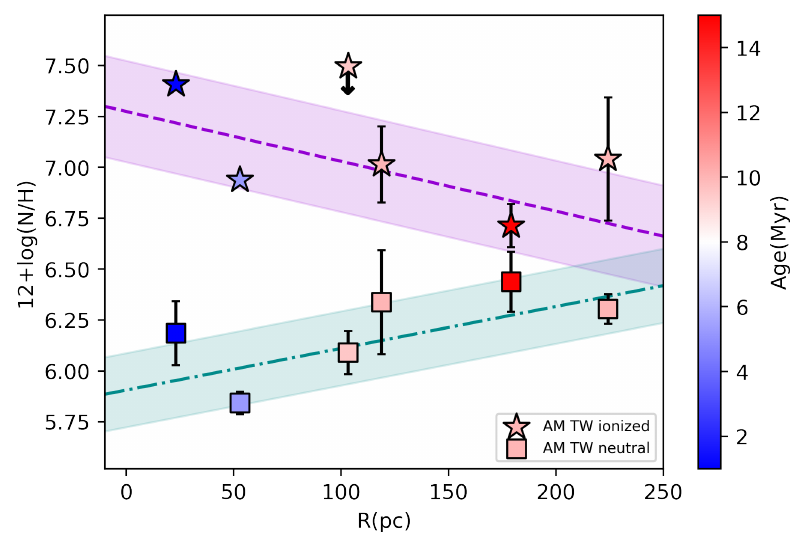}{0.55\textwidth}{}} 
 \vspace{-1cm}
 \gridline{\fig{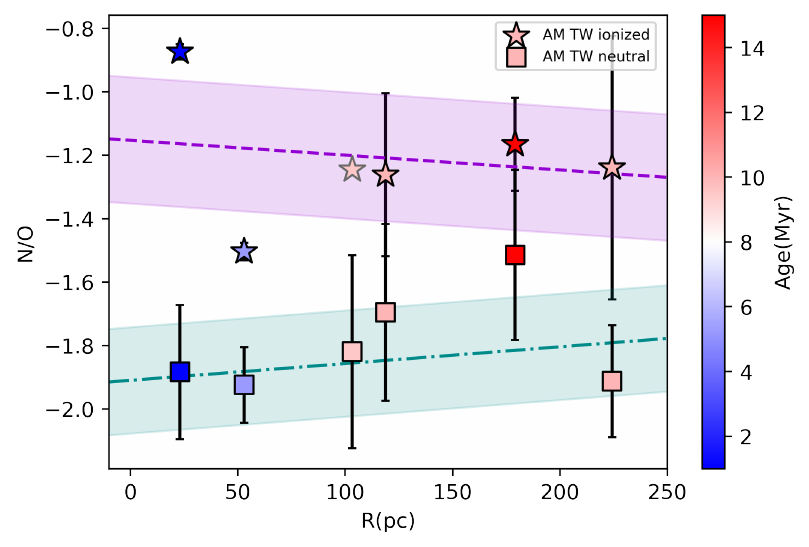}{0.55\textwidth}{}}  
 \vspace{-1cm}

\figcaption{Co-spatial 12+log(O/H) (upper panel), 12+log(N/H) (middle panel) and N/O (lower panel) abundances in the ionized and neutral gas phases, as a function of radius, color-coded as a function of ages of the stellar populations in the targeted clusters. Figure elements are described in the caption of Figure \ref{fig_metal_N_H_N_O_ion_neu_age_radious_cm}. \label{fig_metal_N_H_N_O_ion_neu_radius_age_cm}} 
\end{figure*}

To further explore differences between gas phases we draw from our previously discussed Pearson coefficient analysis and LTS linear regression fits to the O/H, N/H, as well as to the N/O relative abundances in both phases. The best fit parameters for these fits are presented in Table \ref{Table_lin_reg_param_X_Y_ion_neu} showing the ionized and neutral best fit models in contiguous rows to facilitate the comparison. Figure \ref{fig_metal_N_H_N_O_ion_neu_age_radious_cm} shows the abundances obtained for the ionized gas as stars and for the neutral gas as squares, along with the linear regression for both distributions with their corresponding errors, as a function of age and color coded as a function of radius. Figure \ref{fig_metal_N_H_N_O_ion_neu_radius_age_cm}, shows the same elements as Figure \ref{fig_metal_N_H_N_O_ion_neu_age_radious_cm}, but as a function of radius and color coded as a function of age, with their respective linear regression models.  In all instances, the average values for the neutral gas abundances are lower than those for the ionized gas, obtaining differences in the averages for the neutral vs. ionized gas phases of $\sim$0.22 dex, $\sim$0.80 dex and $\sim$0.58 dex, for O/H, N/H and N/O, respectively.

With regards to overall distributions, as described above, Figures \ref{fig_metal_N_H_N_O_ion_neu_age_radious_cm} and \ref{fig_metal_N_H_N_O_ion_neu_radius_age_cm} show negative linear trends for the ionized-gas abundances, and positive linear trends for the neutral-gas abundances (albeit with a lower level of correlation as a function of radius). The ionized and neutral phase abundance distributions converge as a function of age around 15 Myr, as it is shown in Figure \ref{fig_metal_N_H_N_O_ion_neu_age_radious_cm}. This indicates a general trend in which ionized gas abundances decrease as a function of both age and radius, while neutral gas abundances increase until having similar abundance values for the oldest clusters. Regarding N/O we obtain a strong correlation for the neutral gas N/O while no correlation is observed with the ionized gas N/O. The numerical values of the slopes in Table \ref{Table_lin_reg_param_X_Y_ion_neu}, which are all negative for the ionized phase and positive for the neutral phase, indicate the quantification of the observed trend in Figures \ref{fig_metal_N_H_N_O_ion_neu_age_radious_cm} and \ref{fig_metal_N_H_N_O_ion_neu_radius_age_cm}. 

For a cluster-by-cluster case of O/H (upper panel of Figures \ref{fig_metal_N_H_N_O_ion_neu_age_radious_cm} and \ref{fig_metal_N_H_N_O_ion_neu_radius_age_cm}) the smallest differences between the neutral and ionized phases correspond to 0.064 dex for OBJ-7, -0.073 dex for OBJ-8 (the oldest clusters with ages of 10 and 15 Myr, respectively) and 0.21 dex for OBJ-5+11. In the latter case this could be related to the fact that OBJ-5+11 is recently experiencing a starburst ($\sim$1 Myr) so the ionized gas has not yet been enriched with new O and both phases show similar O/H, probably the metallicity of the progenitor cloud of the SF episode. The largest phase abundance differences for O/H correspond to 0.83 dex for OBJ-9+10 and 0.68 dex for OBJ-1234.

For the case of N/H (middle panel of Figures \ref{fig_metal_N_H_N_O_ion_neu_age_radious_cm} and \ref{fig_metal_N_H_N_O_ion_neu_radius_age_cm}) the average abundance discontinuity between the phases is larger than for O/H. As already anticipated in Figures \ref{fig_metal_N_H_N_O_ion_neu_age_radious_cm} and \ref{fig_metal_N_H_N_O_ion_neu_radius_age_cm}, the largest differences are observed for the youngest clusters, OBJ-5+11, OBJ-1234 and OBJ-9+10, with differences of 1.22 dex, 1.10 dex and 1.41 dex respectively. The oldest (10--15 Myr) clusters, OBJ-6, OBJ-7 and OBJ-8 show differences of 0.68 dex 0.74 dex and 0.28 dex, supporting a N production from WN stars on 2--5 Myr timescales and a differential N enrichment between gas phases, mainly depending on mixing timescales due to the layered structure of the transition from \hii\ to \hi\ regions \citep{Berg+2021}. 

The multi-phase relative N/O abundances (lower panel of Figures \ref{fig_metal_N_H_N_O_ion_neu_age_radious_cm} and \ref{fig_metal_N_H_N_O_ion_neu_radius_age_cm}) show the strongest evidence that N enrichment is time dependent. The relative N/O abundance difference between gas phases is maximum for the youngest cluster OBJ-5+11 (1.01 dex) and remains constant for around 5--9 Myr (0.67 dex -- 0.57 dex) suggesting the N-enhanced ionized gas has not had enough time to cool within that timescale. After 10--12 Myr the offset progressively decreases as a function of age until reaching a minimum of 0.35 dex for the oldest target OBJ-8 (15 Myr). This may indicate the time required for the cooling/mixing of the N-enriched material from the stellar winds of WR stars to be observable in the neutral gas.

\begin{figure}[htb!]
\centering
\includegraphics[width=1.05\linewidth]{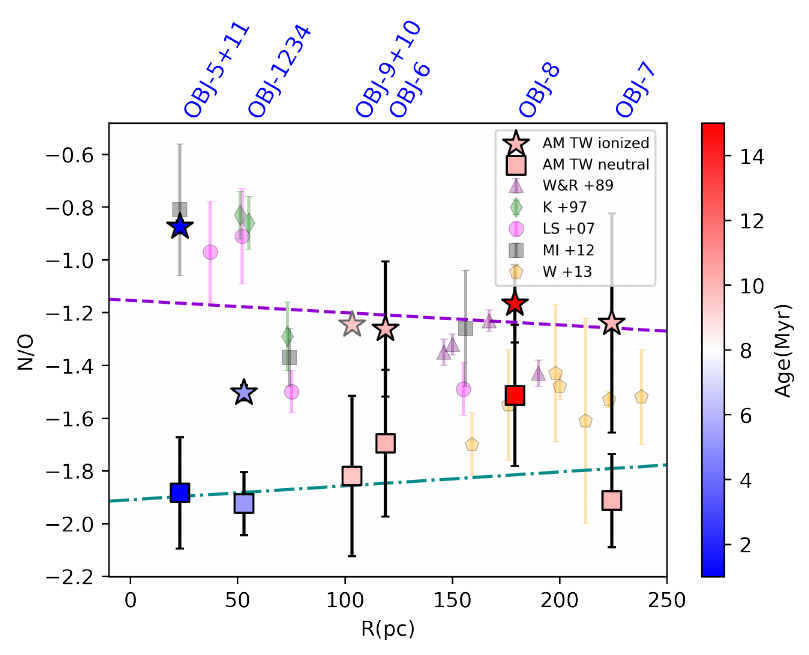}
\figcaption{Co-spatial N/O relative abundances obtained by means of the multi-phase analysis presented in this study compared to the ionized gas N/O values from different studies in the literature: \citealt{Walsh+1989} (purple triangles), \citealt{Kobulnicky+1997} (green diamonds), \citealt{Lopez-Sanchez+2007} (pink circles), \citealt{Monreal-Ibero+2012} (gray squares), \citealt{Westmoquette+2013} (yellow pentagons). The dashed and dot-dashed lines correspond to the best linear regression models we obtained for the observed neutral and ionized N/O abundance distributions. \label{fig_N_O_abu_ion_neu_radius_center_age_lit_cm_A}}

\end{figure}

Within 50 pc from the center, other SF regions sampled in the literature show high N values for the ionized gas where the neutral gas N/O is very low within our sample (a difference of -1.88 dex, see Figure \ref{fig_N_O_abu_ion_neu_radius_center_age_lit_cm_A}). This is in agreement with the lower limit timescale of 5--9 Myr, before the completion of cooling and mixing of the N-enhanced material, as previously discussed. The neutral phase N/O relative abundances are lower than the ionized gas values at almost all locations as a function of radius. The N enrichment has been attributed to the presence of WR stars detected in the center of the galaxy ($\sim$ 50 pc in the location of OBJ-1234) by \cite{Schaerer+1997}, \cite{Westmoquette+2013} and \cite{Monreal-Ibero+2012}. It should be noted that while the values from the literature have radial values computed in a common frame of reference, they may not be spatially located in the same regions as one another so pointing-to-pointing comparisons should be made with caution.


\subsection{The effects of ISM and stellar outflows on gas mixing}

Several factors may contribute to the distribution of metals, or the timescales within which metals are mixed throughout the different phases. One such factor could be the size or magnitude of outflows from ISM gas and/or stellar winds within these clusters. 
We measured gas outflows originated by ISM and/or stellar winds by fitting the asymmetric absorption profiles of the \siIV~$\lambda\lambda$ 1393, 1402 doublet. This feature is a composite of both the ISM absorption and stellar wind profile, which originates in  WR and AGB stars inside the stellar population of the different targets. As such, when fitting it was necessary to add a blue-shifted outflow component to the narrow ISM component we observe for all the targets. Since the clusters studied in this work have different stellar population properties the outflow velocities vary in both age and location of the targets. 

\begin{figure}[htb!]
\centering
\includegraphics[width=1.05\linewidth]{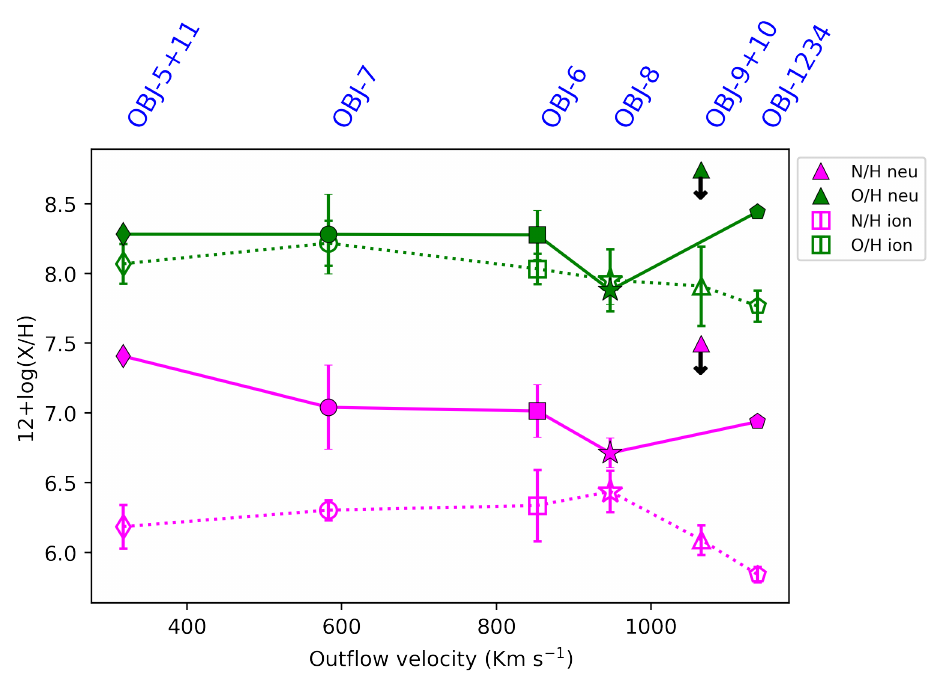}
\figcaption{Oxygen (green) and Nitrogen (magenta) abundances as a function of the ISM and stellar outflow velocities, as measured from the \siIV~$\lambda\lambda$ 1393, 1402 doublet. Dotted and solid lines correspond to the neutral and ionized phase abundances, respectively. \label{fig_Diff_ionized_neutral_abundances_vs_st_wind_vel}}
\end{figure}

In Figure \ref{fig_Diff_ionized_neutral_abundances_vs_st_wind_vel} we present the ionized (solid lines) and neutral abundances (dotted lines) for N and O as a function of outflow velocity of the targets. OBJ-9+10 and OBJ-1234 show highest outflow velocities (with 1140 and 1070 km s$^{-1}$, respectively), corresponding to the SF regions with the highest ionized N and O abundances and the highest abundance offsets. Clusters OBJ-8, OBJ-6 and OBJ-7 have intermediate outflows velocities with values 963, 890 and 604 km s$^{-1}$, respectively. OBJ-7 and OBJ-6 show similar N and O abundances in each gas phase, while a pronounced decrease of the ionized gas abundances is observed for OBJ-8, producing the smallest offsets. OBJ-5+11, the youngest cluster inside the radio nebula structure in the center of the galaxy shows the lowest outflow velocity with 340 km s$^{-1}$; this due to the fact that the stellar population is too young so there is no presence yet of WR or AGB stars. OBJ-5+11 also shows the highest N abundance offset compared to other clusters, 
which show an approximately decreasing abundance offset as a function of outflow velocity (between 300 -- 900 km s$^{-1}$) and then rising slightly for OBJ-1234, which has the highest ionized N abundance and the lowest neutral abundance (OBJ-9+10 abundances are upper limits and not included here). In the case of the O abundance distributions a constant offset is observed between 300 -- 900 km s$^{-1}$, almost equivalent O abundances for OBJ-8 and then rising offsets for outflow velocities $>$ 1000 km s$^{-1}$, corresponding to  OBJ-1234.

These results show the highest outflow velocities for OBJ-9+10 and OBJ-1234 which are the targets whose stellar populations have ages ranging from 5--9 Myr at which WR episodes occur and have been detected for OBJ-1234. These are also the targets with the highest abundance offsets (highest ionized and lowest neutral abundances), which implies that the magnitude of the outflowing gas may have an influence in increasing the abundance offsets, such that the outflow first enriches the warm ionized phase by injecting energy into the ISM through the warm winds of enriched gas from the stars \citep{Chisholm+2017, Chisholm+2018}. The influence on abundance offsets of winds diminishes with the age of the stellar populations, reaching the minimum for OBJ-8 (15 Myr) when the cooling and mixing is completed, showing nevertheless a high outflow velocity of 900 km s$^{-1}$ likely originated by AGB stars, although their effect on abundance enhancement is not yet observed as was discussed in Section \ref{sec:ionized_gas}. A deeper study is necessary to understand the influence of outflow velocities in multi-phase chemical enrichment.

\section{Conclusions} \label{sec:Conclusions}

\begin{deluxetable*}{c|cc|cc|cc|cc|cc|cc}[htb!]
\label{Table_corr_coefficients}									
\tablecaption{Pearson correlation coefficients of the ionized and neutral abundances N/H, O/H and Fe/H as a function of Age, Radius, outflow velocity and \hi\ column density. Strong correlations are highlighted in boldface.}			
\tablehead{										
\colhead{ }& \colhead{N/H$_{\rm ion}$} & \colhead{ }& \colhead{N/H$_{\rm neu}$} & \colhead{ }& \colhead{O/H$_{\rm ion}$} & \colhead{ }& \colhead{O/H$_{\rm neu}$} & \colhead{ }& \colhead{Fe/H$_{\rm ion}$} & \colhead{ }& \colhead{Fe/H$_{\rm neu}$} & \colhead{ } }
\startdata								
	&	$r$	&	$p$	&	$r$	&	$p$	&	$r$	&	$p$	&	$r$	&	$p$	&	$r$	&	$p$	&	$r$	&	$p$	\\
Age	&	\textbf{-0.86}	&	0.06	&	\textbf{0.60}	&	0.21	&	\textbf{-0.72}	&	0.17	&	0.06	&	0.91	&	\textbf{0.65}	&	0.23	&	0.52	&	0.29	\\
Radius	&	-0.57	&	0.32	&	\textbf{0.65}	&	0.16	&	-0.49	&	0.40	&	0.52	&	0.29	&	\textbf{0.77}	&	0.13	&	\textbf{0.79}	&	0.06	\\
Outflow $v$	&	\textbf{-0.82}	&	0.09	&	-0.37	&	0.47	&	-0.05	&	0.94	&	\textbf{-0.78}	&	0.06	&	-0.07	&	0.91	&	-0.46	&	0.36	\\
N(\hi)	&	-0.36	&	0.56	&	\textbf{-0.61}	&	0.19	&	0.23	&	0.71	&	\textbf{-0.91}	&	0.01	&	-0.45	&	0.44	&	\textbf{-0.79}	&	0.06	\\									
\enddata

\end{deluxetable*}	

\begin{deluxetable*}{c|cc|cc|cc}[htb!]													
\label{Table_corr_coefficients_ab_diff}													
\tablecaption{Pearson correlation coefficients of multi-phase abundance differences $\Delta$(N/H), $\Delta$(O/H) and $\Delta$(Fe/H) as a function of Age, Radius, outflow velocity and \hi\ column density. Strong correlations are highlighted in boldface. }	
\tablehead{													
\colhead{}& \colhead{$\Delta$(N/H)\tablenotemark{a}} & \colhead{}& \colhead{$\Delta$(O/H)\tablenotemark{a}} & \colhead{}& \colhead{$\Delta$(Fe/H)\tablenotemark{a}} & \colhead{} }
\startdata													
\hline 													
	&	$r$	&	$p$	&	$r$	&	$p$	&	$r$	&	$p$	\\
Age	&	\textbf{-0.72}	&	0.11	&	-0.34	&	0.51	&	-0.52	&	0.29	\\
Radius	&	\textbf{-0.68}	&	0.14	&	-0.54	&	0.27	&	\textbf{-0.79}	&	0.06	\\
Out. $v$	&	-0.04	&	0.95	&	0.54	&	0.26	&	0.46	&	0.36	\\
N(HI)	&	0.40	&	0.44	&	\textbf{0.81}	&	0.05	&	\textbf{0.79}	&	0.06	\\
\enddata	
\tablenotetext{a}{The notation for the abundance phase difference corresponds to $\Delta(X/H)$ = X/H$_{\rm ion}$ - X/H$_{\rm neu}$ }
\end{deluxetable*}	

\vspace{-1.8cm}

We performed a spatially resolved study with the aim of obtaining the distribution of multi-phase gas chemical abundances in the metal poor BDG NGC~5253. We used six pointings throughout the galaxy disk, targeting SF clusters with a wide range of physical characteristics. We analyzed a unique composite of UV HST/COS and optical VLT/MUSE spectra, to model absorption profiles in the far UV and nebular emission lines in the optical to derive co-spatial abundances in the neutral and ionized gas phases, respectively. We obtained multi-phase abundance distributions as a function of age and radius, that were characterized and quantified by means of a robust LTS linear regression method and a Pearson correlation analysis. Additionally, we evaluated correlation as a function of outflow velocity and \hi\ column densities. All correlation coefficients are summarized in Tables \ref{Table_corr_coefficients} and \ref{Table_corr_coefficients_ab_diff}. 

We summarize our results first presenting our findings regarding the neutral and ionized gas phases separately, and then the multi-phase comparison.

1. Neutral phase abundance results: 

\begin{itemize}
    \item The neutral gas abundances show evidence of correlation as a function of radius, albeit with shallow slopes, indicating a low level of variation with radius. We recover anti-correlations between N(\hi) and neutral gas abundances, while no correlation is found as a function of outflow velocity. 
    
    \item The distributions of neutral abundances as a function of age show slopes that are one order of magnitude larger than those of the radial distributions. The strongest correlation with age is observed for N/H, which is in agreement with the difference of 0.59 dex between the lowest neutral N abundance for OBJ-1234 (5 Myr) and the highest for OBJ-8 (15 Myr). This indicates the increase of N in the neutral phase is linked to the timescale at which N produced in the \hii\ regions cools down and mixes with the cold gas in the \hi\ regions which are farther from the SF regions due to the layered structure of the ISM. N/O also shows a correlation as a function of age, with a total increment (N/O$\rm _{max}$ - N/O$\rm _{min}$) of 0.41 dex. 
    
    \item O/H shows uniform distributions as a function of both age and radius, which indicates O has not yet been produced or O enhancement is not yet detectable in the neutral phase in our sample.
\end{itemize}

2. Ionized phase abundance results:

\begin{itemize}
   \item We observe in-homogeneous abundance distributions for all the studied elements (Fe, S, O and N) in the ionized gas. OBJ-5+11 and OBJ-1234 show the highest N and O abundances respectively, while OBJ-8 shows the lowest abundances for both O and N. The magnitude of the variation for the ionized O/H corresponds to 0.56 dex, whereas N/H shows a total variation of 0.69~dex. The youngest clusters (OBJ-5+11 and OBJ-1234) have the lowest Fe abundances with 0.6 dex of difference with respect to the highest Fe abundance showed by OBJ-6. The distribution of S/H is similar to that of O/H but with a lower abundance difference of $\sim$0.4 dex, with OBJ-1234 and OBJ-8 showing the highest and the lowest S/H abundances, respectively.

   \item While no clear correlation is found for the radial ionized O/H or N/H distribution, a strong anti-correlation is observed as a function of age for both O/H and N/H. No correlations were found for the distribution of N/O as a function of age or radius.
    
\end{itemize}

3. Results from the comparison between ionized and neutral gas abundances: 

\begin{itemize}
    \item From our analysis we found larger abundances in the ionized phase than the neutral phase for all the elements for which we performed the comparison (N, O, S and Fe). We measured average abundances differences of 0.22 dex, 0.80 dex and 0.58 dex, for 12+log(O/H), 12+log(N/H) and N/O respectively (see Figure~\ref{fig_metal_N_H_N_O_ion_neu_radius_age_cm}). Fe shows the smallest offsets, which might be related to the fact that Fe is mostly produced by SNe Ia (SNe II only produce very minimal amounts of Fe), whose timescales are longer than the age of the sampled clusters. 

    \item The magnitude of the N/H abundance offset decreases with age (from 1.22 dex in the youngest clusters to 0.28 dex in the oldest clusters), indicating that the chemical enrichment episodes visible in the ionized gas, require longer timescales to be mixed in the neutral gas phase (see Figure~\ref{fig_metal_N_H_N_O_ion_neu_radius_age_cm}). By comparison, the O/H offset has a more constant variation around the average with no correlation with age.
     
    \item The N/O offset shows the strongest dependency with age, due to N/H enhancement in the ionized gas, with a maximum value of $\Delta$(N/O)$\sim$1.01 dex corresponding to the youngest cluster OBJ-5+11. The N/H enhancement is first observed in the ionized gas around $\sim$5--9 Myr when the N/O distribution remains constant, i.e., \textit{after} the episodes of WN stars ($\sim$2--5 Myr timescales) which may be the main source of N enrichment. We then observe a timescale of $\sim$10--15 Myr when the abundance difference between gas phases reaches a minimum of 0.35 dex, which corresponds to the time required to detect the N enrichment in the neutral gas.

     \item We observe an anti-correlation for the N/H abundance offset ($\Delta$(N/H)) with respect to age and radius, indicating the N/H abundances between gas phases converge to a common value with age for the farthest clusters from the center which are also the oldest. $\Delta$(O/H) and $\Delta$(Fe/H) did not show any correlation with age, however they did show correlation with the \hi\ column densities, suggesting higher abundance offsets for higher values of N(\hi) for which the neutral abundances are the lowest. An anti-correlation is identified for $\Delta$(Fe/H) with radius, showing higher abundances offsets at the center of the galaxy and diminishing towards the outskirts.
    
    \item By modeling the blue-shifted component of the SiVI $\lambda\lambda$ 1393, 1402 doublet, we observe the highest outflow velocities of the gas for OBJ-9+10 and OBJ-1234 (see Figure~\ref{fig_Diff_ionized_neutral_abundances_vs_st_wind_vel}), with ages of 5--9 Myr, which coincide with the periods of occurrence of WR episodes. In turn, these are the regions showing the highest ionized gas O and N abundances and the highest abundance offsets between gas phases, which might imply the outflowing gas is playing a role favoring the injection of metals in the ionized gas and increasing the abundance offsets.
    
\end{itemize}

In summary, our results show that N enrichment happens differentially, first locally in the ionized gas phase with timescales of the order of 5--9 Myr, and then mixing out to greater scales into the cold neutral gas until converging to similar abundance values in both gas phases with timescales around 10--15 of Myr. The presence of enhancement processes like N enrichment by WN-type WR stars can locally increase the N abundance in the ionized gas. Similar processes could be expected for C produced by C rich WR stars (WC type) and released as hot stellar winds into the ISM. For other elements like O, which are produced by evolved medium-mass to massive stars in timescales of 10$^{7}$ -- 10$^{9}$ and by core-collapsed SN explosions on time scales of a few tens of Myr, the cooling and mixing could differ from the N-enrichment we have studied. 

Here we have shown how powerful a multi-phase, spatially-resolved analysis can be in providing a holistic understanding of the interplay between star formation and metal enrichment on different timescales. Performing similar studies on a significant sample of high-z analog BDGs and other types of galaxies with older stellar populations, will be necessary in providing a statistically significant analysis on chemical enhancement and to explore longer time-scales like those corresponding to the production of Fe-peak elements. This study also demonstrates the necessity of UV-IFU instrumentation in deciphering the chemical evolution of galaxies, which we hope to see on future telescopes such as Habitable Worlds Observatories.

\begin{acknowledgments}
    The research project leading to this study was funded by program HST-GO-16240, that was provided by NASA through a grant from the Space Telescope Science Institute, which is operated by the Association of Universities for Research in Astronomy, incorporated, under NASA contract NAS5-26555. The authors thank the anonymous referee for their valuable comments and suggestions that contributed to improve the quality of the paper. V.A.M., B.L.J. and A.A. thank Claus Leitherer for helpful discussions. 
    B.L.J., M.M., S.H., and N.K. are thankful for support from the European Space Agency (ESA). The UV data presented in this article were obtained from the Mikulski Archive for Space Telescopes (MAST) at the Space Telescope Science Institute. The specific observations analyzed in this study can be accessed via \dataset[doi:10.17909/5059-qf82]{https://dx.doi.org/10.17909/5059-qf82}.
       
\end{acknowledgments} 



\bibliography{NGC5253_multi-phase}{}
\bibliographystyle{aasjournal}


\appendix

\section{Column densities from absorption line fitting}\label{Appendix_Table_column_densities}

We present in Table \ref{Table_Column_densities} the column densities derived from line fitting best models on absorption profiles originated by different ions in the neutral phase of the ISM of galaxy NGC~5253. We analysed UV spectroscopy data from HST/COS of 6 different pointings targeting 11 SF regions distributed through the stellar disk. We used the fitting code \texttt{Voigtfit} \citep{Krogager2018} to compute the column densities inside the COS apertures for 13 ions of 9 different elements. For each ion we specify the lines used to perform the fit. Where necessary additional blue-shifted components were used to account for the MW ISM absorption contribution. Line fitting parameters of best models (based on a $\chi^{2}$ minimization): column densities, b-parameter and the systemic velocity are presented in separated columns. Details on the Line fitting procedure are presented in Section \ref{subsec:abs_lines_fit}. In the case of O we present the results derived from line fitting of the \oi\ $\lambda$1302 profile, and the column density derived from the P/S/O proxy method \citep{James+2018} presented in Section \ref{subsec:col_dens_abundances}.

\startlongtable
\begin{deluxetable*}{ccccc}

\tablecolumns{5}
 \tablecaption{Line fitting parameters \label{Table_Column_densities} }
  
\tablehead{
\colhead{Ion} & \colhead{Lines Used} &  \colhead{log[$N$(X)]} & \colhead{$b$} &\colhead{$v_{sys}$} \\
\colhead{} & \colhead{($\rm \AA$)} &  \colhead{(cm$^{-2}$)} & \colhead{(km s$^{-1}$)} &\colhead{(km s$^{-1}$)}
} 
\startdata
\multicolumn{5}{c}{NGC~5253-OBJ-6}\\
\cline{1-5}
\hi	&	1215.67	&	21.22	$\pm$	0.01		&	60.00	$\pm$	$\--$	&	381.10	$\pm$	20.70	\\
\cii	&	1334.53	&	15.19	$\pm$	0.07	$^{\dag}$	&	81.39	$\pm$	9.95	&	397.50	$\pm$	3.50	\\
\ciia	&	1335.66, 1335.71	&	14.90	$\pm$	0.18		&	43.43	$\pm$	12.71	&	386.80	$\pm$	4.50	\\
\nI	&	1134.17, 1134.41, 1134.98	&	15.54	$\pm$	0.26		&	21.05	$\pm$	4.18	&	396.70	$\pm$	5.80	\\
\oi	&	1302.17	&	15.45	$\pm$	0.05	$^{\dag}$	&	70.76	$\pm$	7.94	&	401.20	$\pm$	4.40	\\
\oi$_{\rm P,S}$	&	P,S,O correlation$^{1}$	&	17.25	$\pm$	0.11		&		$\--$		&		$\--$		\\
\siII	&	1304.37	&	14.36	$\pm$	0.03	$^{\dag}$	&	72.36	$\pm$	13.52	&	428.80	$\pm$	6.40	\\
\siIIa	&	1264.74, 1265.002	&	13.06	$\pm$	0.06		&	82.76	$\pm$	28.20	&	396.20	$\pm$	0.00	\\
\siIII	&	1206.50	&	14.13	$\pm$	0.14	$^{\dag}$	&	112.43	$\pm$	34.71	&	388.10	$\pm$	20.20	\\
\pii	&	1152.81	&	13.84	$\pm$	0.09		&	56.31	$\pm$	39.06	&	388.50	$\pm$	16.60	\\
\sii	&	1250.58, 1253.81	&	15.82	$\pm$	0.10		&	23.66	$\pm$	3.75	&	409.20	$\pm$	2.90	\\
\feii	&	\begin{tabular}[c]{@{}c@{}}1096.88, 1121.97, 1125.45, 1127.09,\\  1142.37, 1143.23, 1144.94\end{tabular}	&	15.73	$\pm$	0.08		&	31.33	$\pm$	1.85	&	399.00	$\pm$	2.80	\\
\feiii	&	1122.526	&	15.03	$\pm$	0.05		&	89.03	$\pm$	17.38	&	395.50	$\pm$	10.50	\\
\niII	&	1317.22, 1345.88, 1370.13	&	13.69	$\pm$	0.17		&	33.08	$\pm$	1.01	&	411.70	$\pm$	12.10	\\
\cline{1-5}
\multicolumn{5}{c}{NGC~5253-OBJ-7}\\
\cline{1-5}														
\hi	&	1215.67	&	20.82	$\pm$	0.01		&	60.00	$\pm$	$\--$	&	361.10	$\pm$	16.50	\\
\cii	&	1334.53	&	15.01	$\pm$	0.02	$^{\dag}$	&	81.61	$\pm$	4.78	&	421.60	$\pm$	2.70	\\
\ciia	&	1335.66, 1335.71	&	14.57	$\pm$	0.02		&	71.85	$\pm$	7.49	&	415.00	$\pm$	0.00	\\
\nI	&	1134.17, 1134.41, 1134.98	&	15.12	$\pm$	0.07		&	31.11	$\pm$	5.73	&	400.60	$\pm$	5.80	\\
\oi	&	1302.17	&	15.22	$\pm$	0.04	$^{\dag}$	&	107.64	$\pm$	22.64	&	415.40	$\pm$	10.30	\\
\oi$_{\rm P,S}$	&	P,S,O correlation$^{1}$	&	17.04	$\pm$	0.16		&		$\--$		&		$\--$		\\
\siII	&	1304.37	&	14.38	$\pm$	0.02	$^{\dag}$	&	58.09	$\pm$	7.57	&	467.50	$\pm$	3.50	\\
\siIIa	&	1264.74, 1265.002	&	12.72	$\pm$	0.08		&	37.33	$\pm$	42.41	&	393.30	$\pm$	12.40	\\
\siIII	&	1206.50	&	13.94	$\pm$	0.03	$^{\dag}$	&	114.35	$\pm$	8.55	&	421.90	$\pm$	5.50	\\
\pii	&	1152.81	&	13.72	$\pm$	0.17		&	142.41	$\pm$	91.48	&	465.00	$\pm$	44.20	\\
\sii	&	1253.81	&	15.51	$\pm$	0.12		&	80.55	$\pm$	38.04	&	433.90	$\pm$	22.00	\\
\feii	&	\begin{tabular}[c]{@{}c@{}}1096.88, 1121.97, 1125.45, 1127.09, \\ 1142.37, 1143.23, 1144.94\end{tabular}	&	15.76	$\pm$	0.09		&	24.23	$\pm$	1.51	&	428.50	$\pm$	3.00	\\
\feiii	&	1122.526	&	14.91	$\pm$	0.06		&	66.61	$\pm$	17.22	&	441.60	$\pm$	8.60	\\
\niII	&	1317.22, 1345.88, 1370.13	&	13.91	$\pm$	0.05		&	127.74	$\pm$	22.27	&	441.00	$\pm$	13.00	\\
\cline{1-5}
\multicolumn{5}{c}{NGC~5253-OBJ-9+10}\\
\cline{1-5}															
\hi	&	1215.67	&	21.43	$\pm$	0.01		&	60.00	$\pm$	$\--$	&	345.50	$\pm$	15.00	\\
\cii	&	1334.53	&	15.14	$\pm$	0.02	$^{\dag}$	&	102.37	$\pm$	8.31	&	395.40	$\pm$	3.90	\\
\ciia	&	1335.66, 1335.71	&	14.73	$\pm$	0.03		&	71.83	$\pm$	8.90	&	390.40	$\pm$	5.60	\\
\nI	&	1134.17, 1134.41, 1134.98	&	15.50	$\pm$	0.11		&	25.46	$\pm$	2.45	&	391.90	$\pm$	3.60	\\
\oi	&	1302.17	&	15.26	$\pm$	0.06	$^{\dag}$	&	236.90	$\pm$	29.96	&	476.60	$\pm$	18.30	\\
\oi$_{\rm P,S}$	&	P,S,O correlation$^{1}$	&	17.34	$\pm$	0.28		&		$\--$		&		$\--$		\\
\siII	&	1304.37	&	14.91	$\pm$	0.02	$^{\dag}$	&	80.40	$\pm$	5.72	&	406.30	$\pm$	3.60	\\
\siIIa	&	1264.74, 1265.002	&	13.21	$\pm$	0.03		&	111.57	$\pm$	12.52	&	381.70	$\pm$	7.30	\\
\siIII	&	1206.50	&	14.13	$\pm$	0.09	$^{\dag}$	&	142.15	$\pm$	45.03	&	379.40	$\pm$	26.90	\\
\pii	&	1152.81	&	13.86	$\pm$	0.09		&	46.37	$\pm$	31.85	&	397.60	$\pm$	13.30	\\
\sii	&	1250.58, 1253.81	&	15.98	$\pm$	0.38		&	21.86	$\pm$	9.86	&	403.10	$\pm$	7.10	\\
\feii	&	\begin{tabular}[c]{@{}c@{}}1096.88, 1121.97, 1125.45, 1127.09, \\ 1142.37, 1143.23, 1144.94\end{tabular}	&	15.96	$\pm$	0.08		&	30.81	$\pm$	1.63	&	405.20	$\pm$	2.70	\\
\feiii	&	1122.526	&	15.02	$\pm$	0.08		&	114.85	$\pm$	28.22	&	380.90	$\pm$	22.20	\\
\niII	&	1345.88, 1370.13	&	14.46	$\pm$	0.13		&	122.38	$\pm$	56.88	&	420.50	$\pm$	34.00	\\
\cline{1-5}
\multicolumn{5}{c}{NGC~5253-OBJ-8}\\
\cline{1-5}															
\hi	&	1215.67	&	21.13	$\pm$	0.01		&	60.00	$\pm$	$\--$	&	382.00	$\pm$	0.00	\\
\cii	&	1334.53	&	15.08	$\pm$	0.02	$^{\dag}$	&	102.90	$\pm$	25.69	&	388.20	$\pm$	4.70	\\
\ciia	&	1335.66, 1335.71	&	14.62	$\pm$	0.03		&	48.97	$\pm$	15.49	&	364.30	$\pm$	12.70	\\
\nI	&	1134.17, 1134.41, 1134.98	&	15.54	$\pm$	0.15		&	26.33	$\pm$	5.08	&	391.60	$\pm$	8.60	\\
\oi	&	1302.17	&	15.43	$\pm$	0.03	$^{\dag}$	&	95.51	$\pm$	6.15	&	412.20	$\pm$	3.90	\\
\oi$_{\rm P,S}$	&	P,S,O correlation$^{1}$	&	17.09	$\pm$	0.22		&		$\--$		&		$\--$		\\
\siII	&	1304.37	&	14.48	$\pm$	0.03	$^{\dag}$	&	95.53	$\pm$	11.67	&	355.10	$\pm$	2.90	\\
\siIIa	&	1264.74, 1265.002	&	12.91	$\pm$	0.07		&	47.50	$\pm$	66.24	&	325.70	$\pm$	14.50	\\
\siIII	&	1206.50	&	13.95	$\pm$	0.04	$^{\dag}$	&	104.96	$\pm$	18.92	&	396.30	$\pm$	12.40	\\
\pii	&	1152.81	&	13.69	$\pm$	0.24		&	31.56	$\pm$	81.23	&	393.20	$\pm$	29.60	\\
\sii	&	1250.58, 1253.81	&	15.64	$\pm$	0.18		&	43.26	$\pm$	69.30	&	384.70	$\pm$	29.60	\\
\feii	&	\begin{tabular}[c]{@{}c@{}}1096.88, 1121.97, 1125.45, 1127.09, \\ 1142.37, 1143.23, 1144.94\end{tabular}	&	15.99	$\pm$	0.09		&	28.00	$\pm$	1.83	&	394.80	$\pm$	3.20	\\
\feiii	&	1122.526	&	14.92	$\pm$	0.06		&	91.06	$\pm$	31.34	&	393.00	$\pm$	10.50	\\
\niII	&	1317.22, 1345.88, 1370.13	&	13.80	$\pm$	0.07		&	78.83	$\pm$	135.57	&	445.30	$\pm$	71.35	\\
\cline{1-5}
\multicolumn{5}{c}{NGC~5253-OBJ-5+11}	\\
\cline{1-5}														
\hi	&	1215.67	&	21.00	$\pm$	0.01		&	60.00	$\pm$	$\--$	&	382.00	$\pm$	0.00	\\
\cii	&	1334.53	&	15.30	$\pm$	0.14	$^{\dag}$	&	89.70	$\pm$	25.69	&	388.20	$\pm$	4.70	\\
\ciia	&	1335.66, 1335.71	&	14.67	$\pm$	0.05		&	64.35	$\pm$	15.49	&	364.30	$\pm$	12.70	\\
\nI	&	1134.17, 1134.41, 1134.98	&	15.15	$\pm$	0.16		&	21.66	$\pm$	5.08	&	391.60	$\pm$	8.60	\\
\oi	&	1302.17	&	15.51	$\pm$	0.02	$^{\dag}$	&	113.62	$\pm$	6.15	&	412.20	$\pm$	3.90	\\
\oi$_{\rm P,S}$	&	P,S,O correlation$^{1}$	&	17.07	$\pm$	0.14		&		$\--$		&		$\--$		\\
\siII	&	1304.37	&	14.79	$\pm$	0.20	$^{\dag}$	&	37.75	$\pm$	11.67	&	355.10	$\pm$	2.90	\\
\siIIa	&	1264.74, 1265.002	&	12.81	$\pm$	0.06		&	35.06	$\pm$	66.24	&	325.70	$\pm$	14.50	\\
\siIII	&	1206.50	&	14.16	$\pm$	0.05	$^{\dag}$	&	114.02	$\pm$	18.92	&	396.30	$\pm$	12.40	\\
\pii	&	1152.81	&	13.85	$\pm$	0.12		&	105.82	$\pm$	81.23	&	393.20	$\pm$	29.60	\\
\sii	&	1250.58, 1253.81	&	15.44	$\pm$	0.14		&	60.61	$\pm$	69.30	&	384.70	$\pm$	29.60	\\
\feii	&	\begin{tabular}[c]{@{}c@{}}1096.88, 1121.97, 1125.45, 1127.09, \\ 1142.37, 1143.23, 1144.94\end{tabular}	&	15.60	$\pm$	0.06		&	35.99	$\pm$	1.83	&	394.80	$\pm$	3.20	\\
\feiii	&	1122.526	&	14.99	$\pm$	0.13		&	55.61	$\pm$	31.34	&	393.00	$\pm$	10.50	\\
\niII	&	1317.22, 1345.88, 1370.13	&	14.41	$\pm$	0.20		&	168.12	$\pm$	135.57	&	445.30	$\pm$	71.35	\\
\cline{1-5}
\multicolumn{5}{c}{NGC~5253-OBJ-1234}\\
\cline{1-5}															
\hi	&	1215.67	&	21.43	$\pm$	0.01		&	60.00	$\pm$	$\--$	&	416.80	$\pm$	11.70	\\
\cii	&	1334.53	&	15.20	$\pm$	0.03	$^{\dag}$	&	83.52	$\pm$	4.69	&	408.40	$\pm$	2.00	\\
\ciia	&	1335.66, 1335.71	&	14.84	$\pm$	0.02		&	64.09	$\pm$	4.46	&	394.80	$\pm$	2.50	\\
\nI	&	1134.17, 1134.41, 1134.98	&	15.25	$\pm$	0.04		&	33.33	$\pm$	2.83	&	411.00	$\pm$	3.40	\\
\oi	&	1302.17	&	15.55	$\pm$	0.02	$^{\dag}$	&	112.63	$\pm$	4.18	&	454.70	$\pm$	2.90	\\
\oi$_{\rm P,S}$	&	P,S,O correlation$^{1}$	&	17.20	$\pm$	0.11		&		$\--$		&		$\--$		\\
\siII	&	1304.37	&	14.89	$\pm$	0.01	$^{\dag}$	&	82.02	$\pm$	2.66	&	407.70	$\pm$	1.60	\\
\siIIa	&	1264.74, 1265.002	&	13.38	$\pm$	0.02		&	48.42	$\pm$	6.52	&	408.60	$\pm$	3.00	\\
\siIII	&	1206.50	&	14.13	$\pm$	0.17	$^{\dag}$	&	109.83	$\pm$	55.12	&	390.10	$\pm$	31.50	\\
\pii	&	1152.81	&	13.81	$\pm$	0.13		&	88.51	$\pm$	46.84	&	424.00	$\pm$	25.40	\\
\sii	&	1250.58	&	15.74	$\pm$	0.04		&	58.74	$\pm$	12.74	&	419.90	$\pm$	6.40	\\
\feii	&	\begin{tabular}[c]{@{}c@{}}1096.88, 1121.97, 1125.45, 1127.09, \\ 1142.37, 1143.23, 1144.94\end{tabular}	&	15.64	$\pm$	0.06		&	31.85	$\pm$	1.52	&	415.70	$\pm$	2.10	\\
\feiii	&	1122.526	&	14.89	$\pm$	0.04		&	64.59	$\pm$	16.39	&	400.50	$\pm$	8.80	\\
\niII	&	1317.22	&	13.75	$\pm$	0.05		&	75.90	$\pm$	16.38	&	450.80	$\pm$	9.60	\\
\enddata
\vspace{0.2cm}
\textbf{Notes. }
 Measured column densities, intrinsic dispersion ($b$) and systemic velocities of the different ion ISM absorption lines for each of the six targeted SF regions with COS. The $^{\dag}$ symbol next to the column density values indicate the lines used are affected by saturation.
\end{deluxetable*}

\pagebreak

\section{Oxygen abundances from correlation with P and S abundances}\label{Appendix_O_PS_abundances}

The \oi$\lambda$1302 absorption line is saturated and contaminated by geocoronal emission, so the values directly measured from line fitting on COS observations are considered as a lower limits of the intrinsic O column densities. Neutral O abundances are then computed using the correlation between \oi ,  \pii\ and \sii\ column densities as a proxy to indirectly derive the O/H abundance (the P/S/O proxy method, see Section \ref{subsec:col_dens_abundances}). Those correlations were studied by \cite{James+2018} on a sample of local and intermediate redshift SFGs, covering a wide range of physical properties (z$\sim$0.083--0.321, Z$\sim$0.03 -- 3.2 Z$_{\odot}$ and log[N(\hi)]$\sim$18.44--21.28). We present O/H abundances derived from the column densities using the correlation proxies with \pii\ and \sii\  column densities in Tables \ref{Table_O_P_abundances} and \ref{Table_O_S_abundances}, respectively.

\begin{deluxetable}{lccccc}[h!]							
\label{Table_O_P_abundances}								
\tablecaption{ }												
																		
\tablehead{														
\colhead{Target } 	&	\colhead{ log[N(O$_{\rm P}$)]}			&	\colhead{ log(O$_{\rm P}$/H)}			&	\colhead{ log(O$_{\rm P}$/H)$_{\rm ICF}$}			&	\colhead{ [O$_{\rm P}$/H]}			&	\colhead{ [O$_{\rm P}$/H]$_{\rm ICF}$}		}
\startdata
OBJ-6	&	17.118	$\pm$	0.105	&	-4.102	$\pm$	0.105	&	-4.103	$\pm$	0.105	&	-0.792	$\pm$	0.113	&	-0.793	$\pm$	0.113	\\
OBJ-7	&	16.995	$\pm$	0.184	&	-3.825	$\pm$	0.185	&	-3.828	$\pm$	0.185	&	-0.515	$\pm$	0.189	&	-0.518	$\pm$	0.189	\\
OBJ-9+10	&	17.137	$\pm$	0.106	&	-4.294	$\pm$	0.106	&	-4.295	$\pm$	0.106	&	-0.984	$\pm$	0.113	&	-0.985	$\pm$	0.113	\\
OBJ-8	&	16.965	$\pm$	0.251	&	-4.168	$\pm$	0.251	&	-4.169	$\pm$	0.251	&	-0.858	$\pm$	0.254	&	-0.859	$\pm$	0.254	\\
OBJ-5+11	&	17.131	$\pm$	0.131	&	-3.871	$\pm$	0.131	&	-3.873	$\pm$	0.131	&	-0.561	$\pm$	0.137	&	-0.563	$\pm$	0.137	\\
OBJ-1234	&	17.087	$\pm$	0.139	&	-4.342	$\pm$	0.139	&	-4.343	$\pm$	0.139	&	-1.032	$\pm$	0.145	&  -1.033	$\pm$	0.145	\\
\enddata																		
\vspace{0.2cm}				
\textbf{Notes.}																
Computed Oxygen abundances from the correlation with P abundances (O$_{\rm P}$) studied in \cite{James+2018}.																
\end{deluxetable}

\begin{deluxetable}{lccccc}[h!]													
\label{Table_O_S_abundances}							
\tablecaption{ }																
\tablehead{
\colhead{Target } 	&	\colhead{ log[N(O$_{\rm S}$)]}			&	\colhead{ log(O$_{\rm S}$/H)}			&	\colhead{ log(O$_{\rm S}$/H)$_{\rm ICF}$}			&	\colhead{ [O$_{\rm S}$/H]}			&	\colhead{ [O$_{\rm S}$/H]$_{\rm ICF}$}  }			
\startdata
OBJ-6	&	17.387	$\pm$	0.113	&	-3.833	$\pm$	0.113	&	-3.834	$\pm$	0.113	&	-0.523	$\pm$	0.120	&	-0.524	$\pm$	0.120	\\
OBJ-7	&	17.081	$\pm$	0.133	&	-3.740	$\pm$	0.134	&	-3.743	$\pm$	0.134	&	-0.430	$\pm$	0.140	&	-0.433	$\pm$	0.140	\\
OBJ-9+10	&	17.545	$\pm$	0.388	&	-3.885	$\pm$	0.388	&	-3.887	$\pm$	0.388	&	-0.575	$\pm$	0.390	&	-0.577	$\pm$	0.390	\\
OBJ-8	&	17.207	$\pm$	0.191	&	-3.926	$\pm$	0.191	&	-3.928	$\pm$	0.191	&	-0.616	$\pm$	0.195	&	-0.618	$\pm$	0.195	\\
OBJ-5+11	&	17.013	$\pm$	0.152	&	-3.989	$\pm$	0.152	&	-3.991	$\pm$	0.152	&	-0.679	$\pm$	0.157	&	-0.681	$\pm$	0.157	\\
OBJ-1234	&	17.306	$\pm$	0.073	&	-4.123	$\pm$	0.073	&	-4.124	$\pm$	0.073	&	-0.813	$\pm$	0.084	& -0.814	$\pm$	0.084	\\
\enddata																		
\vspace{0.2cm}				
\textbf{Notes.}				
Computed Oxygen abundances from the correlation with S abundances (O$_{\rm S}$) studied in \cite{James+2018}.																																
\end{deluxetable}	

\pagebreak

\section{ICF total values interpolation}\label{Appendix_ICF_Computation}

In order to account for the ionization effects of highly ionized species inside the neutral phase (ICF$\rm _{neutral}$) and the contaminating ionized gas in \hii regions along the line of sight of our targets (ICF$\rm _{ionizedl}$), we used the grid physical parameters derived from a photo-ionization modeling performed by \citealt{Hernandez+2020},that simulates the physical conditions of a wide range of galactic environments. Several physical properties of the gas (Z, T$_{\rm eff}$, LUV, log[N(\hi)] and \feiii/\feii) were individually used as fixed input parameters to generate a set of ICF values for all the elements. We used the grid models corresponding to the iron ionization factor \feiii/\feii and the \hi\ column density. We performed a simple interpolation to search for the ICFs values matching the properties of each target, and repeated this procedure for each element in the corresponding parameter space (see Section \ref{subesec:ICF_correction}). To compute the final abundances corrected by ionization we use the grid values with log[N(\hi)] as fixed parameter (ICF values in Table \ref{Table_ICF_log(N_HI)}), for all the targets except for OBJ-6 and OBJ-7 for which we use the ICF derived by \cite{Hernandez+2020}.

\begin{deluxetable}{lcccccccccc}[h!]

\label{Table_ICF_FeIII_FeII}
\tablecaption{ }

\tablehead{
\colhead{Target } & \colhead{ \feii / \feiii} & \colhead{H} & \colhead{C} & \colhead{N} & \colhead{O} & \colhead{Si} & \colhead{P} & \colhead{S} & \colhead{Fe} & \colhead{Ni} } 
\startdata
OBJ-6	&	-0.7002	&	0.0016	&	0.0702	&	-0.0178	&	0.0016	&	0.0538	&	0.0748	&	0.0388	&	0.0192	&	0.0280	\\
OBJ-7	&	-0.8453	&	0.0013	&	0.0551	&	-0.0126	&	0.0013	&	0.0442	&	0.0594	&	0.0336	&	0.0172	&	0.0265	\\
OBJ-9+10	&	-0.9397	& 0.0011	&	0.0453	&	-0.0092	&	0.0011	&	0.0380	&	0.0494	&	0.0302	&	0.0158	&	0.0256	\\
OBJ-8	&	-1.0000	&	0.0010	&	0.0390	&	-0.0070	&	0.0010	&	0.0340	&	0.0430	&	0.0280	&	0.0150	&	0.0250	\\
OBJ-5+11	&	-0.6111	&	0.0018	&	0.0794	&	-0.0210	&	0.0018	&	0.0597	&	0.0842	&	0.0420	&	0.0204	&	0.0289	\\
OBJ-1234	&	-0.7553	& 0.0015	&	0.0644	&	-0.0158	&	0.0015	&	0.0502	&	0.0689	&	0.0368	&	0.0184	&	0.0274	\\
\enddata

\vspace{0.2cm}
\textbf{Notes.}
ICF total values computed thorough interpolation using the ICF grid derived from the \feiii/\feii\ ratio in \cite{Hernandez+2020}

\end{deluxetable}

\begin{deluxetable}{lcccccccccc}[h!]

\label{Table_ICF_log(N_HI)}
\tablecaption{ }

\tablehead{
\colhead{Target } & \colhead{ log[N(\hi)]} & \colhead{H} & \colhead{C} & \colhead{N} & \colhead{O} & \colhead{Si} & \colhead{P} & \colhead{S} & \colhead{Fe} & \colhead{Ni} } 
\startdata
OBJ-6	&	21.2200	&	0.0006	&	0.0700	&	-0.0274	&	0.0016	&	0.0418	&	0.0611	&	0.0258	&	0.0114	&	0.0145	\\
OBJ-7	&	20.8200	&	0.0019	&	0.0940	&	-0.0308	&	0.0033	&	0.0654	&	0.0898	&	0.0462	&	0.0235	&	0.0347	\\
OBJ-9+10	&	21.4308	&	0.0001	&	0.0576	&	-0.0211	&	0.0011	&	0.0329	&	0.0479	&	0.0199	&	0.0088	&	0.0111	\\
OBJ-8	&	21.1330	&	0.0007	&	0.0752	&	-0.0300	&	0.0017	&	0.0454	&	0.0666	&	0.0283	&	0.0124	&	0.0159	\\
OBJ-5+11	&	21.0022	&	0.0010	&	0.0829	&	-0.0339	&	0.0020	&	0.0509	&	0.0749	&	0.0319	&	0.0140	&	0.0180	\\
OBJ-1234	&	21.4293	&	0.0001	&	0.0577	&	-0.0211	& 0.0011 &	0.0330	&	0.0480	&	0.0200	&	0.0088	& 0.0111	\\
\enddata
\vspace{0.2cm}
\textbf{Notes.}
ICF total values computed thorough interpolation using the ICF grid derived using the log[N(\hi)] in \cite{Hernandez+2020}. 
\end{deluxetable}

\section{Emission Line Fluxes}

\begin{splitdeluxetable}{lccccccBlcccccc}
\label{tab:emission_fluxes}
\tablecaption{Emission line flux measurements from spectra integrated across a circular 2\farcs{5} aperture centered on each of the clusters, as shown in Figure~\ref{fig_NGC5253_COS_MUSE}. For each line we provide the observed and de-reddened fluxes (calculated using the E(B-V) values listed) are given relative to H$\beta=100$. The integrated H$\beta$ fluxes in units of $\times10^{-15} \rm erg/s/cm^{2}$ are also provided.}
\tablehead{
\colhead{Line ID}	
&	\colhead{F($\lambda$)/F(H$\beta$)}	&		\colhead{I($\lambda$)/I(H$\beta$)} 
&	\colhead{F($\lambda$)/F(H$\beta$)}	&		\colhead{I($\lambda$)/I(H$\beta$)}
&	\colhead{F($\lambda$)/F(H$\beta$)}	&		\colhead{I($\lambda$)/I(H$\beta$)}
& \colhead{Line ID} 
&	\colhead{F($\lambda$)/F(H$\beta$)}	&		\colhead{I($\lambda$)/I(H$\beta$)}
&	\colhead{F($\lambda$)/F(H$\beta$)}	&		\colhead{I($\lambda$)/I(H$\beta$)}
&	\colhead{F($\lambda$)/F(H$\beta$)}	&		\colhead{I($\lambda$)/I(H$\beta$)} }
\startdata
&\multicolumn{2}{c}{OBJ-6}
& \multicolumn{2}{c}{OBJ-7}								 
& \multicolumn{2}{c}{OBJ-9+10}
&
& \multicolumn{2}{c}{OBJ-8}
& \multicolumn{2}{c}{OBJ-5+11}
& \multicolumn{2}{c}{OBJ-1234}\\
H$\beta$	&	100.00	$\pm$	0.28	&	100.00	$\pm$	0.28	&	100.00	$\pm$	0.10	&	100.00	$\pm$	0.10	&	100.00	$\pm$	0.33	&	100.00	$\pm$	0.33	&	H$\beta$	&	100.00	$\pm$	0.44	&	100.00	$\pm$	0.44	&	100.00	$\pm$	0.07	&	100.00	$\pm$	0.07	&	100.00	$\pm$	0.08	&	100.00	$\pm$	0.08	\\
\ffeiii\ 4881	&	0.47	$\pm$	0.02	&	0.47	$\pm$	0.02	&	0.20	$\pm$	0.02	&	0.20	$\pm$	0.02	&	0.46	$\pm$	0.04	&	0.46	$\pm$	0.04	&	\ffeiii\ 4881	&	0.52	$\pm$	0.05	&	0.52	$\pm$	0.05	&	0.30	$\pm$	0.00	&	0.30	$\pm$	0.00	&	0.28	$\pm$	0.01	&	0.28	$\pm$	0.01	\\	
\foiii\ 4959	&	107.67	$\pm$	0.08	&	106.28	$\pm$	0.08	&	115.01	$\pm$	0.37	&	113.96	$\pm$	0.37	&	106.36	$\pm$	0.11	&	105.11	$\pm$	0.10	&	\foiii\ 4959	&	95.49	$\pm$	0.42	&	94.50	$\pm$	0.42	&	216.24	$\pm$	0.11	&	211.44	$\pm$	0.11	&	156.76	$\pm$	0.10	&	156.43	$\pm$	0.09	\\	
\feiii\ 4986	&	2.61	$\pm$	0.02	&	2.56	$\pm$	0.02	&	1.98	$\pm$	0.03	&	1.96	$\pm$	0.03	&	2.93	$\pm$	0.05	&	2.89	$\pm$	0.05	&	\feiii\ 4986	&	3.54	$\pm$	0.06	&	3.50	$\pm$	0.06	&	0.44	$\pm$	0.00	&	0.43	$\pm$	0.00	&	0.69	$\pm$	0.01	&	0.69	$\pm$	0.01	\\	
\foiii\ 5007	&	324.98	$\pm$	0.20	&	318.80	$\pm$	0.20	&	345.99	$\pm$	0.95	&	341.35	$\pm$	0.93	&	318.68	$\pm$	0.90	&	313.14	$\pm$	0.88	&	\foiii\ 5007	&	288.94	$\pm$	1.03	&	284.51	$\pm$	1.02	&	635.83	$\pm$	0.25	&	615.01	$\pm$	0.24	&	458.66	$\pm$	0.28	&	457.08	$\pm$	0.27	\\	
\ffeii\ 5270	&	1.96	$\pm$	0.02	&	1.86	$\pm$	0.02	&	1.44	$\pm$	0.02	&	1.39	$\pm$	0.02	&	2.32	$\pm$	0.04	&	2.21	$\pm$	0.04	&	\ffeii\ 5270	&	2.63	$\pm$	0.05	&	2.52	$\pm$	0.05	&	0.50	$\pm$	0.00	&	0.45	$\pm$	0.00	&	0.53	$\pm$	0.01	&	0.52	$\pm$	0.01	\\	
\fnii\ 5755	&	0.36	$\pm$	0.05	&	0.33	$\pm$	0.05	&	0.30	$\pm$	0.06	&	0.28	$\pm$	0.06	&	0.21	$\pm$	0.03	&	0.19	$\pm$	0.02	&	\fnii\ 5755	&	0.69	$\pm$	0.05	&	0.63	$\pm$	0.05	&	0.52	$\pm$	0.00	&	0.44	$\pm$	0.00	&	0.31	$\pm$	0.01	&	0.31	$\pm$	0.00	\\	
\foi\ 6300	&	6.43	$\pm$	0.04	&	5.48	$\pm$	0.03	&	4.94	$\pm$	0.06	&	4.41	$\pm$	0.06	&	8.41	$\pm$	0.03	&	7.26	$\pm$	0.02	&	\foi\ 6300	&	6.50	$\pm$	0.10	&	5.71	$\pm$	0.09	&	2.67	$\pm$	0.00	&	2.02	$\pm$	0.00	&	3.47	$\pm$	0.01	&	3.36	$\pm$	0.01	\\	
\fsiii\ 6312	&	1.85	$\pm$	0.05	&	1.57	$\pm$	0.04	&	1.82	$\pm$	0.05	&	1.63	$\pm$	0.05	&	1.57	$\pm$	0.08	&	1.35	$\pm$	0.07	&	\fsiii\ 6312	&	1.57	$\pm$	0.09	&	1.38	$\pm$	0.08	&	3.02	$\pm$	0.01	&	2.28	$\pm$	0.01	&	2.39	$\pm$	0.01	&	2.31	$\pm$	0.01	\\	
H$\alpha$	&	343.95	$\pm$	0.41	&	286.02	$\pm$	0.34	&	325.68	$\pm$	0.47	&	286.01	$\pm$	0.41	&	338.56	$\pm$	0.52	&	286.02	$\pm$	0.44	&	H$\alpha$	&	331.75	$\pm$	0.62	&	286.01	$\pm$	0.54	&	393.83	$\pm$	0.10	&	286.03	$\pm$	0.07	&	297.20	$\pm$	0.29	&	286.16	$\pm$	0.28	\\	
\fnii\ 6584	&	25.83	$\pm$	0.07	&	21.44	$\pm$	0.06	&	21.54	$\pm$	0.09	&	18.89	$\pm$	0.07	&	27.50	$\pm$	0.12	&	23.19	$\pm$	0.10	&	\fnii\ 6584	&	28.06	$\pm$	0.14	&	24.15	$\pm$	0.12	&	33.69	$\pm$	0.02	&	24.40	$\pm$	0.01	&	20.34	$\pm$	0.02	&	19.58	$\pm$	0.02	\\	
\fsii\ 6716	&	42.52	$\pm$	0.04	&	34.90	$\pm$	0.03	&	33.87	$\pm$	0.10	&	29.47	$\pm$	0.09	&	47.32	$\pm$	0.05	&	39.50	$\pm$	0.04	&	\fsii\ 6716	&	44.33	$\pm$	0.16	&	37.82	$\pm$	0.13	&	14.91	$\pm$	0.01	&	10.58	$\pm$	0.00	&	23.13	$\pm$	0.02	&	22.21	$\pm$	0.02	\\	
\fsii\ 6731	&	31.70	$\pm$	0.03	&	25.98	$\pm$	0.03	&	24.37	$\pm$	0.09	&	21.19	$\pm$	0.08	&	34.44	$\pm$	0.04	&	28.72	$\pm$	0.04	&	\fsii\ 6731	&	32.29	$\pm$	0.14	&	27.52	$\pm$	0.12	&	13.70	$\pm$	0.01	&	9.70	$\pm$	0.00	&	18.55	$\pm$	0.02	&	17.81	$\pm$	0.02	\\	
\foii\ 7321	&	4.52	$\pm$	0.02	&	3.54	$\pm$	0.01	&	3.81	$\pm$	0.02	&	3.21	$\pm$	0.01	&	4.56	$\pm$	0.03	&	3.65	$\pm$	0.02	&	\foii\ 7321	&	4.45	$\pm$	0.03	&	3.65	$\pm$	0.03	&	3.98	$\pm$	0.08	&	2.60	$\pm$	0.05	&	3.95	$\pm$	0.01	&	3.76	$\pm$	0.01	\\	
\foii\ 7332	&	3.77	$\pm$	0.02	&	2.95	$\pm$	0.01	&	3.13	$\pm$	0.02	&	2.63	$\pm$	0.01	&	3.73	$\pm$	0.03	&	2.98	$\pm$	0.02	&	\foii\ 7332	&	3.42	$\pm$	0.04	&	2.81	$\pm$	0.03	&	3.37	$\pm$	0.11	&	2.20	$\pm$	0.07	&	3.27	$\pm$	0.01	&	3.11	$\pm$	0.01	\\	
\ffeii\ 8619	&	0.53	$\pm$	0.02	&	0.38	$\pm$	0.01	&	0.34	$\pm$	0.02	&	0.27	$\pm$	0.01	&	0.74	$\pm$	0.03	&	0.55	$\pm$	0.02	&	\ffeii\ 8619	&	0.74	$\pm$	0.04	&	0.57	$\pm$	0.03	&	0.32	$\pm$	0.00	&	0.18	$\pm$	0.00	&	0.23	$\pm$	0.00	&	0.21	$\pm$	0.00	\\	
\fsiii\ 9069	&	22.70	$\pm$	0.04	&	15.96	$\pm$	0.03	&	19.24	$\pm$	0.04	&	15.01	$\pm$	0.03	&	19.58	$\pm$	0.05	&	14.19	$\pm$	0.04	&	\fsiii\ 9069	&	17.57	$\pm$	0.05	&	13.23	$\pm$	0.04	&	36.19	$\pm$	0.05	&	19.64	$\pm$	0.03	&	35.16	$\pm$	0.06	&	32.69	$\pm$	0.05	\\ 
F(H$\beta$)	&	46.76	$\pm$	0.13	&	90.45	$\pm$	0.25	&	31.14	$\pm$	0.03	&	49.56	$\pm$	0.05	&	26.08	$\pm$	0.08	&	47.67	$\pm$	0.16	&	F(H$\beta$)	&	14.43	$\pm$	0.06	&	24.53	$\pm$	0.11	&	355.72	$\pm$	0.26	&	1116.45	$\pm$	0.83	&	157.81	$\pm$	0.12	&	180.94	$\pm$	0.14	\\	
E(B-V)	&	0.16			& 				&	0.11	& 	&						0.14	& 						&	E(B-V)	&	0.13	& 	&						0.27	& 	&						0.03	& 					\\		
\enddata          								
\end{splitdeluxetable}


\end{document}